%% file: draft.tex
\documentclass[graybox, secnum]{svmult}

\usepackage{mathptmx}       
\usepackage{helvet}         
\usepackage{courier}        
\usepackage{type1cm}        
\usepackage{makeidx}         
\usepackage{graphicx}        
\usepackage{multicol}        
\usepackage[bottom]{footmisc}
\usepackage{hyperref}        
\usepackage{soul}            
\hypersetup{colorlinks=true,urlcolor=blue}
\usepackage[square,numbers,compress]{natbib}
\usepackage{amssymb}
\usepackage{amsmath}
\usepackage{slashed}
\makeindex             
                       
\usepackage[usenames,dvipsnames]{xcolor}
\usepackage[normalem]{ulem}

\newcommand{\FRT}[1]{\textbf{#1}}
\newcommand{\FR}[1]{\emph{#1}}

\newcommand{\GN}{\ensuremath{G_{\mathrm{N}}}}
\newcommand{\GNast}{\ensuremath{G_{\mathrm{N},\,\ast}}}

\newcommand{\gy}{\ensuremath{g_{\mathrm{y}}}}
\newcommand{\gyast}{\ensuremath{g_{\mathrm{y},\,\ast}}}

\newcommand{\Nscal}{\ensuremath{N_{\mathrm{S}}}}
\newcommand{\Nferm}{\ensuremath{N_{\mathrm{F}}}}
\newcommand{\Nvec}{\ensuremath{N_{\mathrm{V}}}}

\newcommand{\Regk}{\ensuremath{R_k}}

\newcommand{\be}{\begin{equation}}
\newcommand{\ee}{\end{equation}}

\newcommand{\bea}{\begin{eqnarray}}
\newcommand{\eea}{\end{eqnarray}}

\begin{document}

\title*{Asymptotic safety of gravity with matter}

\author{Astrid Eichhorn \thanks{corresponding author} and Marc Schiffer}

\institute{Astrid Eichhorn \at CP3-Origins, University of Southern Denmark, Campusvej 55, 5230 Odense M, Denmark \email{eichhorn@cp3.sdu.dk}
\and Marc Schiffer \at Perimeter Institute for Theoretical Physics, 31 Caroline Street North, Waterloo, ON N2L 2Y5,
Canada \email{mschiffer@perimeterinstitute.ca}}
\maketitle
\abstract{The asymptotic-safety paradigm posits that the symmetry of quantum theories of gravity and matter is enhanced to quantum scale symmetry, i.e., scale symmetry in the presence of quantum fluctuations, at very high energies. To achieve such a symmetry enhancement, the effect of quantum fluctuations must balance out. It is to be expected that such a balance can only be achieved within a set of theories with limited field content and interaction structure. In this chapter, we review how much is known about these limits.\\
From the quantum scale invariant regime, the theory transits to a theory with distinct physical scales -- most importantly masses for various elementary particles -- at low energies. There, quantum scale invariance can leave its imprint in relations between various interactions and mass scales of the theory. These relations can be compared to experimental data, which has two possible implications: first, if the relations do not match the data, the underlying quantum theory of gravity and matter, formulated at and beyond the Planck scale, has been ruled out using experimental data from energies much below the Planck scale. Second, if the relations match the data, the asymptotic-safety paradigm provides a first-principles derivation of free parameters of the Standard Model. Most importantly, this may include the ratios of the Higgs mass to the electroweak scale as well as the value of the finestructure constant. Similarly, theories beyond the Standard Model may come with fewer free parameters than in their effective-field-theory incarnation without gravity. This may lead to an explanation of the smallness of neutrinos masses and predictions for the nature and interactions of dark matter.
}

\section*{Keywords:}Asymptotic safety, quantum gravity and matter, Standard Model, scale invariance,  Beyond Standard Model

\tableofcontents

\input{Input/Motivation}

\input{Input/GravityFP}

\input{Input/GlobalSymm}
\input{Input/Lightfermions}

\input{Input/dgreater4}

\section{
 Towards a UV completion of the Standard Model
}
\input{Input/SMUVcompletion}

\section{Physics beyond the Standard Model}
\input{Input/DarkMatter}
\input{Input/nondarkBSM}
\input{Input/nearperturbative}
\section{Summary, outlook, and open questions}
\input{Input/summary}

\section*{Acknowledgments}
We thank Johanna Borissova for comments on the manuscript.
 A.~E.~is supported by a research grant (29405)
	from VILLUM FONDEN. M.~S.~acknowledges support by Perimeter Institute for Theoretical Physics. Research at Perimeter Institute is supported in part by the Government of Canada through the Department of Innovation, Science and Economic Development and by the Province of Ontario through the Ministry of Colleges and Universities.
\bibliographystyle{spbasic}
\bibliography{references, manuals}
\end{document}

%% file: Input/Motivation.tex
\section{Invitation: Matter matters in quantum gravity}
\subsection{Motivation: Why matter matters}
There are two reasons why matter\footnote{By matter we mean all non-gravitational degrees of freedom, including the scalar, fermionic and gauge fields of the Standard Model as well as potential beyond-Standard-Model fields.} matters in quantum gravity: the first is theoretical and relates to the differences between theories of quantum gravity only versus theories of all fundamental interactions and matter; the second is phenomenological and relates to observational tests of quantum gravity. We expand on both reasons below.

The theoretical search for a quantum theory of gravity is often conducted in a setting without matter. The underlying rationale says that a viable quantum theory of pure gravity can be constructed first, and matter added later, in such a way that key features of the pure-gravity theory remain intact. The rationale would break down in two cases:\footnote{In principle, there is a third scenario, where the pure gravitational theory is not UV-complete, and matter degrees of freedom induce a UV-completion.}
 \\
First, it would break down if key features of matter-gravity theories would be very different from those of pure-gravity theories. An analogy for this case is non-Abelian Yang-Mills theory with and without matter. Without matter, the quantum theory is asymptotically free. With matter, asymptotic freedom can be lost in the quantum theory, changing the very nature of the ultraviolet (UV) completion.  \\
Second, it would break down if the coupling to quantum gravity would not render the matter sector UV complete, because the combined theory would then not be UV complete. In many quantum-gravity approaches, it is argued that UV completeness of the matter sector is not an issue, because there is a fundamental Planck-scale cutoff due to fundamental spacetime discreteness. This is actually insufficient for a properly UV complete theory, because such a theory should not just be free of divergences (in the sense of Landau poles, not in the sense of loop divergences removable by renormalization), but also predictive. An effective field theory of the matter sector which comes with a Planckian cutoff is only predictive at energies much lower than the cutoff. At energies close to the cutoff, an infinite number of interactions, each parameterized by its own coupling, can exist. Unless the combination with quantum gravity provides a predictivity principle, the combined matter-gravity theory is not a proper UV complete theory. Such a predictivity principle either sets all but finitely many couplings to zero at Planckian scales, or provides an infinite number of relations such that a finite number of free parameters remain. 

Observational test of quantum gravity typically rely on the gravitational effect on matter. For instance, potential
quantum-gravity effects  in the very early universe are typically looked for in matter observables, such as the cosmic microwave background. Further, quantum-gravity effects in particle physics or even table-top-experiments rely on the interplay of quantum gravity with matter. Finally, even tests relying on putative pure-gravity observables, such as gravitational waves, are ultimately only accessible to us in experimental setups that rely on the interplay of matter and spacetime, see \cite{Addazi:2021xuf} for an overview.\\
Additionally, many more observational tests become available at low energies, if one explores matter-gravity systems. For pure gravity, the only requirement is that at low energies, it reduces to General Relativity, with higher-order corrections to it being sufficiently small. Current observations constrain curvature-squared couplings to be smaller than $10^{60}$ \cite{Berry:2011pb}, which indicates that those are currently only very weakly constrained. In terms of free parameters, one is essentially left with the Newton coupling (and the cosmological constant) as potentially predictable quantities from a quantum theory of gravity.

For gravity-matter theories, there is an additional requirement, namely that the matter sector reduces to the  Standard Model (SM), plus potential dark matter and other beyond Standard Model (BSM) fields. This provides many more and stronger constraints than the pure-gravity-setting does. In terms of free parameters, one gains the additional 19 free parameters of the SM as potentially predictable quantities from a quantum theory of gravity with matter.

\subsection{Matter matters in asymptotically safe gravity}
Here we focus on the asymptotic-safety paradigm. Its starting point is the perturbative nonrenormalizability of gravity, which means that gravity loses predictivity at the Planck scale, because the couplings of all possible interactions are free parameters. Asymptotic safety restores predictivity because an additional symmetry, not easily seen in standard perturbation theory (see, however, \cite{Niedermaier:2009zz}), holds above the Planck scale: Quantum scale symmetry means that all couplings, made dimensionless by division through an appropriate power of a scale, are constant. This is referred to as a fixed point in the Renormalization Group flow, which describes how the theory changes with respect to an energy scale. As an analogy, one may view the Renormalization Group flow as the mathematical counterpart of a microscope, with which one can change the resolution scale at which a system is considered. At a fixed point, changes of the resolution scale do not result in changes of the system, i.e., scale symmetry -- a form of self-similarity -- is achieved.\\
Just like any symmetry in a QFT, quantum scale symmetry relates the values of couplings to each other. The special aspect of quantum scale symmetry is that relations continue to hold at lower energy scales/larger distances scales than the Planck scale, where quantum scale symmetry is no longer realized. The reason for these relations is that departure from quantum scale symmetry is only achievable in a QFT, if particular interactions (so-called relevant ones) are present. These relations restore predictivity and make a fundamental QFT of gravity conceivable.
Examples of relevant interactions and the resulting relations will be shown in \autoref{sec:SMUVcompletion} below.\\

In the asymptotic-safety paradigm, evidence is starting to accumulate for the following scenario:
\begin{itemize}
\item An asymptotically safe pure-gravity fixed point can be step by step deformed to an asymptotically safe fixed point in theories which contain matter degrees of freedom, most importantly the SM. We discuss this in detail in \autoref{sec:matterongrav}.
\item Gravitational fluctuations induce new interactions in the matter sector. All induced interactions respect the symmetries of the kinetic terms, which includes global symmetries. Hence, the properties of the asymptotically safe fixed point may be in part determined by those global symmetries.\footnote{In the SM, the global symmetries of the kinetic terms are typically broken by the marginal interactions. The presence of global symmetries may therefore be more relevant for BSM physics, e.g., in a dark sector.} We discuss this in detail in \autoref{sec:GlobalSymms}. 
\item Under the impact of gravitational fluctuations, the SM becomes UV complete. The Landau poles that the SM on its own contains are substituted by an asymptotically safe, quantum scale invariant regime. 
\item In the infrared (IR),  some of the free parameters of the SM, i.e., some of its perturbatively renormalizable couplings, become calculable quantities that can be predicted from first principles. The technical reason is that they are irrelevant couplings in the asymptotically safe regime.
This provides observational tests of asymptotically safe matter-gravity systems. We discuss this in detail in \autoref{sec:SMUVcompletion}.
\item Beyond the SM, not all theories are asymptotically safe. This provides predictions for ongoing and future searches for new physics, including the nature of dark matter. We discuss this in detail in \autoref{sec:DarkBSM}.
\end{itemize}

\subsection{Interplay of quantum gravity and matter and structure of this chapter}

\begin{figure}[!t]
	\centering
	\includegraphics[width=0.8\linewidth, clip=true, trim=2cm 7cm 10cm 6cm]{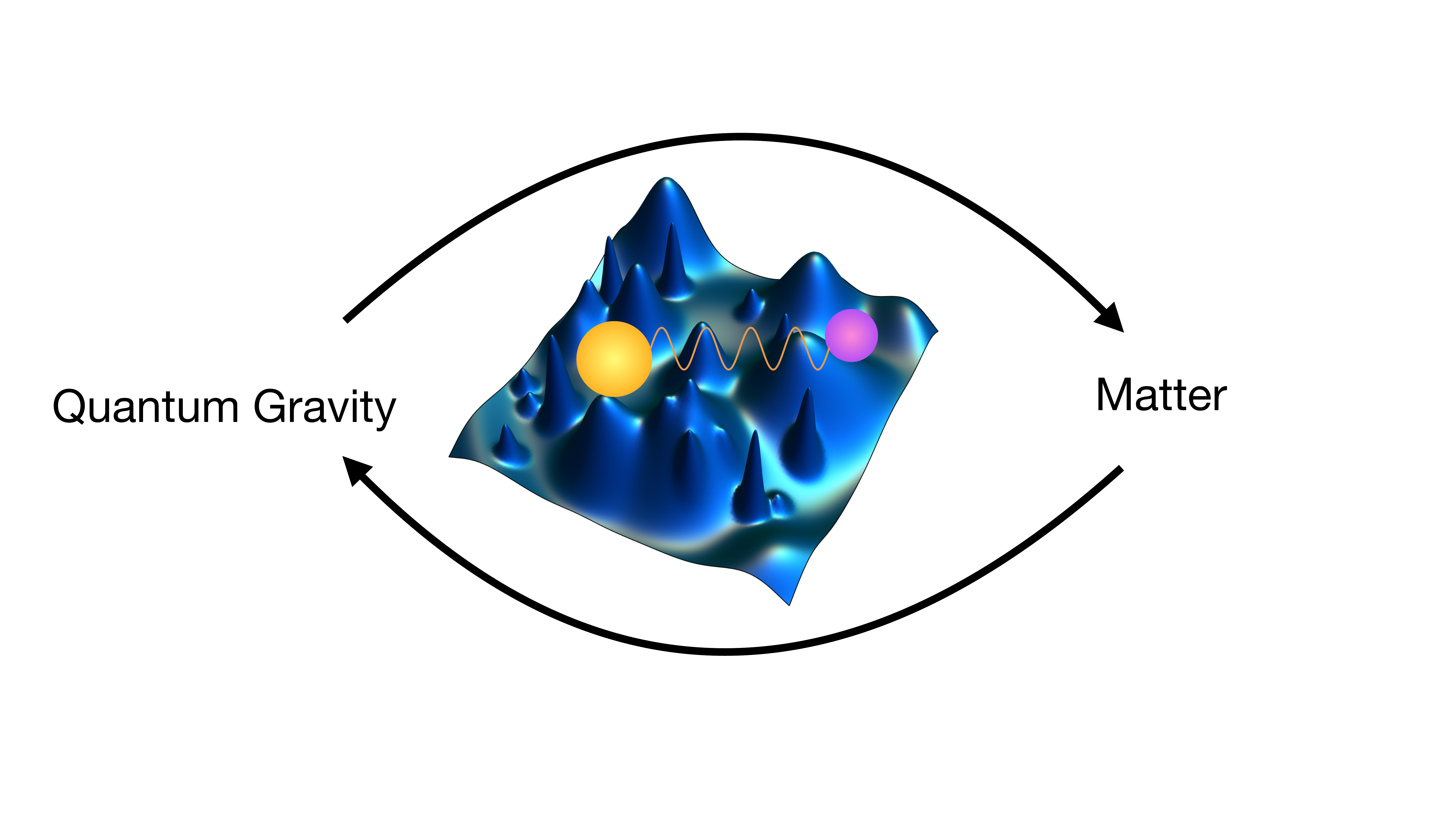}
	\caption{\label{fig:interplay}The interplay of quantum gravity and matter can as a first step be  approximately ``factorized", and the effect of matter on the gravitational fixed point be considered separately from the effect of quantum gravity on matter. Therefore, we construct this chapter by considering the impact of matter on quantum gravity first, and the effect of quantum gravity on matter second, adding more and more realistic interactions and field content as we go along.}
\end{figure}

This chapter is structured as follows: First, we introduce key concepts of asymptotically safe quantum gravity, as well as the most important methods to explore it. Then, we start from investigating asymptotic safety in a system with few interactions and step by step add interactions, and later also fields: First, we review how matter that is non-interacting impacts a gravitational fixed point. Then, we discuss the role of symmetries and how they determine which interactions of matter are unavoidably present. Finally, we add those interactions that need to be present in order to obtain a viable phenomenology, including the SM and some physics beyond the  SM.\\ 
It is in fact a nontrivial consequence of the methodology used, namely the functional Renormalization Group, that the two sides of the interplay of quantum gravity and matter can be ``factorized" at the level of calculations, at least approximately:\footnote{These approximations consist in neglecting the effect of non-minimal interactions, as well as the impact of the anomalous dimensions of matter fields on the gravitational couplings, and vice versa.} Within approximations, one can consider first the impact of matter fields on quantum gravity and second the effect of quantum gravity and matter, cf.~\autoref{fig:interplay}. These two independent studies are combined in a second step, where the fully coupled system is investigated.

We keep this introduction as non-technical and pedagogical as possible, and highlight the basic mechanisms behind the results. In several sections we provide \emph{Further reading} paragraphs, where we discuss some more technical details or the relation to other results.

\subsection{Methods to investigate asymptotically safe gravity-matter systems}
\label{sec:methods}
Since asymptotic safety makes an interacting theory scale-invariant, it is characterized by an interacting fixed point of the Renormalization Group flow, where all scale-dependence is lost. A fixed point is called interacting, if interactions are present, i.e., (some of) the couplings are non-zero.
Methods to explore such a scale invariant regime for quantum gravity and matter include 
\begin{enumerate}
\item[i)] perturbative methods,
\item[ii)] lattice methods,  
\item[iii)] functional methods.
\end{enumerate}
Due to the interacting nature of the fixed point, couplings are generically nonzero. Thus, perturbative methods, which explore the theory in the vicinity of the non-interacting limit, can only be used to limited extent, see \cite{Niedermaier:2009zz, Niedermaier:2010zz} for examples. \\
However, an interacting theory can often become perturbative, when a parameter in the theory is taken to a particular limit, e.g., a limit of many fields, or of a special spacetime dimensionality, see, e.g., \cite{Eichhorn:2018yfc} for examples. In gravity, $d=2$ is a special dimensionality, because it makes the Newton coupling dimensionless. In the language of statistical physics, we therefore call $d=2$ the critical dimension $d_{\rm c}$ for the Newton coupling.
It is a general feature that theories that are asymptotically free in their critical dimension $d_{\mathrm{c}}$ are asymptotically safe in $d=d_{\mathrm{c}}+\epsilon$. Because gravity is also topological in $d=2$, the situation is somewhat special. However, calculations in $d=2+\epsilon$ show a beta function with a negative quadratic term, i.e., the limit $\epsilon \rightarrow 0$ gives a beta function that exhibits asymptotic freedom. At $\epsilon>0$ it therefore features an asymptotically safe fixed point. Whether or not  $\epsilon$ can be extended to $\epsilon=2$ is an open question \cite{Gastmans:1977ad, Christensen:1978sc, Kawai:1992np, Kawai:1993mb, Aida:1996zn}.\\

Lattice methods are powerful tools to discover and characterize scale invariant regimes. On the lattice, an asymptotically safe fixed point generates a higher-order phase transition in the phase diagram (spanned by the couplings of the system). At the phase transition, the system looses any memory of scales due to diverging correlation lengths, which is characteristic for critical phenomena. \\
For gravity, one faces the challenge that the lattice itself has to become dynamical, because gravity is not a theory \emph{on} spacetime, but a theory \emph{of} spacetime, where the path integral sums over different configurations of the lattice.
In addition, one may face the challenge that a lattice can (but need not) break spacetime symmetries, e.g., a regular lattice breaks local Lorentz invariance.
In dynamical triangulations, the $d$-dimensional spacetime is discretized in terms of $d$-dimensional simplices, and the path integral for quantum gravity is expressed as a sum over all possible combinatorics of such triangulations. These are weighted by the exponential of their respective action. In the Euclidean setting of Dynamical Triangulations, one can run Monte-Carlo simulations of the path integral directly; in the Lorentzian setting of Causal Dynamical Triangulations, one builds configurations so that they can be Wick-rotated, thereby transforming the Lorentzian path integral to a statistical generating functional that can be explored with Monte-Carlo techniques.  Evidence for the existence of a higher-order phase transition, which is necessary to take the continuum limit, has been collected in the case of Causal Dynamical Triangulations \cite{Loll:2019rdj}, see also \cite{Catterall:2018dns, Dai:2021fqb, Bassler:2021pzt, Asaduzzaman:2022kxz} for recent studies in Euclidean Dynamical Triangulations.  For further discussions, we refer the reader to the section on Dynamical Triangulations of this handbook. \\
In addition, lattice simulations may be based on Regge calculus, which varies not the triangulation, but the edge lengths of the building blocks to sum over all spacetime configurations \cite{Hamber:2009mt}.
Further, a combinatorial approach based on random graphs may also feature a second-order phase transition \cite{Kelly:2018diy, Kelly:2019rpx}, as could another combinatorial approach based on tensor models \cite{Eichhorn:2019hsa}. \\
These phase transitions need not necessarily constitute the same universality class\footnote{ The terminology ``universality class" is taken from statistical physics/condensed matter, where interacting fixed points have long played an important role, because they characterize continuous phase transitions. A universality class is determined by the dimensionality, field content and symmetries, and quantitatively described by the set of critical exponents, which describe the scaling behavior of physical quantities in the vicinity of the phase transition.} that is commonly known as ``asymptotically safe gravity", i.e., while they may be asymptotically safe in a technical sense, their emergent physics may differ from that encoded in the continuum functional approach we discuss below. The decision whether or not the physics agrees between such different approaches which all search for a second-order phase transition, can be based on sufficiently precise calculations of the critical exponents, which uniquely characterize the universality class of a phase transition.\\
Finally, it has been proposed in \cite{Eichhorn:2017bwe, Eichhorn:2019xav} that causal sets, reviewed in another section of this book, may also shed light on asymptotic safety: it is usually assumed that the discreteness scale in causal sets is fixed. However, if one can take it to zero at a higher-order phase transition, one obtains a continuum limit in a genuinely Lorentzian setting.\\
In summary, while current explorations of asymptotically safe gravity are mainly based on functional methods reviewed below, there is certainly scope to extend the toolbox and achieve complementary insights based on other methods or by ``repurposing" other approaches, such as the causal-set approach.
\\

\begin{figure}[!t]
\includegraphics[width=\linewidth]{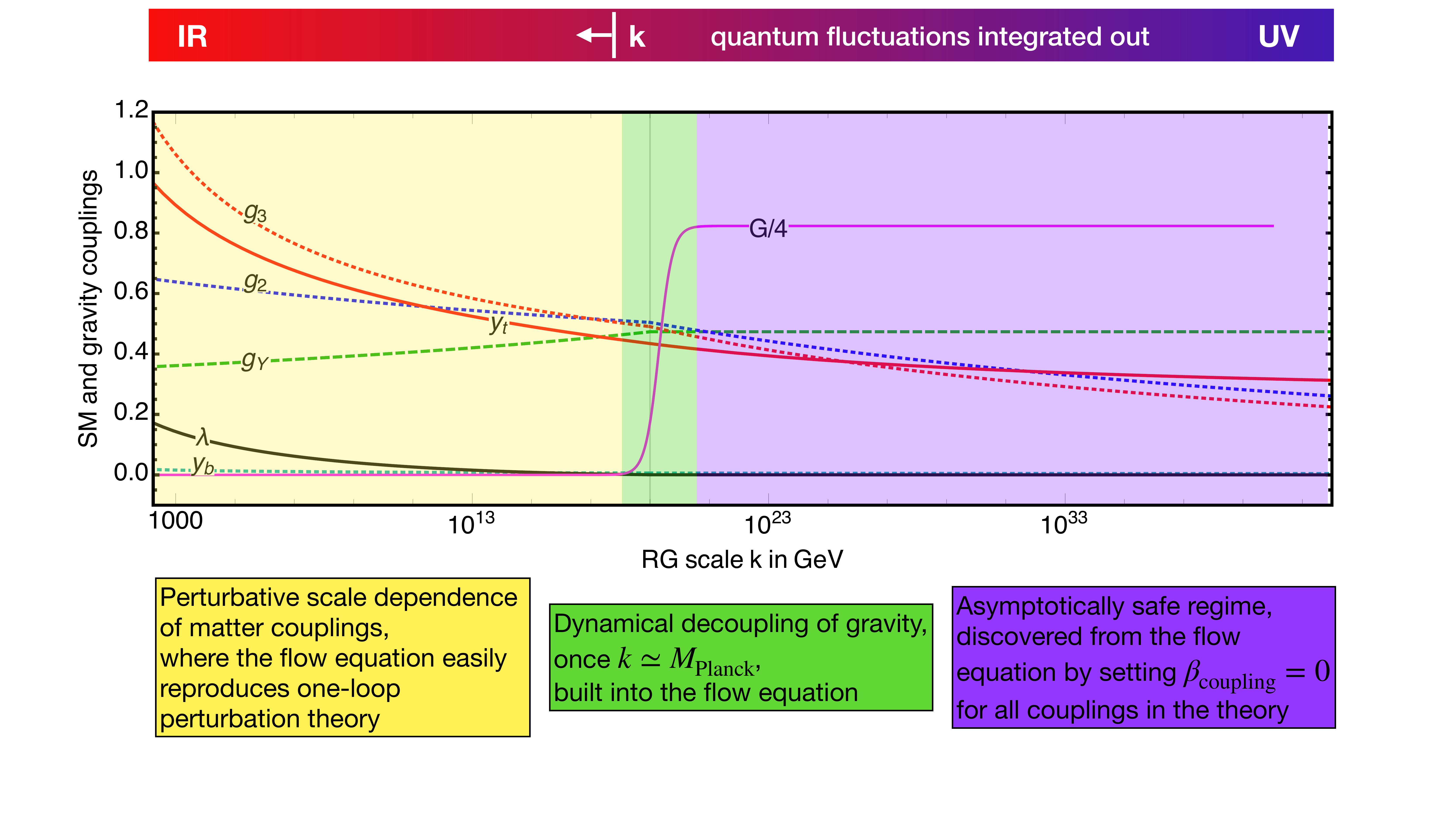}
\caption{\label{fig:RGflowschematic} We illustrate the functional RG flow of gravity coupled to the SM, which has three regimes as a function of $k$: in the UV, asymptotic safety is realized and discoverable through the flow equation by setting $\beta=0$ for all couplings. At the Planck scale, gravity decouples dynamically. Below the Planck scale, the SM couplings (here shown are the three gauge couplings $g_i$, the two largest Yukawa couplings $y_t$ and $y_b$ and the Higgs quartic self-interaction $\lambda$) exhibit a perturbative scale dependence and the flow equation easily reproduces one-loop perturbation theory.  (At even lower scales, in the very deep IR, perturbation theory no longer describes QCD; the FRG can also be used in that regime successfully, see, e.g., \cite{Dupuis:2020fhh} for reviews. )}
\end{figure}

The ideal tool to probe an asymptotically safe theory can do two things: first, it can probe the  UV regime to search for scale symmetry. Second, it can connect a scale-symmetric regime in the UV to emergent phenomenology in the IR.\\
Functional methods are such a tool, because they allow to extract the scale dependence of a system within and beyond perturbation theory. Since most research on asymptotically safe gravity-matter systems relies on functional methods, in particular on the functional renormalization group (FRG), we will briefly introduce the method and some notation in the following.\\
The key object in the FRG is the scale-dependent effective action $\Gamma_k$. It is a scale-dependent counterpart of the classical action, i.e., it gives rise to the equations of motion for the expectation value of the field. 
As a function of the RG scale $k$, it interpolates between the microscopic action $\Gamma_{k\to\infty}$ when no quantum fluctuations are integrated out, and the full quantum effective action $\Gamma_{k\to0}$ when all quantum fluctuations are integrated out.\footnote{The microscopic action $\Gamma_{k\to\infty}$ is sometimes also referred to as the classical action. This is technically not completely accurate, see \cite{Manrique:2009tj, Morris:2015oca, Fraaije:2022uhg} for the relation between bare (or classical) action and $\Gamma_{k\to\infty}$. Further, the term ``classical action" can be conceptually confusing in the gravitational context: Observationally, we know that the action that describes gravity at low curvature scales is $S_{\rm EH}$, the Einstein-Hilbert action of GR and we usually refer to it as the classical action, given that GR does not contain quantum effects. However, in the context of asymptotic safety it is not correct that $S_{\rm EH}$ is the action that is ``quantized" in the sense of a path integral $Z= \int \mathcal{D}g_{\mu\nu}\, e^{i\, S_{\rm EH}}$. Instead, $S_{\rm EH}$ should be recovered as the leading approximation to $\Gamma_{k\rightarrow 0}$ in the limit of low curvature.} We are interested in the scale-derivative of $\Gamma_k$, i.e., in $k\, \partial_k \Gamma_k$, because it allows us to do the two things we are interested in: first, finding whether there is a scale-invariant regime, related to  $k\, \partial_k \Gamma_k=0$, and second, integrating $k\, \partial_k \Gamma_k$ from the scale-invariant regime in the limit $k \rightarrow \infty$ to $k=0$ to investigate the phenomenology of asymptotic safety.\\
The FRG indeed provides a flow equation for $\Gamma_{k}$, which reads \cite{Wetterich:1992yh, Morris:1993qb, Ellwanger:1993mw, Reuter:1996cp}
\begin{equation}
\label{eq:floweq}
k\partial_k\,\Gamma_k=\frac{1}{2}\mathrm{sTr}\left[\left(k\partial_k\,\Regk\right)\left(\Gamma_k^{(2)}+\Regk\right)^{-1}\right]\,.
\end{equation}
Here, the right-hand-side integrates over quantum fluctuations, with those fluctuations with momenta of the order of $k$ contributing most to the change of $\Gamma_k$ at $k$. The technical ingredients of the right-hand side are:
the second functional derivative of $\Gamma_k$ with respect to all fields of the system, $ \Gamma_k^{(2)}$, the regulator functional $\Regk$  and a supertrace $\mathrm{sTr}$ which sums/integrates over all discrete/continuous indices. 
The combination $\left(\Gamma_k^{(2)}+\Regk\right)^{-1}$ is the regularized propagator. In the propagator, $\Regk$ acts akin to a scale-dependent mass-term, because it appears just like a standard mass term would, together with the momentum, in the schematic form $p^2+\Regk$ or $p^2+m^2$, respectively. The difference to a standard mass term is that it is not constant, but only present for low-energy modes, $p^2<k^2$. Therefore, these are suppressed; and a $\sim 1/p^2$ divergence, i.e., an IR divergence, is avoided. Thus, technically speaking, the regulator ensures the IR-finiteness of the flow equation. In addition, the physical masses of modes also enter the propagator and ensure that a mode decouples dynamically, once $k$ falls below the mass scale. This is relevant even for massless fluctuations, which couple through a mass-like term: gravity decouples automatically, once $k \simeq M_{\rm Planck}$, see \autoref{fig:RGflowschematic} for a schematic illustration of the functional RG flow and the different regimes for gravity-matter theories.
In the numerator, the scale derivative $k\partial_k\,\Regk$ suppresses quantum fluctuations of high momenta.
The two occurrences of $\Regk$ therefore realize the Wilsonian idea of integrating out quantum fluctuations according to their momentum in a step-wise fashion.
To achieve this, the regulator has to satisfy several conditions, most importantly $\Regk(p^2)>0$ for $p^2< k^2$ (where $p$ denotes a four-momentum) to suppress low-energy modes, and $\Regk (p^2)=0$ for $p^2>k^2$, such that the high-energy modes are also suppressed in the flow equation and only modes with $p^2 \approx k^2$ remain. Since quantum fluctuations are integrated out according to their four-momentum squared, the FRG is best employed in Euclidean settings, where a cutoff on the four-momentum squared indeed distinguishes UV and IR. For steps towards a generalization to Lorentzian spacetimes in the context of quantum gravity, see \cite{Manrique:2011jc, Bonanno:2021squ, Fehre:2021eob}.

Structurally, the flow equation \eqref{eq:floweq} is of one-loop form, but in terms of the full and regularized propagator, such that it is valid beyond perturbation theory. In fact, $\Gamma_k^{(2)}$ is not just the perturbative expression $p^2+m^2$ (or the appropriate version for fields which are not scalar), but contains higher-order terms, i.e., it is the inverse propagator fully dressed by quantum fluctuations. Despite this difference, calculations share similarities with perturbative loop calculations, and, most importantly, are feasible for a wide range of theories, including gravitational ones.\\

The flow equation is successfully employed in various physical scenarios which are governed by an interacting fixed point, see, e.g., \cite{Dupuis:2020fhh} for an overview.\\
Quantum fluctuations generate all interactions compatible with the symmetries of a system. Thus,
the scale dependent effective action $\Gamma_k$ contains all these interactions, and the scale-dependence of the corresponding couplings can be extracted by projecting the left and right-hand sides of the flow equation \eqref{eq:floweq} on the corresponding interaction. In practise this is done by taking functional derivatives with respect to the fields. In this way, one obtains the beta function for the coupling of an interaction term.\\
In practise, only a subset of interactions can be accounted for.
Practical computations therefore have to restrict the set of terms -- kinetic terms and interactions  --  that enter $\Gamma_k$ to a typically finite subset. This constitutes a truncation in the space of all interactions and introduces a systematic uncertainty in the results obtained within the method. Extending the truncation by adding more and more interactions into the system decreases this systematic uncertainty, see \cite{Balog:2019rrg} for an example.\\
Crucially, completely random choices of interactions do of course not lead to robust results. Instead, reliable truncations are based on physical insight into the nature of the system, for example, regarding the degree of a system's non-perturbativeness. Systematic expansion schemes can be employed, e.g., a derivative expansion (including all orders in the field, but subsequent orders in derivatives), a vertex expansion (including all orders in derivatives, but subsequent orders in fields), or an expansion based on canonical dimension (including the most relevant interactions first). Calculations in gravity-matter systems are typically based on the last scheme; based on the assumption that the fixed point is near-perturbative, i.e., the canonical dimension remains a useful ordering principle. We will get back to this point at the very end of this chapter, see \autoref{sec:ner-pert}, to review whether this assumption is self-consistent and supported by the results obtained by basing truncations on it.\\

In the discussion above, we have referred to the four-momentum of modes, which is of course a notion closely tied to flat spacetime. On a curved background, it can be generalized: just as $p^2$ are the eigenvalues of the flat-spacetime d'Alembertian, $\lambda_p$ are the eigenvalues of a suitable curved-spacetime d'Alembertian $\Box.$ However, if spacetime itself is fluctuating, the definition of a suitable generalization of momenta is non-trivial. Therefore, 
when applying the FRG to gravity, one has to specify an auxiliary background metric, with respect to which the momenta of quantum fluctuations can be measured \cite{Reuter:1996cp}. This is best implemented via the background-field method, see \cite{Martini:2022sll} for details, because this method also allows to preserve a background diffeomorphism invariance, which becomes full diffeomorphism invariance when the auxiliary background metric is removed.\\
Therefore, we split the metric into background metric $\bar{g}_{\mu\nu}$ and a fluctuation field $h_{\mu\nu}$, for example with a linear\footnote{Other choices are possible, most popular among them an exponential split.} split
\begin{equation}
\label{eq:linsplit}
g_{\mu\nu}=\bar{g}_{\mu\nu}+h_{\mu\nu}\,,
\end{equation}
where the metric fluctuation $h_{\mu\nu}$ is not restricted to be perturbatively small. \\
We emphasize that the background is merely a technical ingredient and does not even need to be specified in calculations (although in practise, many studies choose to specify it for reasons of technical simplicity). Given the background metric, the regulator can then be introduced as a mass-like term for $h_{\mu\nu}$, while $\bar{g}_{\mu\nu}$ is specified to measure the ``momentum" of $h_{\mu\nu}$. Due to this regularization, $\Gamma_k$ depends on $\bar{g}_{\mu\nu}$ and $h_{\mu\nu}$ independently. To zeroth order in $h_{\mu\nu}$, i.e., $\Gamma_k[\bar{g};h=0]$ features the so-called \emph{background couplings}, which  are the physical ones which enter observable quantities. However, due to the regularized propagator in the flow equation, their scale dependence is driven by those couplings of $\Gamma_k[\bar{g};h]$ appearing at higher orders in an expansion in $h_{\mu\nu}$, the so-called \emph{fluctuation couplings}. In a wide-spread approximation-scheme, the \emph{background-field approximation}, this difference is neglected when computing the scale dependence of background couplings. We will provide further details in the \emph{further reading} part of this section, and refer the reader to the chapter on the vertex expansion.\\

Due to the formulation as a QFT of the metric, it is rather straightforward to couple matter degrees of freedom to gravity in asymptotic safety. When investigating asymptotically safe quantum gravity within the FRG framework, this is especially true, since the continuum formulation allows using standard formulations of scalars, gauge fields and fermions. This is in contrast with other approaches to quantum gravity, where the mere definition of matter fields can be more involved.\\

In summary, this provides a flow equation for gravity-matter systems, which allows to first search for scale-symmetry, and second, start from a scale-symmetric regime in the UV and integrate out all quantum fluctuations to obtain the resulting effective dynamics $\Gamma_{k\rightarrow 0}$, which can be compared to observations.
\\

\FRT{Further reading:}\\
\FR{Background field approximation and fluctuation computations}\\
Due to the regulator, (and because one has to gauge-fix $h_{\mu\nu}$,) $\Gamma_k$ depends on $\bar{g}_{\mu\nu}$ and $h_{\mu\nu}$ independently, i.e., $\Gamma_k=\Gamma_k[\bar{g};h]$.
$\Gamma_k[\bar{g};h]$ can be expanded as a series in metric fluctuations $h_{\mu\nu}$, with different couplings at each order in $h_{\mu\nu}$. The zeroth order in this expansion, i.e., $\Gamma_k[\bar{g};h=0]$, contains the so-called background couplings, whose scale dependence is driven by the couplings appearing in higher-order terms of the expansion, the so-called fluctuation couplings. It is a wide-spread approximation to neglect the difference between background and fluctuation couplings, when computing the scale dependence of $\Gamma_k[\bar{g};h=0]$. We call this the background-field approximation. Computations which go beyond the zeroth order in the expansion in metric fluctuations will be referred to as fluctuation computations, see also \cite{Pawlowski:2020qer}.\\
Non-trivial symmetry identities, the so-called Nielsen, or split-Ward identities, restore background independence, see, e.g., \cite{Manrique:2009uh, Pawlowski:2020qer}. These identities encode the difference between correlation functions of the background field and correlation functions of the fluctuation field.\\
In the presence of a fluctuation field $h_{\mu\nu}$, the scale-dependent effective action can be expanded in terms of a vertex expansion as
\begin{equation}
	\label{eq:vertexp}
	\Gamma_k[\bar{g},h]=\Gamma_k[\bar{g},0]+\sum_{n=1}^{\infty}\left(\frac{\delta^n \Gamma_k[\bar{g},h]}{\delta h_{\gamma_1\delta_1}\dots \delta h_{\gamma_n\delta_n}}\bigg|_{h=0}\right)h_{\gamma_1\delta_1}\dots\, h_{\gamma_n\delta_n}\,.
\end{equation}
The first term in this expansion depends only on the background metric, and we refer to the couplings in this term as background couplings. These couplings are the physical couplings that eventually enter observable quantities. The second term in Eq.~(\ref{eq:vertexp}) is the sum over $n$-point vertices, which generally have different scale dependences than the background couplings. We refer to the couplings appearing in these $n$-point vertices as fluctuation couplings. We can see from Eq.~(\ref{eq:vertexp}) and the flow equation Eq.~(\ref{eq:floweq}) that the scale dependence of the $n$-point vertex depends on the $n+1$ and the $n+2$ point vertex. In practical computations, the tower of scale-dependent couplings is therefore truncated by identifying the couplings appearing in the $n+1$ and $n+2$-point vertices with those of the $n$-point vertex.\\
As an alternative to the expansion in background metric and fluctuation field, one can also work in a bimetric setting, with background metric and full metric. For gravity-matter systems, this has been implemented in \cite{Manrique:2010mq}, but a majority of works in this context uses the fluctuation field instead. A main reason for this choice is that the scale dependence of the fluctuation couplings can be extracted by choosing a flat background metric, which is technically advantageous.\\
While the physical couplings are only contained in $\Gamma_k[\bar{g},0]$, their scale dependence is driven by the fluctuation couplings. The \emph{background field approximation} only computes the scale dependence of these physical couplings and identifies all fluctuation couplings with the corresponding background coupling. In particular, the momentum-independent part of the two-point function is identified with the cosmological constant $\bar{\Lambda}$. This approximation scheme allows computing the scale-dependence of curvature operators of high powers, see, e.g., \cite{Falls:2013bv, Falls:2017lst, Kluth:2020bdv}, and even of form factors, see, e.g.,  \cite{Knorr:2019atm, Knorr:2022dsx}. It also allows to extract the scale dependence on non-trivial and even arbitrary backgrounds, see, e.g., \cite{Benedetti:2010nr, Falls:2020qhj, Sen:2021ffc}.
The background field approximation has the virtue of relying on background diffeomorphism invariance by extracting the scale-dependence at vanishing fluctuation field $h$. However, it neglects the difference between the fact that the full effective action individually depends on the background metric $\bar{g}$ and metric fluctuations $h$ and therefore assumes a trivial realization of the so-called Nielsen identities.\\
Fluctuation computations focus on the computation of the scale dependences included in the second term of Eq.~(\ref{eq:vertexp}). The tower of $n$-point vertices is typically truncated at some order $m$, and the couplings appearing in the $m+1$ and $m+2$-point vertices are identified with those of the $m$-point vertex.

\subsection{Fundamentals}
There are a few notions that will be key to this whole chapter. We discuss these fundamentals here.\\

\emph{The direction of the RG flow}\\
Which direction of the RG flow should we think of, when talking about asymptotic safety? It is tempting to think about the RG flow from IR to UV, because in this way we extrapolate from known and measured physics into the unknown. Indeed, asymptotic safety is sometimes discussed in this way.
However, this direction is not physically meaningful, because in nature, microphysics determines macrophysics and not the other way around. Thus, a meaningful RG flow always starts at high energies and goes towards low energies, where the high-energy physics has low-energy consequences.
\\

\begin{figure}[!t]
\begin{center}
\includegraphics[width=0.9\linewidth]{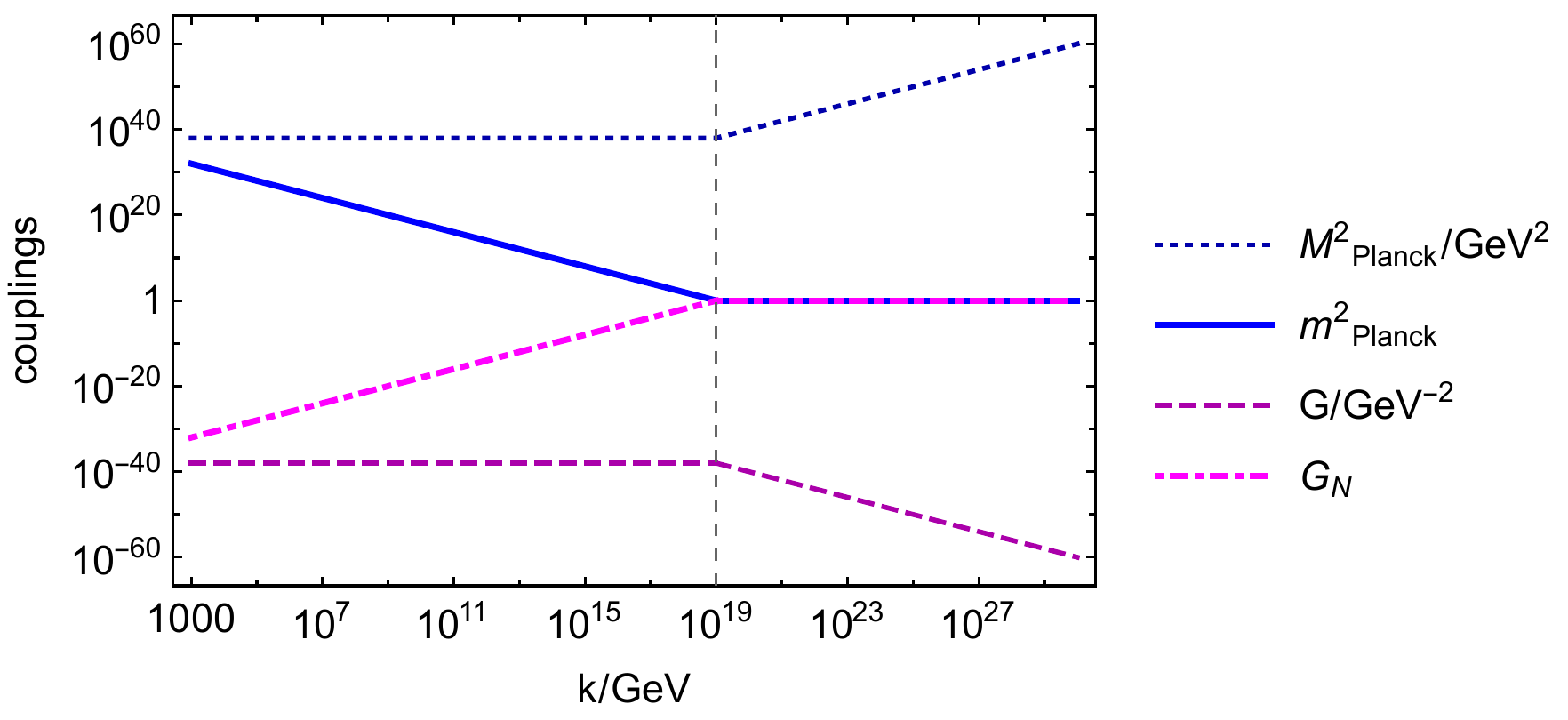}
\end{center}
\caption{\label{fig:Gdimfulldimless} We show the dimensionful Planck mass ($M_{\rm Planck}^2$ and its dimensionless counterpart $m_{\rm Planck}^2 = M_{\rm Planck}^2/k^2$), as well as the dimensionful Newton coupling $G$ and its dimensionless counterpart $G_N= G\cdot k^2$. Beyond the Planck scale, at $10^{19}\, \rm GeV$, the dimensionless quantities are constant and the dimensionful ones show scaling behavior. In the limit $k \rightarrow \infty$, the theory therefore emerges from a regime without scales, because all masses scale to infinity.\\
Thinking about the theory from the IR to the UV, as is sometimes done, one expects a strong-coupling regime to set in at the Planck scale. Instead, a scaling regime sets in, which is characterized by $G_N$ decreasing. Thus, in asymptotic safety, gravity is more weakly coupled than expected, which can be viewed as a reason why the continuum spacetime picture for the gravitational interaction continues to hold and does not need to be substituted by a description of radically different character. Similarly, the Planck mass, at which one expects gravity to become strongly coupled, ``runs away", once $k^2=10^{19}\, \rm GeV$ is reached, i.e., the theory may dynamically protect itself from strong-coupling phenomena.}
\end{figure}

\emph{Quantum scale symmetry -- what is constant?}\\
Asymptotic safety is an enhancement of the symmetry of the QFT. The added symmetry is quantum scale symmetry, which means that couplings are non-vanishing and constant when the energy scale is changed. In a classical field theory,  requiring constant couplings would be a trivial requirement; in a quantum theory, it is not. In a quantum theory, quantum fluctuations screen or anti-screen couplings and thereby generate a scale dependence. Asymptotic safety means that this scale dependence vanishes \emph{in the dimensionless counterparts of couplings}. This is a key difference to what is often referred to as scale invariance in the literature. Because scale invariance is, loosely translated, the absence of distinct physical scales, scale invariance is often taken to mean that there cannot be dimensionful couplings in the theory. This statement is true in quantum scale symmetry, but in a more subtle way: when a dimensionful coupling vanishes in the UV, $\bar{g}=0$, then its dimensionless counterpart, $g = \bar{g}\, k^{-d_{\bar{g}}}$, need not vanish: the RG scale $k$ is taken to make $g$ dimensionless; and when, e.g., the dimension of the coupling, $d_{\bar{g}}$, is negative, then $\bar{g}$ vanishes in the limit $k \rightarrow \infty$, even if $g$ is nonzero in that limit. This is the sense in which quantum scale symmetry is a scale symmetry: the dimensionful quantities scale to zero (or infinity) if one takes the formal limit $k \rightarrow \infty$, so that no scales are present, cf.~\autoref{fig:Gdimfulldimless}. Nevertheless, the dimensionless counterparts of these dimensionful quantities remain non-vanishing and constant. In this sense, asymptotic safety is a generalization of asymptotic freedom, which is familiar from Quantum Chromodynamics. In asymptotic freedom the coupling vanishes in the $k \rightarrow \infty$ limit. In those systems classical scale invariance is restored, since the classical theory only contains dimensionless quantities, and quantum fluctuations vanish for $k \rightarrow \infty$.\\

\begin{figure}[!t]
\includegraphics[width=0.45\linewidth]{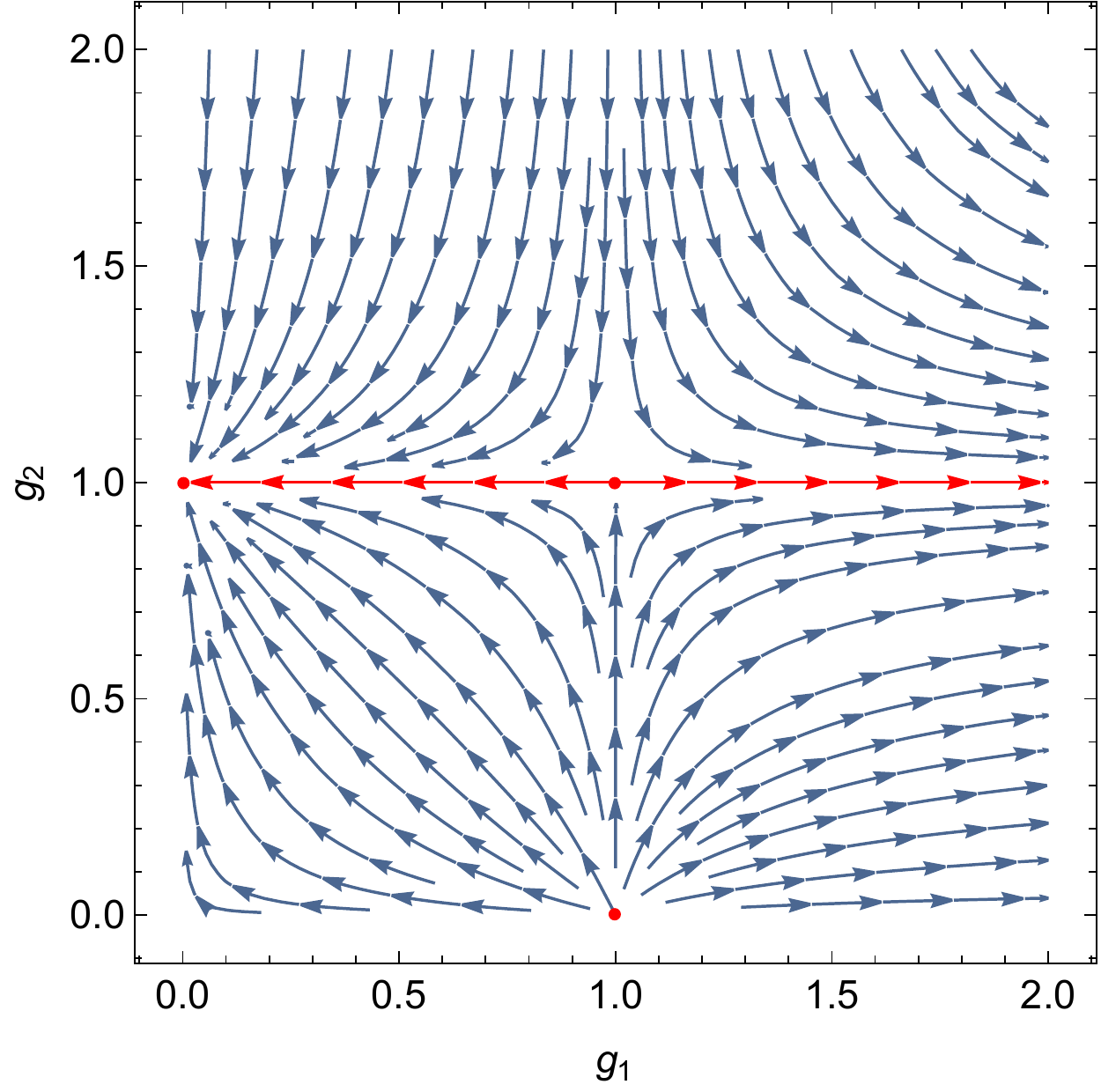}\quad\includegraphics[width=0.45\linewidth]{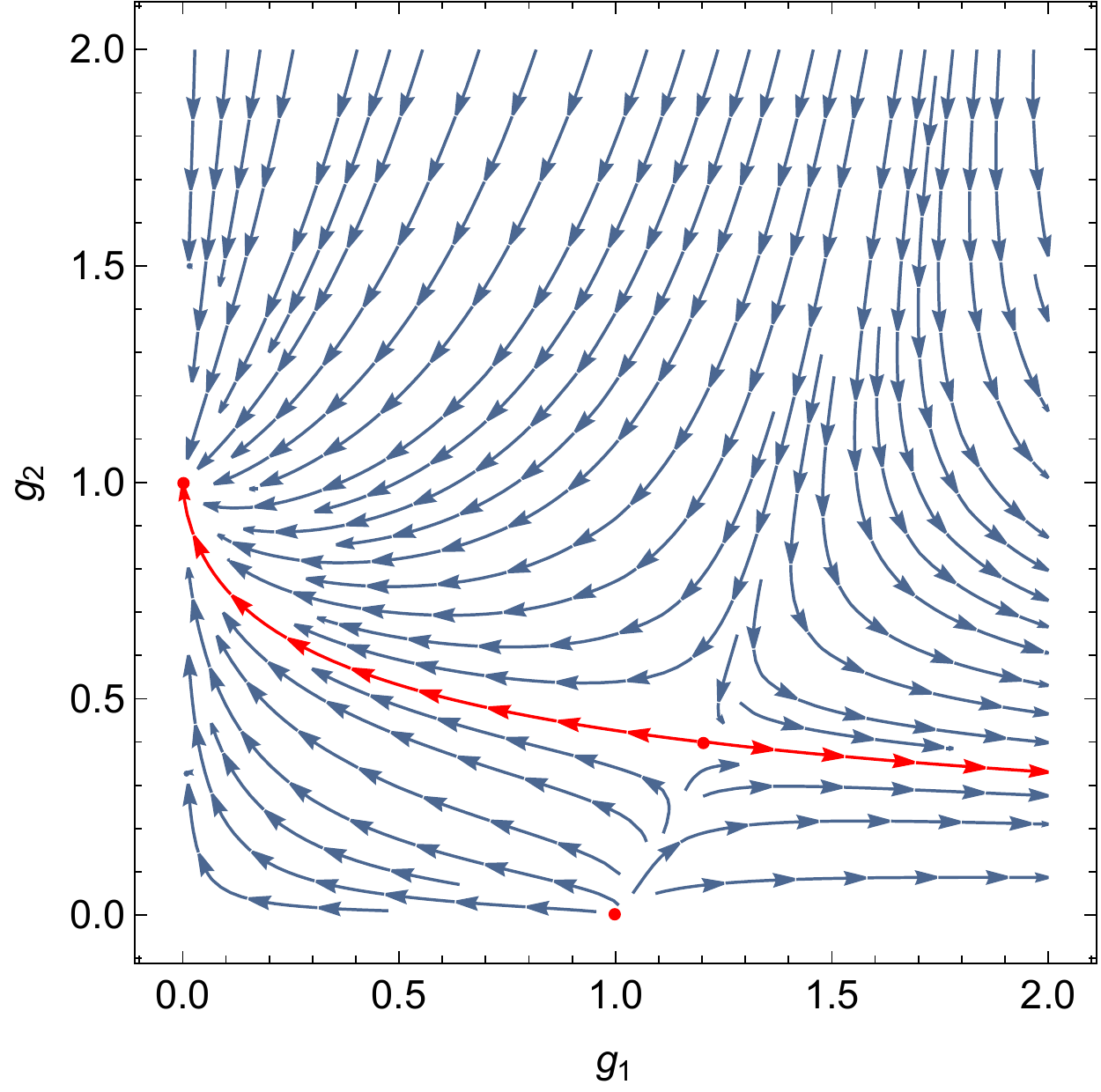}
\caption{\label{fig:predictions_schematic} The schematic RG flow features three fixed points in both panels. The one at $(g_1=1, g_2=0)$ acts as a source in both directions and thus does not generate predictions. The one at $(g_1=0, g_2=1)$ acts as a sink in both directions. It therefore generates two predictions, but also requires scale symmetry at all scales, because the RG flow cannot depart from it. Finally, the point at $(g_1=1, g_2=1)$ (left panel) and $g_1=1.2, g_2=0.4$ (right panel) has one "source-direction" and one "sink-direction". In both cases, it generates a prediction of the value of $g_2$. In the simple case (left panel), the prediction is the fixed-point value. In the more realistic case (right panel) the prediction is a function of $g_1$.}
\end{figure}

\emph{Predictions from asymptotic safety}\\
Because asymptotic safety is an enhancement of the symmetry of the QFT, one may expect that it relates different interactions to each other, because this is what happens in a QFT, when an additional symmetry is imposed. However, the special property of quantum scale symmetry is that such relations between couplings can remain intact, even if the theory departs from quantum scale symmetry under the RG flow towards the infrared (IR). In fact, the simple observation that there are distinct scales in nature, e.g., those of the masses of elementary particles, means that the RG flow must leave the asymptotically safe fixed point regime at some scale, i.e., quantum scale symmetry can only hold in the UV, not the IR, see, though, \cite{Wetterich:2020cxq} for an alternative. Nevertheless, the IR physics can still carry imprints of quantum scale symmetry in relations between couplings. We now exemplify this property. In short, it stems from the fact that the RG flow has "sources" and "sinks" -- in technical language, relevant and irrelevant directions. It can depart from the asymptotically safe fixed-point regime along a "source-direction", but not a "sink-direction". In "sink-directions", quantum fluctuations generate scale symmetry on the way to the IR, i.e., even if one chooses the value of a coupling slightly away from the fixed point, quantum fluctuations drive the coupling back to the fixed point. Therefore, along a "sink-direction", there is only one value that the coupling can take in the IR, namely its fixed-point value. This is true even if the RG flow has departed from the fixed-point value along a "source direction", see left panel in \autoref{fig:predictions_schematic}.\\
In practise, the simple picture we just sketched out is slightly modified in terms of the quantitative aspects of the predictions: first, when a relevant coupling departs from its fixed-point value, it can pull an irrelevant coupling along with it. The special value that the irrelevant coupling is fixed to is then no longer the fixed-point value, but instead a value that depends on the relevant coupling. In technical language, this means that the critical hypersurface of the fixed point (spanned by all its "source directions" is curved). Second, at an asymptotically safe fixed point, it is often a superposition of couplings that corresponds to a "source" or a "sink-direction".\\
Both aspects combined means that one often gets \emph{relations between couplings} that are predicted at low energies. In practise, these can often only be calculated numerically.  

In more technical language, predictions arise from irrelevant directions of the RG flow, which are encoded in negative critical exponents. The critical exponents measure the changes of the flow, i.e., they are related to first derivatives of the beta function. Thereby they encode whether a direction corresponds to a source or sink. The critical exponents are calculated as
\be
\theta_I =- {\rm eig} \left( \frac{\partial \beta_{g_i}}{\partial g_j}\right)\Big|_{g_i = g_{i\, \ast}},
\ee
where $g_{i\, \ast}$ is the fixed-point value of a coupling. Because of the extra negative sign, $\theta_I>0$ corresponds to a source (a relevant direction) and $\theta_I<0$ to a sink (an irrelevant direction).\\

\emph{Partial vs. full fixed points}\\
A fixed point is a point where the scale dependence of all couplings $g_i$ vanishes, i.e., $\beta_{g_i}=0$ for all couplings $g_i$ of the system. However, deciding whether a fixed point exists is a major challenge, because it requires knowledge of all beta functions. In practise, we will nevertheless refer to, e.g., the gravitational fixed point, because a large number of couplings have been included in studies and there are no indications that the remaining couplings would spoil the fixed point in gravity.\\
In the following, and in particular in \autoref{sec:GlobalSymms} and \autoref{sec:SMUVcompletion}, we will at times investigate the fixed-point structure of a subsystem, while treating other couplings as external parameters. This allows us to factorize the search for an asymptotically safe fixed-point for gravity and matter into different sub-systems. Strictly speaking, we then search for \emph{partial} fixed points within these different subsystems. The fixed points are partial, because some couplings are treated as external parameters, instead of also being evaluated at their respective fixed point. \\
The existence of a partial fixed point is a necessary, but not sufficient condition to find a fixed point of the full system. We will refer to partial fixed points as fixed points for coupling $g_i$, while full fixed points will be referred to as fixed points of the full system.\\

\emph{The robustness of results}\\
How sure are we about results in asymptotically safe gravity-matter systems? The answer differs, depending on which result we have in mind. There is no result about the existence of an asymptotically safe fixed point in gravity-matter systems which has been proven in a strict mathematical sense. However, for some results, so much evidence has been accumulated that they can be assumed to hold beyond reasonable doubt. For others, the uncertainty is larger, because the effect of approximations and assumptions is less well understood. We try to highlight when this is the case, without making our text too cluttered by constantly repeating phrases like "within an approximation", "under certain assumptions", etc.~.\\
There is a way of testing the robustness of results that we will refer to: if an approximation is advanced enough, then unphysical choices (for instance, choices of gauge parameters) do not change the value of physical quantities at all or not much. Conversely, if an approximation is not advanced enough, the choice of unphysical parameters can start to matter. 

%% file: Input/GravityFP.tex
\section{Gravitational fixed point under the impact of (minimally coupled) matter}
\label{sec:matterongrav}
There is compelling evidence for an asymptotically safe fixed point in pure gravity, see, e.g., other sections in this handbook. This fixed point is the starting point for our discussion. We will explore whether and how this fixed point continues to exist, when more and more matter fields of different spins are added.  In this section, the self-interactions of matter are ignored, because they do only indirectly impact the gravitational fixed point, although they may or may not feature a fixed point themselves.  

\subsection{Screening and anti-screening effects of matter on the Newton coupling}\label{sec:mattereffectsonGN}
\emph{Synopsis:
Matter fields of different spins have different effects on the fixed point in the Newton coupling: Scalars and fermions disfavor it; gauge fields favor it. This can be understood in terms of screening and anti-screening contributions, i.e., weakening and strengthening of gravity.
}

Asymptotic safety arises when there is an overall anti-screening contribution of quantum fluctuations of all fields -- matter and gravitational. An anti-screening contribution is one with a negative sign in the beta function. Such a contribution is necessary to achieve quantum scale symmetry, because the canonical dimension of the Newton coupling generates a contribution with positive sign. An anti-screening contribution can compensate this, such that asymptotic safety is present.\footnote{There is a way of understanding why an anti-screening contribution is necessary which uses continuation of the theory across dimensions: In $d=2$, the Newton coupling is dimensionless, which is similar to the gauge coupling in $d=4$: just like screening effects mean that the Abelian gauge coupling is screened to zero in four-dimensional QED, screening effects would mean a non-gravitating gravity theory. In contrast, just like anti-screening effects mean that the non-Abelian gauge coupling is anti-screened to a nonzero value in four-dimensional QCD, anti-screening effects yield a gravity theory with nonvanishing gravitational interaction. In the UV, this theory is asymptotically free, i.e., the coupling starts out at zero in the UV and is anti-screened to nonzero in the IR.
Going from $d=2$ to $d>2$, the anti-screening contribution can remain, but must compete with a contribution from a positive sign from the canonical dimension of the coupling, such that asymptotic freedom is no longer available, but asymptotic safety is.}
\\
When gravity is coupled 
to $\Nscal$ scalar fields, $\Nferm$ Dirac fermions and $\Nvec$ vector fields, the scale dependence of the Newton coupling can be schematically written as
\begin{equation}
\beta_{\GN}=2\GN-\GN^2 \left( b_{\mathrm{grav}}+ a_{\mathrm{S}}\,  \Nscal+ a_{\mathrm{F}}\, \Nferm+ a_{\mathrm{V}}\, \Nvec\right)+\mathcal{O}(\GN^3)\,,\label{eq:betaGN}
\end{equation}
where the first term encodes the scale-dependence due to the canonical mass dimension of the Newton coupling which leads to perturbative nonrenormalizability. Gravitational fluctuations and Faddeev-Popov ghosts generate $b_{\mathrm{grav}} >0$, see, e.g., \cite{Reuter:2001ag, Lauscher:2001ya, Lauscher:2002sq, Litim:2003vp, Niedermaier:2006wt, Codello:2008vh, Manrique:2009uh, Benedetti:2009rx, Benedetti:2009iq, Manrique:2010am, Groh:2011vn, Rechenberger:2012pm, Donkin:2012ud, Christiansen:2012rx, Benedetti:2012dx, Dietz:2012ic, Falls:2013bv, Christiansen:2014raa, Becker:2014qya, Falls:2014tra, Gies:2015tca, Christiansen:2015rva, Demmel:2015oqa, Ohta:2015fcu, Gies:2016con,  Denz:2016qks, Christiansen:2017bsy, Knorr:2017fus, Gonzalez-Martin:2017gza, Falls:2018ylp, DeBrito:2018hur, Kluth:2020bdv, Falls:2020qhj, Knorr:2021slg}, as well as \cite{Bonanno:2020bil} and references therein. The coefficients $a_{\mathrm{S}}$ ($a_{\mathrm{F}}$, $a_{\mathrm{V}}$) encode whether scalar (fermionic, vector) fields screen ($a_i<0$) or anti-screen ($a_i>0$) the gravitational coupling\footnote{More precisely matter fields also impact all other gravitational couplings, including, e.g., the cosmological constant, see \autoref{sec:matterbounds}.}. For the case of minimally coupled matter fields, the $a_i$ are numerical factors \cite{Narain:2009fy, Dona:2012am, Dona:2013qba, Percacci:2015wwa, Meibohm:2015twa, Labus:2015ska, Dona:2015tnf, Meibohm:2016mkp, Biemans:2017zca, Christiansen:2017cxa, Alkofer:2018fxj, Eichhorn:2018akn, Eichhorn:2018ydy, Eichhorn:2018nda, Burger:2019upn, Daas:2020dyo, Daas:2021abx}; going beyond minimal coupling, the $a_i$ become functions of the couplings, see, e.g., \cite{Oda:2015sma, Eichhorn:2016vvy, Hamada:2017rvn, Eichhorn:2017sok, Eichhorn:2018nda, Laporte:2021kyp, Knorr:2022ilz} and have to be evaluated at the corresponding fixed-point values.

The resulting gravitational fixed-point value is
\be
\GNast = \frac{2}{b_{\mathrm{grav}}+ a_{\mathrm{S}}\,  \Nscal+ a_{\mathrm{F}}\, \Nferm+ a_{\mathrm{V}}\, \Nvec}.\label{eq:gravityFP}
\ee
Eq.~\ref{eq:gravityFP} shows that a screening contribution ($a_i<0$) increases the fixed-point value of the Newton coupling $\GNast$, while an anti-screening contribution ($a_i>0$) decreases it. If the screening contributions dominate over the anti-screening contributions, then $\GNast \rightarrow \infty$ (and subsequently $\GNast<0$) and the theory is not asymptotically safe. 
Thus, screening contributions destabilize the asymptotically safe gravity system, because they can overcome the gravitational contribution $b_{\mathrm{grav}}$ and thereby remove the interacting fixed point. Conversely, an anti-screening contribution $a_i>0$ stabilizes the system, and drives the system to $\GNast \rightarrow 0$, when increasing the  corresponding $N_{i}$.\footnote{This does not necessarily indicate that a perturbative fixed point is approached, because fixed-point values of couplings can always be made arbitrarily small by an appropriate rescaling. Instead, the critical exponents are a meaningful measure of perturbativity: if they approach the canonical values, the fixed point becomes perturbative. In the present case, the critical exponent of the Newton coupling in the approximation (\ref{eq:betaGN}) is always 2, independent of the fixed-point value.}

In principle, there are different ways of coupling matter to gravity -- minimally and non-minimally. For minimally coupled fields, the interaction with gravity lies in the kinetic term of the matter fields. For non-minimally coupled fields, explicit couplings with curvature terms are present. In classical or phenomenological studies, one can usually choose which coupling to include. In asymptotically safe gravity-matter systems, there is no such choice to make: all interactions compatible with the symmetries are generically present, and this includes some non-minimal interactions. Nevertheless, their effect does not need to be significant, because the  nonminimal couplings may be small at a fixed point. Indeed, in studies to date, the minimal coupling determines whether matter fields screen or anti-screen the gravitational fixed point.

Minimally coupled scalars screen the gravitational coupling, $a_{\mathrm{S}}<0$, see \cite{Narain:2009fy, Dona:2013qba, Percacci:2015wwa, Labus:2015ska, Meibohm:2015twa, Dona:2015tnf, Biemans:2017zca, Alkofer:2018fxj, Eichhorn:2018akn,Wetterich:2019zdo, Laporte:2021kyp, Sen:2021ffc}. These studies cover a range of different approximations, and technical choices, e.g., regulator function and gauge parameters. The sign of $a_{\rm S}$ can thus be regarded as settled.
A screening contribution was also found using using other methods, namely perturbative heat-kernel methods \cite{Kabat:1995eq, Larsen:1995ax} and an $\epsilon$ expansion around $2+\epsilon$ dimensions \cite{Christensen:1978sc}.

Minimally coupled fermions also screen the gravitational coupling, $a_{\mathrm{F}}<0$. This was found in various studies which cover different approximations, regulator functions and gauge parameters \cite{Dona:2012am, Dona:2013qba, Meibohm:2015twa, Meibohm:2016mkp, Alkofer:2018fxj, Eichhorn:2018ydy, Eichhorn:2018nda, Wetterich:2019zdo, Daas:2020dyo, Daas:2021abx, Sen:2021ffc}, and also in perturbative studies \cite{Kabat:1995eq, Larsen:1995ax}. 

Minimally coupled gauge fields anti-screen the gravitational coupling, $a_{\mathrm{V}}>0$, as was found in various studies \cite{Dona:2013qba, Biemans:2017zca, Christiansen:2017cxa, Alkofer:2018fxj, Wetterich:2019zdo, Sen:2021ffc}, in agreement with perturbative studies \cite{Kabat:1995eq, Larsen:1995ax}.
Because gauge fields anti-screen the Newton coupling, one may expect that the gravitational fixed point becomes the free fixed point for $\Nvec \rightarrow \infty$. Indeed,
the fixed-point value $\GNast$ approaches zero for increasing $\Nvec$. However, the fixed-point value for the cosmological constant remains non-zero \cite{Dona:2013qba, Christiansen:2017cxa}; therefore the limit $\Nvec \rightarrow \infty$ does not lead to an asymptotically free fixed point.\\

\FRT{Further reading:}\\

\FR{On the impact of non-minimal couplings}\\
Some studies have gone beyond minimal coupling, and included explicit couplings between matter fields and curvature terms, see Tab.~\ref{tab:mattermatterstruncations} for an overview and references.

A non-minimal coupling between scalars and gravity of the form $R^{\mu\nu}\,D_{\mu}\phi D_{\nu}\phi$ slightly increases the amount of screening \cite{Laporte:2021kyp}, while a non-minimal coupling of the form $R\,D_{\mu}\phi D^{\mu}\phi$ slightly reduces the amount of screening\cite{Laporte:2021kyp}. The non-minimal coupling $\phi^2 R$, which is canonically marginal, breaks shift symmetry; therefore it vanishes at the gravity-matter fixed point \cite{Narain:2009fy, Narain:2009gb, Percacci:2015wwa, Oda:2015sma, Labus:2015ska}, see also \autoref{sec:GlobalSymms}, unless shift-symmetry is broken through the presence of, e.g., Yukawa interactions, see \cite{Eichhorn:2020sbo}.

A non-minimal coupling between fermions and gravity of the form $R\,\psi\bar{\psi}$ slightly increases the amount of screening \cite{Eichhorn:2016vvy}. On the other hand, a non-minimal coupling  of the form $R^{\mu\nu} \bar{\psi}\gamma_{\mu}\nabla_{\nu}\psi$ reduces the amount of screening and therefore stabilizes the system \cite{ Eichhorn:2018nda}. (We motivate the importance of such a non-minimal coupling in \autoref{sec:GlobalSymms}.)
However, even in the non-minimally coupled fermion-gravity systems, fermions screen the Newton coupling.

For vectors, nonminimal interactions have not yet been investigated.\\

\begin{table}[!t]
\begin{tabular}{c|c|c|c|c|c|c|}
gravitational int.'s & non-minimal int.'s &  non-minimal int.'s &  non-minimal int.'s & $N_{\rm S}$ & $N_{\rm F}$ & $N_{\rm V}$ \\ 
 & of scalars & of fermions & of gauge fields& & & \\
\hline
$\sqrt{g}$,\, $\sqrt{g}\, R$ & -& - & - & arb. & arb. & arb.\\
\cite{Dona:2013qba, Biemans:2017zca, Wetterich:2019zdo}& & & & & & \\ \hline
$\sqrt{g}$,\, $\sqrt{g}\, f(R)$ \cite{Alkofer:2018fxj}& - &-&- & arb. & arb. & arb.\\ \hline
$\sqrt{g}$,\, $\sqrt{g}\, R$,\,$\sqrt{g}R^2$ & - &-& - & arb.&arb.&arb.\\
 $\sqrt{g} C_{\mu\nu\kappa\lambda} C^{\mu\nu\kappa\lambda}$\,\,\cite{Sen:2021ffc}& & & & & & \\ \hline\hline
 $\sqrt{g}$,\, $\sqrt{g}\, R$ & $\sqrt{g}\,R\, \phi^2$& - & - & 1 & arb. & 0\\ 
 $\sqrt{g}\, R^2$,\, $\sqrt{g}\, R_{\mu\nu}R^{\mu\nu}$ \cite{Hamada:2017rvn} & & & & & &\\
 \hline
  $\sqrt{g}$,\, $\sqrt{g}\, R$ \cite{Oda:2015sma} & $\sqrt{g}\,R\, \phi^2$& - & - & 1 & arb. & 0\\ \hline
 $\sqrt{g}$,\, $\sqrt{g}\, R$ \cite{Eichhorn:2017sok} &$\ast$ $\sqrt{g}\,R^{\mu\nu}\, \partial_{\mu}\phi \partial_{\nu}\phi$& - & - & 1 & 0 & 0\\\hline
  $\sqrt{g}$,\, $\sqrt{g}\, R$ \cite{Laporte:2021kyp} &$\ast$ $\sqrt{g}\,R^{\mu\nu}\, \partial_{\mu}\phi \partial_{\nu}\phi$& - & - & 1 & 0 & 0\\
   & $\ast$ $\sqrt{g}\, R g^{\mu\nu} \partial_{\mu}\phi \partial_{\nu}\phi$ & - & - & & &  \\ \hline\hline
  $\sqrt{g}$,\, $\sqrt{g}\, R$ \cite{Eichhorn:2016vvy} &-&  $\sqrt{g}\,R\, \bar{\psi}\psi$ & - & 0 & arb. & 0\\ \hline
    $\sqrt{g}$,\, $\sqrt{g}\, R$ \cite{Eichhorn:2018nda} &-& $\ast$ $\sqrt{g}\,R^{\mu\nu}\, \bar{\psi}\gamma_{\mu}\nabla_{\nu}\psi$ & - & 0 & arb. & 0\\ \hline
\end{tabular}
\caption{\label{tab:mattermatterstruncations}We list the interactions that were included in the studies in the corresponding references. We only list the most comprehensive studies at each set of interactions, i.e., for instance for the Einstein-Hilbert truncation, on which the impact of all three species of matter fields was studied, we do not separately list studies which only take into account one or two species of matter fields.\\
Those non-minimal couplings marked by an $\ast$ cannot be set to zero at an interacting gravitational fixed point, see \autoref{sec:GlobalSymms}. In contrast, those non-minimal couplings not marked by an $\ast$ can be set to zero for reasons of symmetry; thus it is consistent to neglect them at a minimally coupled fixed point. 
We also indicate the numbers of matter fields that were studied, where "arb." stands for an arbitrary number of fields of the corresponding species, but does not necessary imply that an asymptotically safe fixed point exists for arbitrarily high numbers of the corresponding field. We omit matter self-interactions in this table; those are discussed in \autoref{sec:GlobalSymms}. }
\end{table}

\FR{Fermions in the background field approximation}\\
For fermions, there is a  technical subtlety: if one chooses a regulator function which does not regularize the modes of the Dirac operator, and works in the background field approximation, the sign of $a_{\rm F}$ can be flipped to $a_{\mathrm{F}}>0$ \cite{Dona:2012am,Biemans:2017zca, Alkofer:2018fxj, Daas:2020dyo}. %

\subsection{Impact of Standard Model fields}
\emph{Synopsis: The four scalar components of the Higgs field and 45 Weyl fermions of the  SM only partially counteract the anti-screening effect of gravitational modes and the 12 gauge fields; therefore, an asymptotically safe fixed point for the Newton coupling persists under the impact of SM matter. The fixed point also exists when further gravitational couplings are included.}

It is an important observational consistency test for any model of quantum gravity, whether the observed matter fields of the SM can exist within the model. 
It is evidently not a given that asymptotically safe gravity passes this consistency test, given the results from the previous \autoref{sec:mattereffectsonGN}. First, passing the test requires that the screening effect of the scalars and fermions in the SM does not overwhelm the anti-screening effect of gravitational and gauge fields on the Newton coupling. Only then can an asymptotically safe fixed point for the Newton coupling exist.
Second, passing the test requires that the fixed point continues to exist when further couplings beyond the Newton coupling are considered on the gravitational and the matter side. Matter couplings are discussed in \autoref{sec:GlobalSymms}.

All computations so far agree on the following important result: the asymptotically safe fixed point of pure gravity continues to exist, when the matter content of the SM is accounted for. The fixed-point values and critical exponents depend on the number of matter fields, but, if the number of matter fields of the three species, (scalars, fermions, vectors) are treated as continuous parameters, the fixed-point values and critical exponents   at $N_{\rm S}= 0, \, N_{\rm F} =0, \, N_{\rm V}=0$ can continuously be connected to those at to $N_{\rm S}= 4,\, N_{\rm F} = 22.5,\, N_{\rm V}= 12$ \cite{Dona:2013qba, Biemans:2017zca, Alkofer:2018fxj, Wetterich:2019zdo, Sen:2021ffc, Pastor-Gutierrez:2022nki}.\footnote{
If a fixed point can be deformed continuously through increasing a continuous parameter from an initial to a final value, one can think of the final fixed point as a deformation of the original universality class. In contrast, if the fixed point at the final value of the parameter is not obtained through a deformation of the fixed point at the initial value, one cannot understand it as a deformation of the original universality class, but instead has to think of these as two different universality classes. For this second case, the existence of the first universality class (at the initial value of the parameter) does not matter for the existence of the second universality class (at the final value of the parameter). Translated to gravity-matter systems this second possibility would mean that a second asymptotically safe universality class is unrelated to the pure-gravity universality class. While this is a logical possibility, this does not appear to be the case.
}
Therefore asymptotically safe quantum gravity passes a crucial observational consistency test.\\ Some extensions of the SM  may also admit a fixed point. The most important extension is probably the addition of three right-handed neutrinos, which are required to explain neutrino-oscillations (unless the neutrinos are Majorana). Further extensions may be necessary to accommodate dark matter, e.g., in the form of an axion or axion-like particle, or a gauge-singlet scalar. Such extensions by 3 Weyl fermions (3/2 Dirac fermions) and one or two scalars are indeed possible in all studies to date.\\
As a first step towards a supersymmetric setting, the inclusion of a gravitino, i.e., a spin 3/2 field, has been studied in \cite{Dona:2014pla}.\\

\begin{figure}[!t]
\includegraphics[width=0.45\linewidth,clip=true,trim=7cm 2.5cm 13cm 4cm]{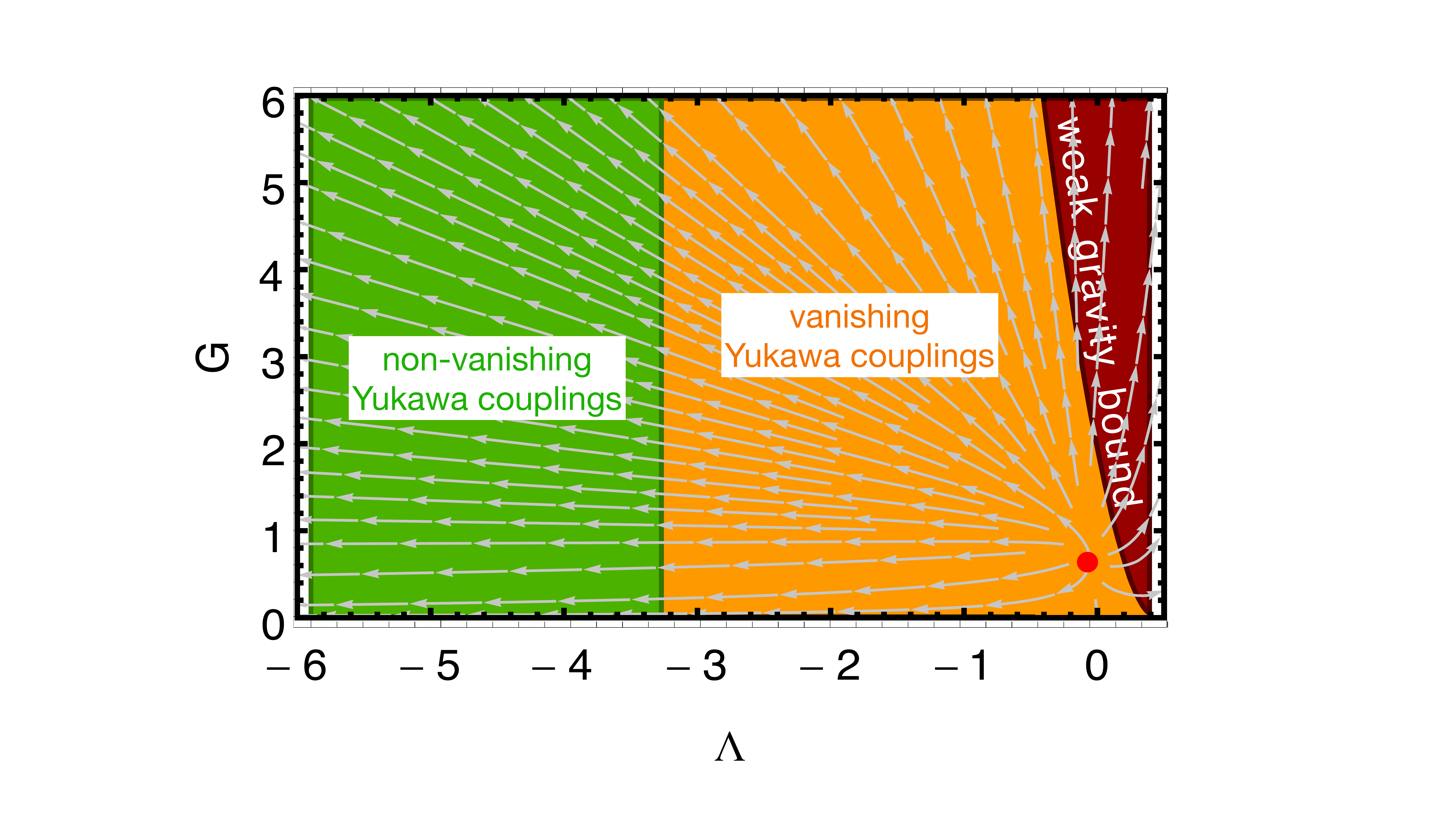}\quad\includegraphics[width=0.45\linewidth,clip=true,trim=7cm 2cm 13cm 3.8cm]{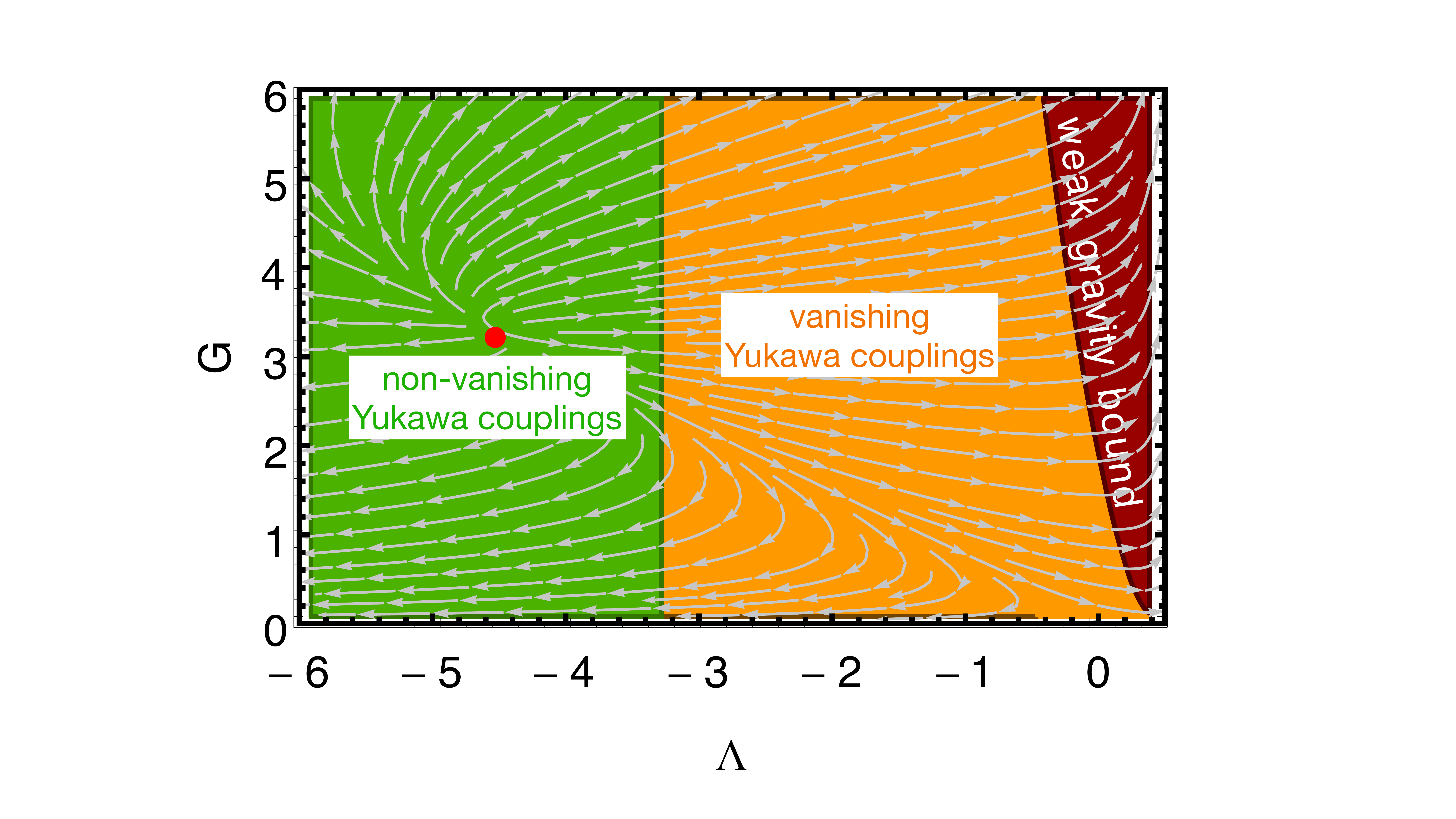}
\caption{\label{fig:mattermattersinterplay} We show the plane spanned by the dimensionless Newton coupling $\GN$ and the dimensionless cosmological constant $\Lambda$. We also indicate the boundary between the region with non-vanishing Yukawa couplings (green) and the region with vanishing Yukawa couplings (orange). Additionally, the weak-gravity bound, see \autoref{sec:WGB}, is indicated. The fixed-point value (red dot) and the RG flow towards the IR in the background approximation are shown for one generation of Standard-Model fermions (left panel) and three generations (right panel). In both panels, 12 vectors and 4 scalars are included. See \cite{Eichhorn:2017ylw} for the corresponding reference. }
\end{figure}

\emph{Preview on phenomenological consequences:}\\
There are indications for a mechanism that renders the SM plus gravity not only asymptotically safe, but also more predictive than the SM on its own. This mechanism relies on matter fields changing  gravitational fixed-point values. Here, we preview this mechanism and get back to it in more detail in \autoref{sec:SMUVcompletion}.
Yukawa interactions which the SM needs for its fermions to be massive, vanish, unless conditions on the gravitational fixed-point values are fulfilled \cite{Oda:2015sma, Eichhorn:2016esv, Eichhorn:2017ylw}, see also \autoref{sec:SMUVcompletion}. These conditions are not fulfilled by the gravitational fixed-point values, as they come out in studies with no or few matter fields \cite{Eichhorn:2016esv,Eichhorn:2017eht,Eichhorn:2017ylw}. However, once three generations of SM fermions are added, the gravitational fixed-point values satisfy the conditions according to the study in \cite{Dona:2013qba}, see \autoref{fig:mattermattersinterplay}. 
Thus, fermions push the gravitational fixed point into a region of values, in which fermion mass generation through Yukawa couplings to the Higgs field becomes possible.\\
We caution that one needs to know the gravitational fixed-point values with relatively high accuracy to determine whether or not this mechanism is indeed at work. Current studies have not yet achieved the accuracy to comprehensively confirm the scenario in \cite{Eichhorn:2017ylw}.\\
That the Yukawa couplings of the SM cannot be accommodated automatically is an important result in quantum gravity phenomenology, because it means that the asymptotically safe model is testable \emph{at SM scales}: if the fixed-point value falls outside the green region in \autoref{fig:mattermattersinterplay}, the model is ruled out by an experimental result, namely the measurement of nonvanishing Yukawa couplings at the LHC \cite{CMS:2018uxb,ATLAS:2018mme,CMS:2018nsn,ATLAS:2018kot,ATLAS:2015xst,CMS:2017zyp}.

\subsection{
 The effective strength of gravity under the impact of matter
}
\label{sec:matterbounds}
\emph{Synopsis:
 The effective strength of gravity can be encoded in a combination of couplings, in which these appear whenever gravitational fluctuations contribute to a system. This effective strength depends on the number of matter fields. Generically, scalar fields drive the effective strength up and fermions and vectors lower it.
}

\begin{figure}[!t]
	\centering
	\includegraphics[width=0.8\linewidth]{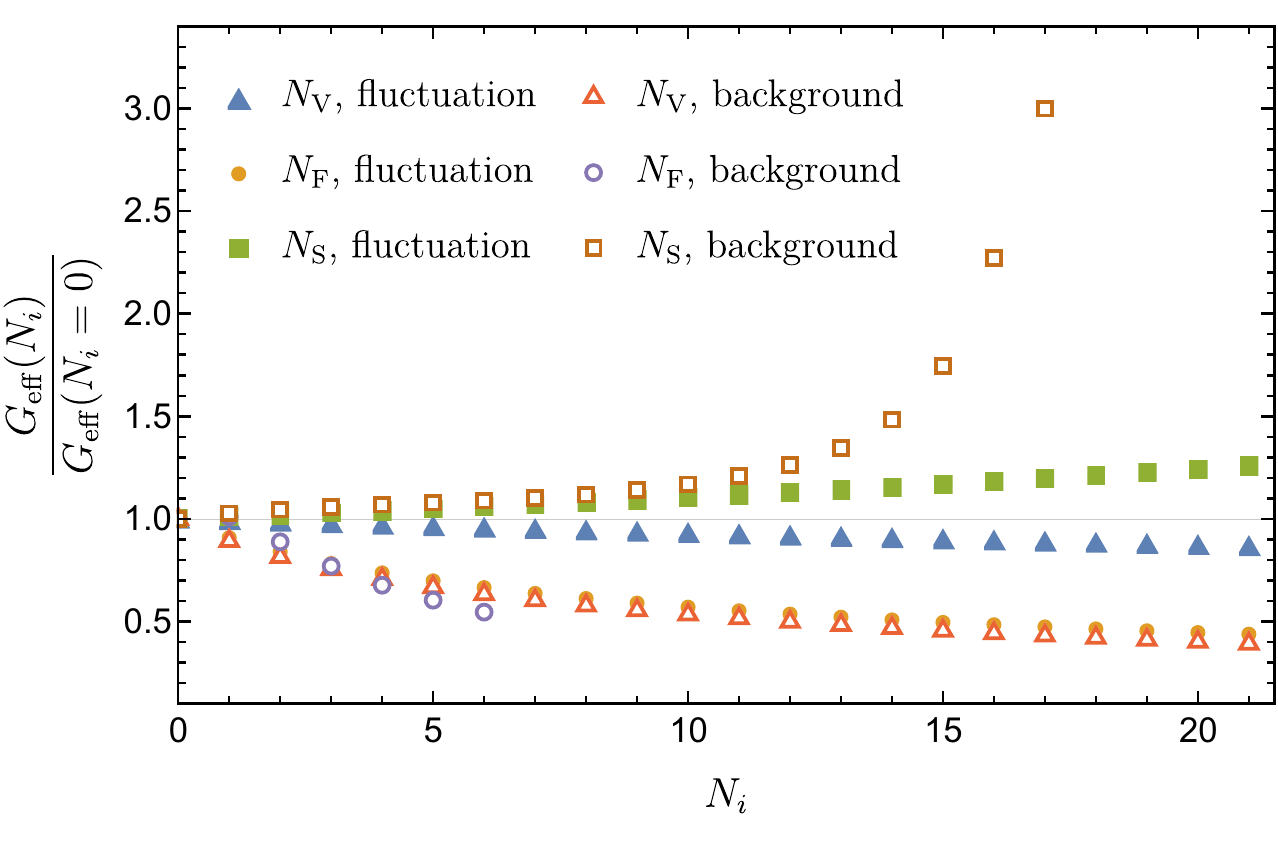}
	\caption{\label{fig:GeffMatterplot}
		 We show the fixed-point values of the effective gravitational coupling $G_{\rm eff}$ \eqref{eq:Geff}, when increasing the number of scalars, fermions or vector fields for minimally coupled matter. The left column refers to fixed-point values obtained within fluctuation computations, taken from \cite{Meibohm:2015twa}, and the right column refers to fixed-point values obtained within the background field approximation, taken from \cite{Dona:2013qba}. We see that the qualitative behavior of $G_{\rm eff}$ agrees for all matter fields between the two different  methods.}
\end{figure}

We introduce an effective strength of gravity, first discussed in \cite{Eichhorn:2017eht}:
\be
G_{\rm eff}= \frac{\GN}{1-2 \Lambda}\, \label{eq:Geff}.
\ee
Herein, $\Lambda$ is the cosmological constant. $G_{\rm eff}$ is a combination of $\GN$ and $\Lambda$, in which the two couplings enter in beta functions, therefore this combination should have an asymptotically safe fixed point.

$G_{\rm eff}$ increases, when additional scalar fields are added to the system, see \autoref{fig:GeffMatterplot}.  Therefore, we expect that the system becomes increasingly non-perturbative, when scalars are added.
Beyond a critical value of $\Nscal$, the asymptotically safe fixed point ceases to exist. We caution that the approximation in which studies are performed may break down at lower $\Nscal$ than the critical value, see \cite{Meibohm:2015twa, Eichhorn:2018akn, Burger:2019upn}. Nevertheless, a fixed point at small values of $G_{\rm eff}$, where calculations are more easily controlled, may only exist at relatively small values of $\Nscal$.

For fermions, the situation differs, because $G_{\rm eff}$ decreases, when additional species are added \cite{Eichhorn:2018nda}, see \autoref{fig:GeffMatterplot}. In fact, the system may become increasingly perturbative, when fermions are added.\footnote{To make a robust statement on this possibility, other couplings need to be analyzed as well.} This may be important to connect asymptotically safe quantum gravity to the SM: it is known that the SM is perturbative at the Planck scale; thus, a UV completion with gravity has to be able to reproduce this perturbative regime. This is most likely achievable if the UV completion itself is (near-) perturbative in nature,  cf.~\autoref{sec:ner-pert}. The SM has 22.5 Dirac fermions, which is sufficient to drive the effective gravitational strength to significantly lower values that for the pure-gravity case.

$G_{\rm eff}$ decreases, when additional vectors fields are added to the system, see \autoref{fig:GeffMatterplot}. Theories, which add additional gauge fields to the SM may therefore be compatible with asymptotic safety. For Grand Unified Theories (GUT), the situation is not so clear, because they need additional scalars to spontaneously break the large gauge group. Different studies find differences on whether the matter content of popular GUT models is compatible with asymptotic safety, e.g.,  \cite{Dona:2013qba, Wetterich:2019zdo}, indicating that further studies are necessary.\\

\FRT{Further reading:}\\

\FR{Integrating out matter fields}\\
In \cite{Christiansen:2017cxa}, the authors argue that truncations with minimally coupled matter fields are insufficient to infer whether or not bounds on the number of matter fields exist: They show that bounds disappear, if quantum fluctuations of matter fields are integrated out first, and of gravity last, instead of both simultaneously. If a truncation is large enough, the order in which fields are integrated out should be irrelevant.\\

\FR{Additional effective gravitational couplings}\\
The definition in Eq.~\eqref{eq:Geff} can be generalized to
\be
\label{eq:Geffn}
G_{\rm eff}^{(n)} = \frac{\GN}{(1-2\Lambda)^n}\,.
\ee
For $n>1$, these couplings also make an appearance in beta functions. For fermions, they have been compared in \cite{Eichhorn:2018nda}.\\

\FR{Comparing background approximation and fluctuation computations}\\
For fermions, the decrease of $G_{\rm eff}$ comes about in different ways depending on the choice of approximation: In \cite{Dona:2013qba}, employing the background field approximation, $\Lambda$ becomes large and negative, which decreases $G_{\rm eff}$, in \cite{Eichhorn:2018nda}, which constitutes a fluctuation computation, $\Lambda$ stays approximately constant, while $G_N$ decreases. Such differences between approximations do not matter at the level of $G_{\rm eff}$, which is  a more useful quantity to consider, both for its higher degree of robustness, and because $G_{\rm eff}$, not $G_N$, enters beta functions and thus determines the strength of gravity fluctuations. More generally speaking, $G_{\rm eff}$ behaves qualitatively similar for computations employing the background field approximation, and for fluctuation computations for all three types of matter, i.e., scalars, fermions and gauge fields, see \autoref{fig:GeffMatterplot}.\\

\begin{figure}[!t]
	\centering
	\includegraphics[width=0.8\linewidth]{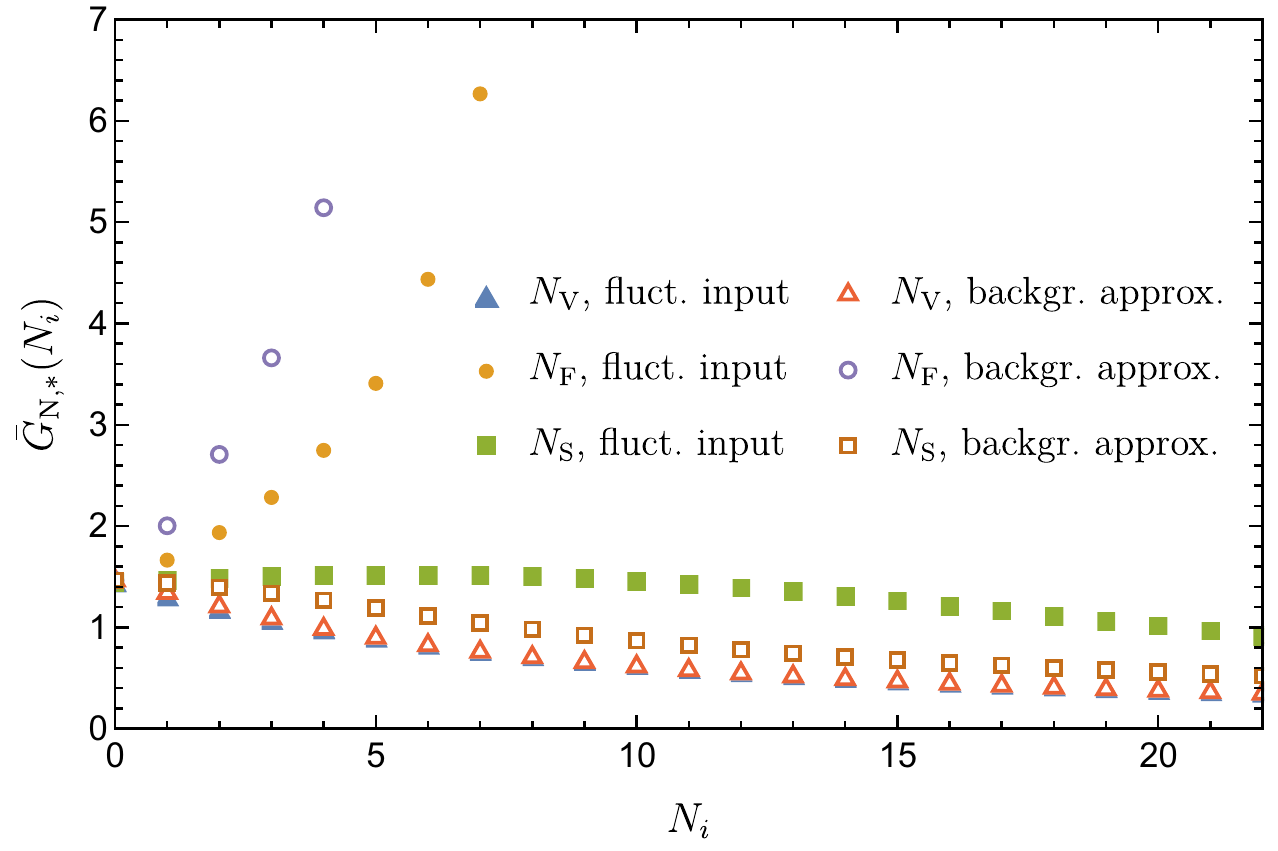}
	\caption{\label{fig:GMatterplot}
		We show the fixed-point values of the background Newton coupling $\bar{G}_{\mathrm{N}}$ as a function of different matter fields. Full markers indicate the background field approximation. Empty markers indicate an approximation, where the fluctuation couplings enter the right-hand side of the flow equation, and therefore the beta-functions of the background couplings.}
\end{figure}

\FR{Background field approximation and fluctuation input}\\
In the background field approximation, the difference of background and fluctuation couplings is neglected, see \autoref{sec:methods}. A first step to lift this approximation can be achieved by taking into account the input of fluctuation couplings in the scale-dependence of the background couplings. In this way, the fixed-point values for the fluctuation coupling enter the beta-functions of the background couplings. In \autoref{fig:GMatterplot} we compare these two setups with each other. While the inclusion of fluctuation couplings changes the fixed-point values for $\bar{G}_{\mathrm{N}}$ on a quantitative level, the qualitative features remain rather robust. 

%% file: Input/GlobalSymm.tex
\section{Global Symmetries persist and have phenomenological consequences}
\label{sec:GlobalSymms}
\emph{Synopsis: Global symmetries, such as shift symmetry for scalar fields or chiral symmetries for fermions, are left intact by gravitational fluctuations. In turn, they determine through which terms matter fields interact at an asymptotically safe fixed point. These interactions may lead to a bound on the strength of gravity, the weak-gravity bound. Further, these interactions find a way to circumvent a mechanism for chiral symmetry breaking which would leave fermions with Planck-scale masses and would rule out asymptotic safety.}

\subsection{The status of global symmetries in asymptotic safety}
\emph{Synopsis: There is a general argument suggesting that there cannot be global symmetries in quantum theories of gravity. We review the argument and point out its assumptions, which may not hold in asymptotically safe gravity. We then review explicit calculations that show that quantum fluctuations of gravity generate new interactions for matter fields. These interactions respect the maximum set of global symmetries of the corresponding matter fields. In contrast, interactions which violate the maximum set of global symmetries of the matter fields are not generated, i.e., can consistently be set to zero.}

In quantum field theory, symmetries play a central role by dictating the interactions of the fields in the theory.
Global symmetries play different roles -- and are understood to varying degrees -- at different scales. At  condensed-matter scales, many different global symmetries occur, under which order-parameter fields transform. The corresponding theories describe phases and phase transitions through the spontaneous breaking of these global symmetries.
  Moving towards smaller length scales, namely those of particle physics, there is only a single global symmetry in the SM, namely a global $U(1)_{\rm B-L}$, where $\rm B$ and $\rm L$ stand for baryon- and lepton-number. Additional global symmetries are present, if interactions are switched off. Beyond the SM, global symmetries play an important role, e.g., in dark-matter models, where they may ensure the stability of dark matter.
 Finally, moving towards even smaller scales, in the quantum-gravity regime, what is the fate of global symmetries?\\
There is a general argument that states that any global symmetry should be broken in quantum gravity \cite{Banks:1988yz,Banks:2010zn}, however, as any such arguments, it relies on assumptions which may or may not hold in a given setting.\footnote{In the context of string theory, this is known as the no-global-symmetries swampland conjecture, see, e.g., \cite{Daus:2020vtf} and references therein.} We will now present the argument and explain where assumptions are being made.\\ 
The argument relies on virtual black-hole configurations in the gravitational path-integral, together with the assumption that global charges are not preserved by black holes. It says that among the various spacetime configurations that the gravitational path integral sums over, there are configurations that correspond to black holes. In turn, if one extrapolates Hawking radiation all the way to zero mass (i.e., through the semi-classical and also the quantum regime), then black holes destroy information on global charges: for instance, one can imagine building a black hole from only protons (so it has a well-defined baryonnumber and leptonnumber), but it evaporates not just into baryons, but into various elementary particles, completely destroying any memory of the initial global charge. Therefore, the argument concludes, there is a contribution in the gravitational path integral that destroys global charges and thus global symmetries cannot be preserved in quantum gravity.\\
However, first, the actual contribution of virtual black-hole configurations to the gravitational path-integral is not known (and indeed depends on the microscopic dynamics -- so may be different in an asymptotically safe setting than, e.g., when using the Einstein action). One can imagine settings where the microscopic dynamics $S$ is such that the phase factor $e^{iS}$, when evaluated on black-hole configurations, leads to destructive interference, see, e.g., \cite{Borissova:2020knn}. Second, whether or not global charges are conserved also depends on whether it is true that there are no black-hole remnants, for which, indeed, there are counter-indications in asymptotic safety \cite{Bonanno:2006eu, Falls:2012nd}. If there are black-hole remnants, i.e., the Hawking evaporation process stops at a finite mass of the black hole, then the original information on global charges may be stored by the remnant.\\
In turn, there is a general argument against the existence of remnants \cite{Susskind:1995da}, which itself relies on assumptions about the behavior of those remnants in scattering processes.\\
In summary, one cannot in general conclude that global symmetries are broken in quantum gravity without knowing more about the specific properties of the theory.
\\

To settle the question whether or not global symmetries are conserved in asymptotically safe gravity, or whether indeed only local symmetries may exist, it is necessary to calculate the gravitational effect on matter systems with global symmetries. Indeed, many such calculations have been performed, which we review below. As an upshot, none of the calculations indicates that global symmetries are broken by asymptotically safe quantum-gravity effects. One reason may be that gravity-matter systems may be near-perturbative in asymptotic safety (see  \autoref{sec:ner-pert}), and thus nonperturbative contributions in the path integral, which may break global symmetries, are negligible. An important caveat to this is that calculations are done in a Euclidean regime. There, analytic continuations of black-hole spacetimes exist, but are physically quite distinct from black holes in a Lorentzian regime, because Lorentzian signature is necessary for the existence of causal relations and thus horizons. Thus, while no indications for global symmetry breaking exist in asymptotic safety to date, the question is not fully settled yet and a different result may be found in Lorentzian signature.\footnote{ First studies of asymptotic safety in Lorentzian gravity exist, which yield a fixed point similar to the Euclidean one \cite{Manrique:2011jc,Fehre:2021eob}.}\\

\begin{figure}[!t]
\begin{center}
\includegraphics[width=0.45\linewidth]{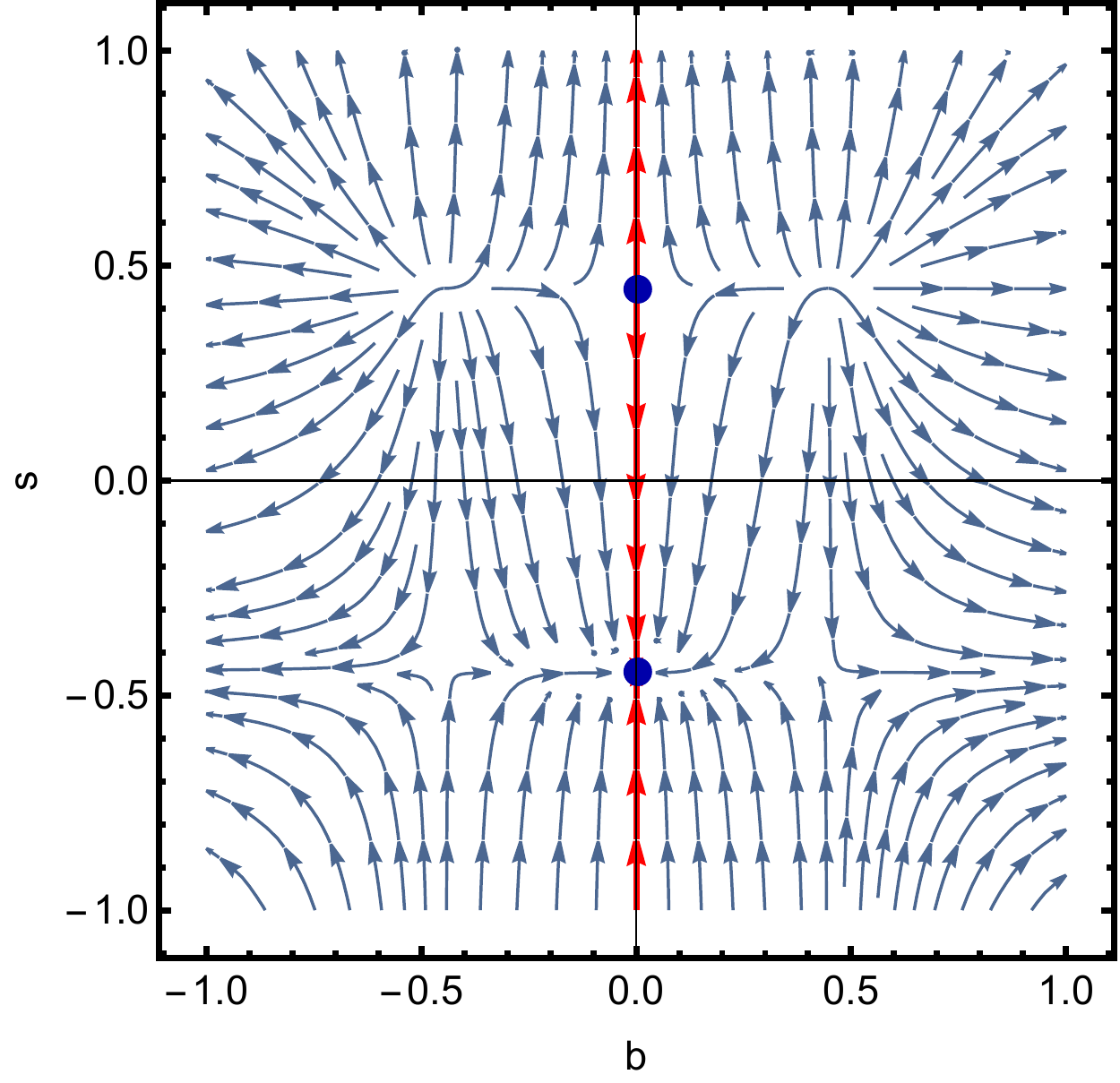}
\newline
\includegraphics[width=0.45\linewidth]{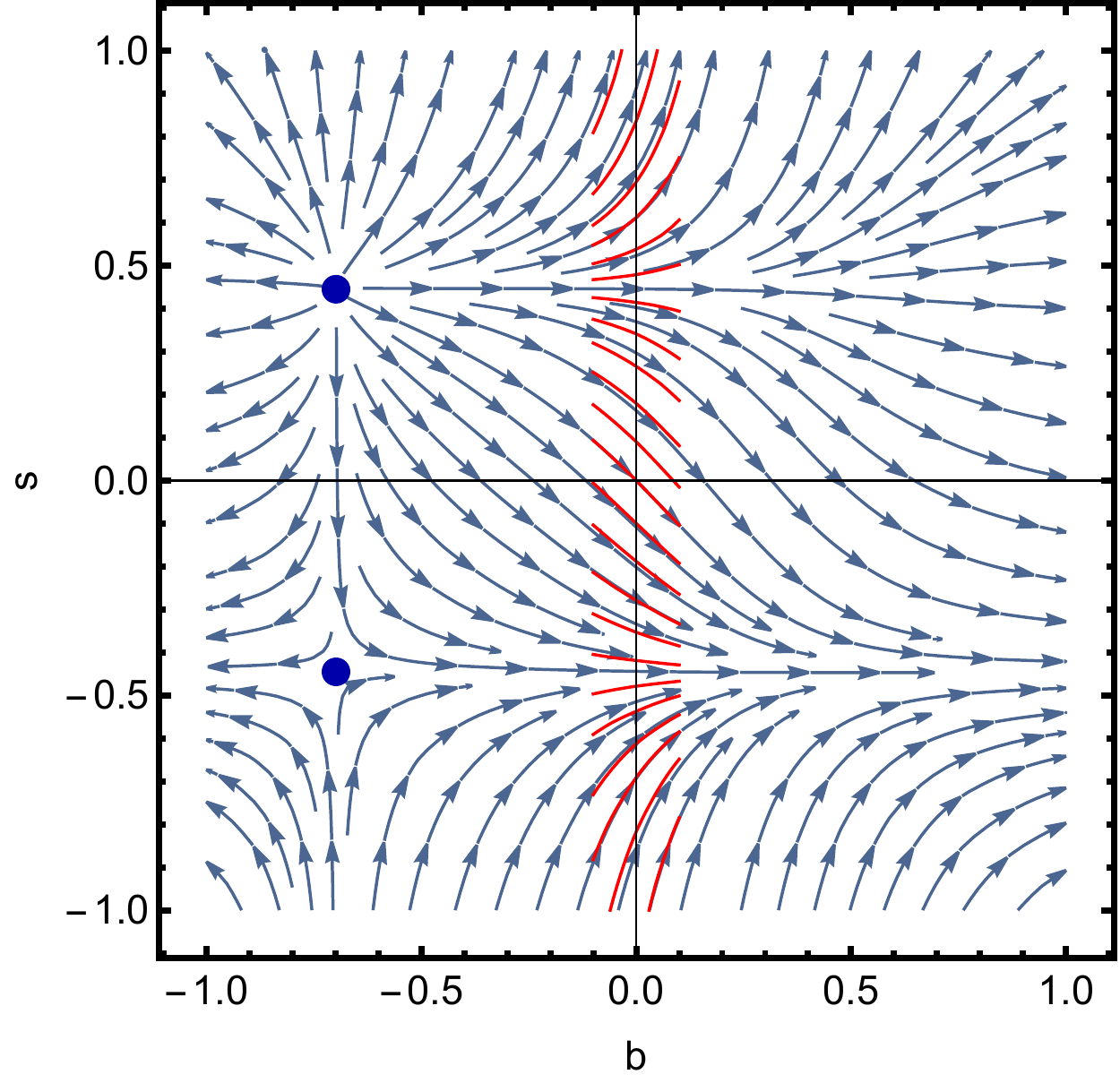}
\quad
\includegraphics[width=0.45\linewidth]{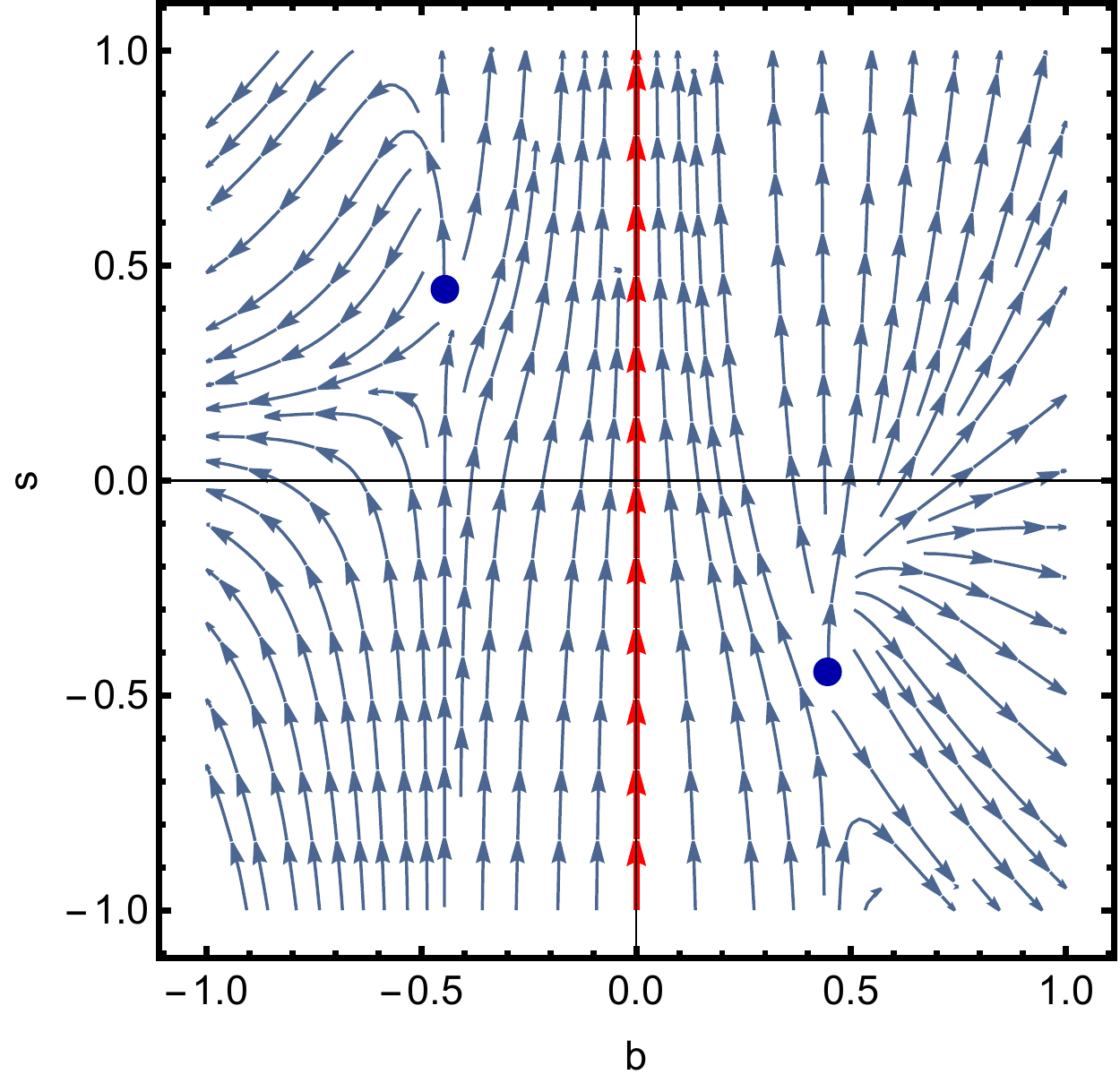}
\end{center}
\caption{\label{fig:globalsym} We illustrate the three distinct possibilities for the status of global symmetries in asymptotically safe gravity. Here, $s$ is a coupling that respects the global symmetry, whereas $b$ breaks it. In the upper figure, the global symmetry is left intact, because quantum gravity does not generate the symmetry-breaking interaction: if $b=0$ is chosen, the RG flow remains at $b=0$ (red line); and because there are fixed points in the theory space with the global symmetry: they lie at $s\neq 0$ and $b=0$ (blue dots). In both lower panels, the global symmetry is broken: in the left panel, gravity fluctuations induce $b$, even if it is set to zero (red lines). In the right panel, gravity fluctuations do not generate the symmetry-breaking interactions. However, there is no fixed point in the theory space with the global symmetry; the only fixed points lie at nonzero $s$ and $b$.}
\end{figure}

There are two ways in which asymptotically safe gravity could reduce the global symmetries of a matter system: the first is by explicitly generating new interaction terms for matter with a lower degree of symmetry. The second is by preventing an asymptotically safe fixed point in the theory space with the maximum global symmetry. For clarity, we illustrate the two possibilities in \autoref{fig:globalsym}.

The simplest setting in which the conservation of global symmetries can be investigated in asymptotic safety through explicit calculations looks as follows: 
A non-interacting field is considered, i.e., only a kinetic term is specified. This minimal coupling to gravity suffices for quantum gravity fluctuations to generate interaction terms.
For a given field, the kinetic term has a maximum set of global symmetries -- e.g., for $\Nscal$ real scalar fields, there is an O($\Nscal$) symmetry, in addition to $\Nscal$ separate shift symmetries (for each of the scalar fields). Under the impact of quantum gravity fluctuations, the field does not remain non-interacting, i.e., quantum gravity fluctuations generate interactions.
If the first possibility for symmetry-breaking is realized, then gravitational fluctuations generate interactions which break the global symmetries explicitly. 
If the second possibility is realized, then no such symmetry-breaking interactions are generated, but there is also no fixed point among the symmetry-preserving interactions.
In contrast, if quantum gravity does not break global symmetries, then the interactions which are generated share the symmetries of the kinetic term and these interactions admit an asymptotically safe fixed point.
\\
In terms of the beta function for a coupling $b$ that breaks a global symmetry, the question can be investigated as follows: 
The most general form of such a beta function is
\be
\label{eq:symmbreak}
\beta_b = \beta_0\, G_N^{\alpha_1} + \left(\beta_1+ \beta_{1\, G_N} G_N^{\alpha_2} \right) b+ \mathcal{O}(b^2).
\ee
If quantum gravity breaks the global symmetry through the generation of symmetry-breaking interactions, then $\beta_0\neq 0$ must hold, so that any (partial or full) fixed point must have the symmetry-breaking interaction present.\\
Conversely, if $\beta_0=0$, gravity does not generate the symmetry-breaking interaction. Then, $b=0$ is a zero of the beta function and the symmetry-breaking coupling can  consistently be set to be zero, i.e., it is a partial fixed point for any value of $\GN$. Of course, this is not sufficient to guarantee that asymptotically safe gravity respects the corresponding global symmetry, because there may not be a full fixed point in the theory space with the global symmetry.\\
To show that asymptotically safe gravity respects global symmetries, one therefore has to do two things: first, one has to show that symmetry-breaking interactions are not generated ($\beta_0=0$ in all corresponding beta functions). This means that the symmetry-breaking interactions feature a partial fixed point at vanishing coupling values.
Second, one has to show that the theory space with the maximum symmetry contains an asymptotically safe fixed point. This means that the partial fixed point extends to a full fixed point.

\subsubsection{Step 1: No symmetry-breaking interactions are generated by gravitational fluctuations}\label{sec:nosymbreak}
\emph{Synopsis: In all examples with global continuous symmetries for matter fields which have been studied, these symmetries are preserved by quantum gravitational fluctuations, under the assumptions spelled out in the corresponding papers, which include Euclidean signature.} 

Continuous global symmetries that have explicitly been investigated include:\\
\begin{itemize}
	\item For scalar fields:
	\begin{itemize}
		\item O($\Nscal$) symmetry, under which an $\Nscal$-component scalar field transforms in the fundamental representation; $\phi^a \rightarrow O_{\Nscal}^{ab}\phi_b$, where $O_{\Nscal}^{ab}$ is a generator in the fundamental representation. This symmetry would be broken, for example, if gravity generated distinct anomalous dimensions for some of the $\Nscal$ scalars, or if gravity generated interaction terms with uneven numbers of scalars field, or if it generated different masses or interaction terms for some of the $\Nscal$ scalars. Neither of these possibilities is realized in \cite{Labus:2015ska,deBrito:2021pyi}.
		\item Shift symmetry, under which  $\phi \rightarrow \phi + \rm const$. This symmetry would be broken by a scalar potential. It was found in \cite{Narain:2009fy,Percacci:2015wwa,Eichhorn:2017als} that a scalar potential is not generated at the asymptotically safe fixed point. In contrast, it was found in \cite{Eichhorn:2012va, Eichhorn:2013ug, Eichhorn:2017sok, deBrito:2021pyi, Laporte:2021kyp} that shift-symmetric interactions (which are proportional to derivatives of the scalar field) are induced at an asymptotically safe fixed point. 
		\item A complex scalar, which has a global U(1) symmetry, was studied in \cite{Ali:2020znq}, where quantum gravity does not generate terms that would break the global U(1) symmetry to a discrete $\mathbb{Z}_n$ symmetry.
	\end{itemize}
	Thus we find that quantum gravity generates interaction terms for scalar matter. These respect the maximum set of symmetries of the kinetic term, irrespective of the number of scalar fields. In addition, these interactions feature a fixed point, if the weak-gravity bound is respected, see \autoref{sec:WGB} below.\\
	\item For fermion fields:
	\begin{itemize}
		\item $SU(\Nferm)_L \otimes SU(\Nferm)_R$ symmetry, under which the left- and the right-handed components of $\Nferm$ Dirac fermions transform separately. This is called a chiral symmetry, because it refers to the chiral components (the left- and right-handed Weyl spinors) of a Dirac fermion.
		This is a symmetry of the kinetic term, because a kinetic term for $\Nferm$ Dirac fermions decomposes into a kinetic term for $\Nferm$ right-handed and $\Nferm$ left-handed Weyl spinors, $\bar{\psi}^i \slashed{\nabla}\psi^i =\bar{\psi}_L^i \slashed{\nabla}\psi_L^i+  \bar{\psi}_R^i \slashed{\nabla}\psi_R^i $ (where $i=1,...,\Nferm$). This symmetry is broken by a mass term, $m \bar{\psi}^i\psi^i = m \bar{\psi}^i_R\psi^i_L + m \bar{\psi}^i_L\psi^i_R$, a non-minimal term of the form $R \bar{\psi}^i\psi^i = R \bar{\psi}^i_R\psi^i_L + R \bar{\psi}^i_L\psi^i_R$, a four-fermion interaction of the form $\bar{\psi}^i \psi^i\, \bar{\psi}^j\psi^j$ and others. Neither of these interactions is generated\footnote{A breaking of chiral symmetry can be introduced explicitly by choosing a regulator that breaks chiral symmetry \cite{Daas:2020dyo}; then such chiral-symmetry breaking interactions are generated.} in the studies in \cite{Eichhorn:2011pc, Eichhorn:2016vvy, Eichhorn:2017eht, deBrito:2020dta}. In contrast, it was found in \cite{Eichhorn:2011pc, Eichhorn:2016vvy, Eichhorn:2017eht, Eichhorn:2018nda} that chirally symmetric four-fermion and non-minimal interactions are generated and feature an asymptotically safe fixed point.
	\end{itemize}
	Thus we find that quantum gravity generates interaction terms for fermionic matter. These respect the maximum set of symmetries of the kinetic term, irrespective of the number of fermion fields. In addition, these interactions feature a fixed point under the inclusion of quantum gravity, without additional conditions, see \autoref{sec:lightfermions} below. The phenomenological consequences of this result entail that fermions can stay light in the presence of quantum gravity, as we will discuss below.\\
	\item For gauge fields:
	\begin{itemize}
		\item O(\Nvec) symmetry, under which an $\Nvec$-component gauge field transforms in the fundamental representation; $A_{\mu}^a \rightarrow O_{\Nvec}^{ab}A_{\mu}^b$, where $O_{\Nvec}^{ab}$ is a generator in the fundamental representation. Similar to the case for scalar fields, this symmetry is broken if gravitational fluctuations induce distinct anomalous dimensions for some of the gauge fields, or interactions that only involve some of the $\Nvec$ gauge fields. These possibilities are not realized \cite{Eichhorn:2021qet}, indicating that gravitational interactions do not break the global $O(\Nvec)$ symmetry. 
		\item Shift symmetry in a gauge field is nothing but the global part of the Abelian gauge symmetry, which is of course also preserved by gravitational fluctuations.
	\end{itemize}
	Thus we find that quantum gravity generates interaction terms for vector fields. These respect the maximum set of symmetries of the kinetic term, irrespective of the number of vector fields. In addition, these interactions feature a fixed point, if the weak-gravity bound is respected, see \autoref{sec:WGB} below.
\end{itemize}

\subsubsection{Step 2: The symmetry-preserving interactions which are generated by gravity feature an asymptotically safe fixed point}
\label{sec:WGB}

Here we proceed in two steps. We first explore, which interactions are generated by gravity. Second, we  explore under which conditions these interactions feature an asymptotically safe fixed point. 
If these conditions are fulfilled by the gravitational fixed-point values, then, together with step 1, the result suggests that global symmetries do indeed remain intact in asymptotically safe gravity matter steps (with the caveats discussed above).\\

\paragraph{\textbf{Step 2 a: Gravity generates new interactions for matter}}

\emph{Synopsis: There are interactions for matter which are necessarily generated by gravity, i.e., which cannot be set to zero consistently at an asymptotically safe fixed point. These interactions satisfy the symmetries of the kinetic term.}

\begin{figure}[!t]
	\centering
	\includegraphics[width=0.35\linewidth,clip=true,trim=4.3cm -.5cm 4.3cm 0cm]{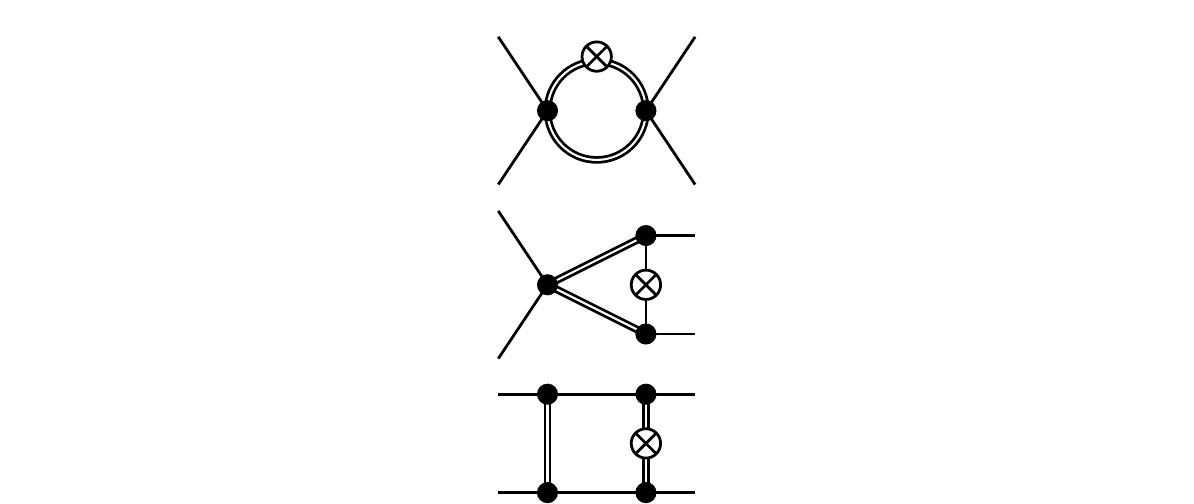}
	\quad
	\includegraphics[width=0.6\linewidth]{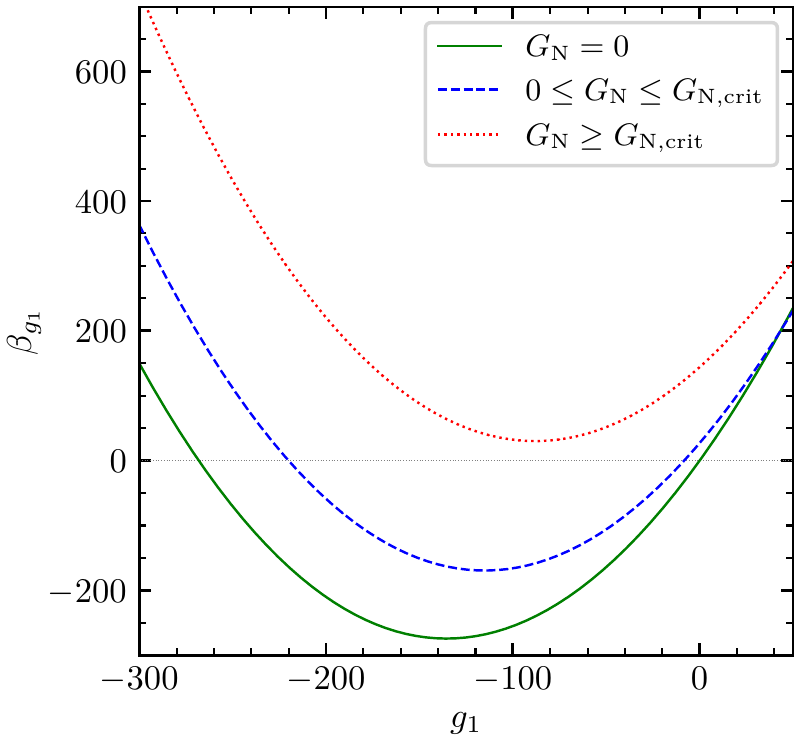}
	\caption{\label{fig:scalpar} Left panel: We show the diagrams through which gravitational fluctuations induce a non-vanishing fixed-point value for $g_1$. The depicted diagrams contribute to the $g_1$ independent contribution $C_0$ of $\beta_{g_1}$, see \eqref{eq:IndSchem}. The circle with cross indicate the regulator insertion $k\partial_k\, \Regk$ of the flow equation. The depicted diagrams with the regulator insertion on all other internal lines, which are not shown separately, also contribute to $C_0$. Right panel: We show the $\beta$-function for the induced coupling $g_1$ defined in \eqref{eq:ScalInd} as a function of $g_1$. For vanishing gravitational fluctuations (green solid line), $g_1=0$ is a fixed point, and $g_1$ can consistently be set to zero. For sufficiently small but non-zero values of the Newton coupling (blue dashed line), $g_1=0$ is not a (partial) fixed point anymore. Increasing the Newton coupling further, the two (partial) fixed points of $\beta_{g_1}$ might collide, such that beyond a critical value of the Newton coupling $\beta_{g_1}$ might not feature any fixed point anymore (red, dotted line), see \autoref{sec:WGB}. }
\end{figure}

At an asymptotically safe fixed point, gravity is interacting. This implies that matter must also have interactions: because gravity couples to any form of energy and matter, it couples to any two free fields, and generates an interaction between them -- already classically. At the quantum level, the same statement is true, i.e., gravity induces interactions also at the loop level. The only way to switch off these induced interactions is to turn off the gravitational coupling, $\GN$. This is not possible when gravity is asymptotically safe; thus asymptotically safe gravity-matter systems necessarily contain interactions for matter. \\
To see this explicitly, let us start with a scalar field $\phi$ which is minimally coupled to gravity and does not have any interactions. The minimal coupling is encoded in the kinetic term of the scalar field
\begin{equation}
\Gamma_{k\, \rm scal}
= \frac{1}{2}\int\mathrm{d}^4x \sqrt{g}\, g^{\mu\nu}\partial_{\mu} \phi \partial_{\nu} \phi\,.
\end{equation}
At the classical level, the presence of the metric in this kinetic term gives rise to gravity-mediated scattering via a tree-level diagram. At the quantum level, loop diagrams generate interaction terms for matter.
In particular, the minimal coupling between the scalar field and gravity gives rise to scalar-gravity vertices, by expanding the kinetic term of the scalar in terms of metric fluctuations. Since we started from the kinetic term, the only coupling appearing in these vertices is the Newton coupling $\GN$. 
We use these vertices in one-loop diagrams with four external scalar fields and a loop of gravitational fluctuations as well as diagrams with gravitational fluctuations and scalars in the loop. Such diagrams generate a scalar self-interaction $g_1$ 
\begin{equation}
\label{eq:ScalInd}
S_{\mathrm{Scal, int.}}=\frac{g_1}{8 k^{4}}\,\int\mathrm{d}^4x \sqrt{g}\, g^{\mu\nu}g^{\rho\sigma}\partial_{\mu} \phi \partial_{\nu} \phi\partial_{\rho} \phi \partial_{\sigma} \phi\,.
\end{equation}
We show the corresponding diagrams in the left panel of \autoref{fig:scalpar}, stressing that all vertices are independent of $g_1$.
These diagrams contribute to the beta function of $g_1$, which is schematically given by
\begin{equation}
\label{eq:IndSchem}
\beta_{g_1}=C_0 + C_1\, g_1 +C_2\, g_1^2 +\mathcal{O}(g_1^3)\,,
\end{equation}
where $C_0$ and $C_1$ are functions of the gravitational couplings. The coefficient $C_0$ contains the contribution of diagrams that do not contain a vertex with four scalar fields and come with $\GN^2$, i.e., those in \autoref{fig:scalpar}, such that $C_0\to 0$ when $\GN\to0$. Therefore, for vanishing gravitational fluctuations, i.e., $\GN=0$, $g_1=0$ is a fixed point, cf.~ the blue line in \autoref{fig:scalpar}. However, when gravitational fluctuations are present, the coefficient $C_0$ is non-vanishing, such that a (partial) fixed point for $g_1$  is necessarily non-zero, i.e., $g_{1,\,*}(\GN  \neq 0)\neq0$. Hence, in the presence of gravitational fluctuations, the coupling $g_1$ is necessarily \emph{induced}, since it cannot be consistently set to zero\footnote{Note that this is not a unique feature of gravity: for a scalar field that is charged under an Abelian gauge field, a finite value of the gauge coupling also induces a four-scalar interaction which is similar to Eq.~(\ref{eq:ScalInd}). In the case of charged matter, however, these interactions are induced  under the RG flow towards the IR, and not necessarily at the fixed point.}. Instead, the (partial) Gaussian fixed point is shifted and becomes an interacting (partial) fixed point, which we will call \emph{shifted Gaussian fixed point} (sGFP) in the following, cf.~the green line in \autoref{fig:scalpar}. Since the sGFP is continuously connected to the Gaussian fixed point, where $g_1$ is irrelevant, the induced coupling also corresponds to an irrelevant direction at the sGFP. Hence, gravitational fluctuations induce new interactions in the matter sector, but do not necessarily introduce new relevant directions, and hence do not reduce the predictivity of the theory.

Crucially, the interaction in Eq.~\eqref{eq:ScalInd} respects the symmetries of the kinetic term, because it essentially corresponds to a square of the kinetic term. Similarly, one can show that higher-order induced interactions, or induced non-minimal interactions, satisfy the same  symmetry-requirement. This completes step 2a: we have reviewed that gravity generates matter interactions, and those satisfy the global symmetries of the kinetic term.

While we introduced the induced coupling in the scalar sector, the same mechanism also takes place in the fermionic \cite{Eichhorn:2011pc} and the gauge sector \cite{Christiansen:2017gtg}, and terms with a larger number of fields. In Tab.~\ref{tab:inducedints} we provide a list of induced interactions that have been explicitly studied in the literature.
Generically, we expect all interactions that respect the symmetries of the kinetic terms, to be induced by gravitational fluctuations.\\
 
\begin{table}[!t]
	\begin{tabular}{c|c|c|c|c|}
		field & global symmetry & selfinteraction & non-minimal inter. & ref.\\\hline\hline
		single scalar & shift & $\left(\partial_{\mu} \phi \partial^{\mu}\phi\right)^2$ &- & \cite{Eichhorn:2012va} \\\hline
		single scalar & shift &- & $\partial_{\mu}\phi \partial_{\nu}\phi R^{\mu\nu}$& \cite{Eichhorn:2017sok} \\\hline
		single scalar & shift & $\left(\partial_{\mu} \phi \partial^{\mu}\phi\right)^2$ & $\partial_{\mu}\phi \partial_{\nu}\phi R^{\mu\nu}$  &\\
		& &  & \& $\partial_{\mu} \phi \partial^{\mu}\phi R$ &  \cite{Laporte:2021kyp, Knorr:2022ilz}\\\hline
		$N_S$ scalars & $N_S$ shift symmetries & $\partial_{\mu}\phi^a \partial^{\mu}\phi^a\,\partial_{\nu}\phi^b \partial^{\nu}\phi^b$ &- & \\
		& \& $O(N_S)$ symmetry & $\partial_{\mu}\phi^a \partial^{\mu}\phi^b\,\partial_{\nu}\phi^a \partial^{\nu}\phi^b$ & -&\cite{deBrito:2021pyi} \\ \hline\hline
		single vector & shift  & $\left(F_{\mu\nu}F^{\mu\nu}\right)^2$ & - & \cite{Christiansen:2017gtg, Eichhorn:2019yzm}\\ \hline
		single vector & shift  & $\left(F_{\mu\nu}F^{\mu\nu}\right)^2$ & - & \\
		& & \& $\left(F_{\mu\nu}\tilde{F}^{\mu\nu}\right)^2$& - & \cite{Eichhorn:2021qet}\\  \hline
		$N_V$ vectors & shift  & $\left(F^a_{\mu\nu}F^{a\,\mu\nu}\right)^2$  & - & \\
		& \& $O(N_V)$ symmetry &  \& $\left(F^a_{\mu\nu}\tilde{F}^{b\,\mu\nu}\right)^2$ & - & \cite{Eichhorn:2021qet}\\ \hline
		\hline
		$\Nferm$ fermions & chiral & $\left(\bar{\psi}^i\gamma_{\mu}\psi^i\right)\left(\bar{\psi}^j\gamma^{\mu}\psi^j\right)$  & - & \\
		& & \&$ \left(\bar{\psi}^i\gamma_{\mu}\gamma_5\psi^i\right)\left(\bar{\psi}^j\gamma^{\mu}\gamma_5\psi^j\right)$ &-& \cite{Eichhorn:2011pc, Meibohm:2016mkp, deBrito:2020dta}\\\hline
		$\Nferm$ fermions & chiral & - & $R^{\mu\nu}\, \bar{\psi}^i\gamma_{\mu}\nabla_{\nu}\psi^i$ & \cite{Eichhorn:2018nda} \\ \hline \hline
		single scalar & shift \& chiral & $\left(\bar{\psi}\gamma^{\mu}D_{\nu}\psi\right)\left(\partial_{\mu}\phi\partial^{\nu}\phi\right)$  & - & \\
		\& single fermion & & \&$ \left(\bar{\psi}\slashed{D}\psi\right)\left(\partial_{\nu}\phi\partial^{\nu}\phi\right)$ &-& \cite{Eichhorn:2016esv, Eichhorn:2017eht}\\\hline
	\end{tabular}
	\caption{\label{tab:inducedints}We list the interactions divided by $\sqrt{g}$ (and corresponding references) that were explicitly shown to be generated by quantum gravity. They all satisfy continuous global symmetries which are the maximum continuous global symmetries of their respective kinetic terms.
	}
\end{table}

\paragraph{\textbf{Step 2 b: The weak-gravity bound as a condition under which a symmetry-preserving fixed point exists}}

\emph{Synopsis: It is non-trivial to satisfy the condition that all generated interactions from Step 2a feature an asymptotically safe fixed point. While some of them feature a  partial fixed point for any value of gravitational couplings, others only feature a  partial fixed point if gravity is sufficiently weakly coupled, i.e., if $G_{\rm eff}$ and its generalizations are sufficiently small. Such  a weak-gravity-bound (WGB) has been discovered for scalars, vectors, and for scalars coupled to fermions, although not for fermions on their own.}

The WGB arises, because the beta function in Eq.~\eqref{eq:IndSchem} only has real zeros under certain conditions on the coefficients $C_i$. If we neglect contributions $\mathcal{O}(g_i^3)$, the sGFP is only real for
\begin{equation}
\label{eq:WGBCond}
4C_0\,C_2\leq C_1^2\,.
\end{equation}
For some systems, such as four-fermion couplings, this condition is satisfied automatically, see \autoref{sec:lightfermions}. For others, such as four-scalar couplings, this condition is only fulfilled up to critical values of the gravitational couplings. In fact, if gravity is too strongly coupled, i.e., simply put, $G_{\rm eff}$ (the strength of metric fluctuations) is too large, then the condition is violated. Therefore, the corresponding bound on the couplings is called the \emph{weak-gravity bound}.

\begin{figure}[!t]
	\centering
	\includegraphics[width=0.45\linewidth]{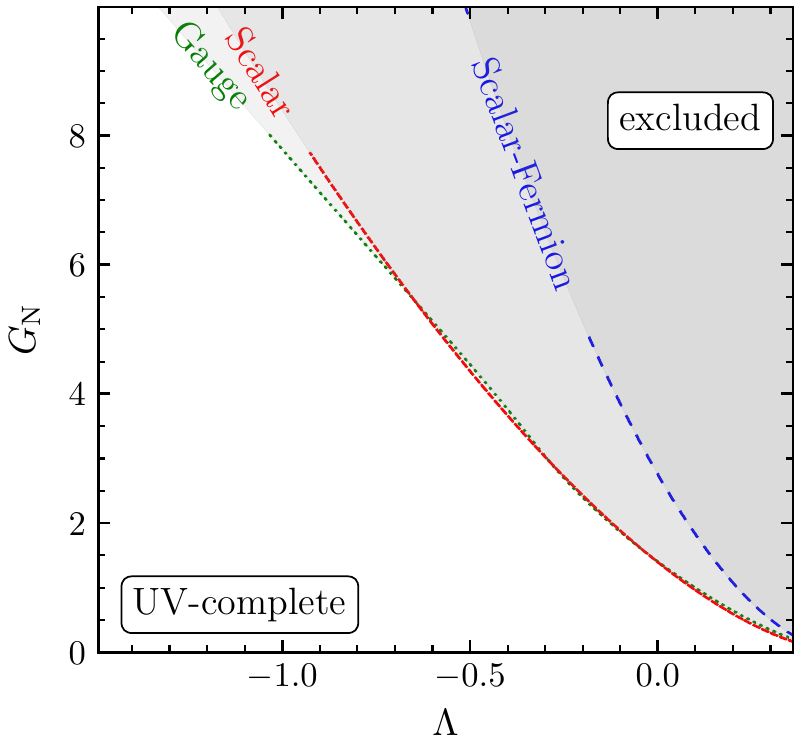}\quad
	\includegraphics[width=0.45\linewidth]{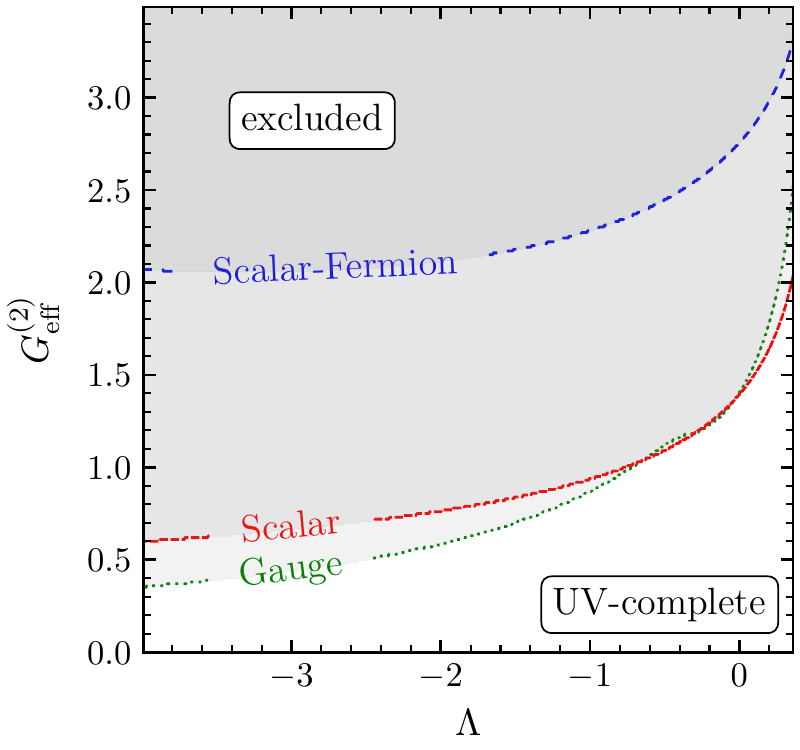}
	\caption{\label{fig:WGBScalars} We show the WGB for a shift symmetric scalar, see \eqref{eq:ScalInd}, an Abelian gauge field, see \eqref{eq:GaugeInd}, and for a scalar-fermion system \eqref{eq:ScalFermInd}. In the gray region, no partial fixed point for the respective induced coupling exists, such that asymptotic safety cannot be realized in the respective region of the gravitational parameterspace.. It is therefore excluded from the viable parameterspace. In the left panel we show the WGB in the plane spanned by the gravitational coupling $\Lambda$ and $\GN$, indicating that for any value of $\Lambda$, there is a critical value ${\GN}_{,\mathrm{crit}}$ at which the asymptotically safe partial fixed point for the induced coupling vanishes. In the right panel we show the same WGB, but in a plane spanned by $\Lambda$ and the effective gravitational coupling $G_{\rm eff}^{(2)}$, see \eqref{eq:Geffn}.}
	\end{figure}

When studying induced interactions and the WGB, one usually proceeds order by order in canonical dimension. All induced interactions are canonically irrelevant, because they are essentially the square, or higher powers, of the kinetic terms.
For a single shift symmetric scalar field, the scalar self interaction (\ref{eq:ScalInd}) 
is the only self-interaction up to this level of the canonical mass dimension. 
The coupling $g_1$ is indeed induced by gravitational fluctuations \cite{Eichhorn:2012va, deBrito:2021pyi, Laporte:2021kyp}, since the coefficient $C_0$ in Eq.~(\ref{eq:IndSchem}) is non-zero in general. This induced interaction gives rise to a WGB \cite{Eichhorn:2012va, deBrito:2021pyi}, which excludes a part of the plane spanned by the Newton coupling $\GN$ and the cosmological constant $\Lambda$ from the viable gravitational parameter-space, see the left panel of \autoref{fig:WGBScalars}. 
In $G_N$ and $\Lambda$ it is less straightforward to see where the strong-coupling-regime is, because $\Lambda \rightarrow 1/2$ is also a strong-coupling limit, not just $G_N \gg 1$. We therefore work in terms of the effective strength of metric fluctuations, $G_{\rm eff}$ and $G_{\rm eff}^{(2)}$, defined in Eq.~\eqref{eq:Geff}, and Eq.~\eqref{eq:Geffn}, respectively.
 Both $G_{\rm eff}$ and $G^{(2)}_{\rm eff}$ can be thought of as measures of the strength of metric fluctuations and dominate the scale dependence of induced matter interactions. As we can see in the right panel of  \autoref{fig:WGBScalars}, the WGB is described by a rather constant value of $G^{(2)}_{\rm eff}$ , which enters the diagrams in \autoref{fig:scalpar} at leading order, for some range of $\Lambda$. Deviations from the constant appear due to dependencies on $G_{\rm eff}$.

When adding more scalar fields to the system, only those interactions that respect the $O(\Nscal)$ symmetry of the kinetic term are induced by gravitational fluctuations \cite{deBrito:2021pyi}. Increasing the number of scalar fields $\Nscal$ makes the WGB stronger, such that more gravitational parameterspace is excluded.

	For fermions, the leading-order generated interactions are special, because they contain information not just about global symmetries. Because the maximum symmetry for fermions is chiral, i.e., distinguishes left- and right-handed fermions, the generated interactions also contain information about fermion masses. Thereby, they provide an important observational consistency test of asymptotic safety, because fermion masses in the SM are measured. We therefore devote \autoref{sec:lightfermions} to the discussion of generated fermion interactions and their phenomenological consequences. For the present section, it is only relevant that no WGB has been discovered yet in purely fermionic systems.\\
	The situation is different for systems with fermions and scalars. The leading-order induced interactions are
	\begin{equation}
	\begin{aligned}
	\label{eq:ScalFermInd}
	S_{\mathrm{Scal-Ferm ,int.}}=\frac{i}{k^{4}}\,\int\mathrm{d}^4x&\left(\chi_1[\bar{\psi}\gamma^{\mu}D_{\nu}\psi-(D_{\nu}\bar{\psi})\gamma^{\mu}\psi]\partial_{\mu}\phi\partial^{\nu}\phi\right.\\
	&+\left. \chi_2[\bar{\psi}\gamma^{\mu}D_{\mu}\psi-(D_{\mu}\bar{\psi})\gamma^{\mu}\psi]\partial_{\nu}\phi\partial^{\nu}\phi\right)\,,
	\end{aligned}
	\end{equation}
	which indeed give rise to a WGB \cite{Eichhorn:2016esv, Eichhorn:2017eht}.

	For a single gauge field and at lowest order in the canonical mass dimension there are two linearly independent induced four-vector interactions, namely
	\begin{equation}\label{eq:GaugeInd}
	S_{\mathrm{Vector ,int.}}=\frac{1}{8 k^{4}}\,\int\mathrm{d}^4x\left(w_2 (F_{\mu\nu}F^{\mu\nu})^2+\kappa_2(F_{\mu\nu}\tilde{F}^{\mu\nu})^2\right)\,,
	\end{equation}
	with the field strength tensor $F$ and the dual field strength $\tilde{F}$. These interactions give rise to a WGB \cite{Christiansen:2017gtg, Eichhorn:2021qet}, which generalizes to the case of several gauge fields \cite{Eichhorn:2021qet}.

	Comparing the WGBs from the different sectors -- scalar, scalar-fermion and gauge -- we find a nearly universal curve, see \autoref{fig:WGBScalars}. This indicates that the different sectors are equally sensitive to gravitational fluctuations.\\
	
	\FRT{Further reading:} \\
	
	\FR{Preservation of global symmetries: beyond truncations}\\
	As in any computation relying on the FRG, practical computations require choosing a truncation, i.e., only take a finite subset of interactions into account. All statements above on global symmetry breaking are therefore made within a truncation. However, from the structure of the flow equation, one can infer that the specific statements above, which pertain to the non-generation of symmetry-breaking interaction terms, generalize beyond truncations, see \cite{Eichhorn:2020mte} for a discussion. Further, \cite{Laporte:2021kyp} contains a proof of shift-symmetry-preservation in scalar-gravity systems.\\
	
	\FR{WGB: condition to prevent symmetry breaking and constraint on gravitational parameter-space}\\
	There are two points of view on the WGB: the first is as a necessary condition to prevent the breaking of global symmetries; the second is as a condition on microscopic gravitational couplings that arises because matter-gravity-theories should feature fixed points. This second view has mostly been discussed in the literature, see \cite{Eichhorn:2011pc, Eichhorn:2012va, Meibohm:2016mkp, Eichhorn:2016esv, Eichhorn:2017eht, Eichhorn:2017sok, Christiansen:2017gtg, Eichhorn:2018nda, Eichhorn:2019yzm, deBrito:2020dta, deBrito:2021pyi,  Laporte:2021kyp, Eichhorn:2021qet, Knorr:2022ilz}.\\
	
	\FR{The WGB and asymptotically safe gravity-matter systems}\\
	Ultimately, we would like to know if scalar-gravity, or more generally, gravity-matter systems can be asymptotically safe with the maximum set of symmetries. This is the case if gravitational fixed-point values in the presence of matter satisfy the WGB. For systems with $\Nvec$ Abelian gauge fields  this is the case in all studies to date. For scalars, the answer changes, when nonminimal interactions are accounted for: For $\Nscal$ minimally coupled scalars, there is no fixed point of the full scalar-gravity systems, i.e., the WGB is violated \cite{deBrito:2021pyi}. At nonminimal coupling, the WGB holds for a single scalar \cite{Laporte:2021kyp, Knorr:2022ilz}. It is an open question whether this result may change under the inclusion of further interactions and whether the WGB is violated at $\Nscal>1$.\\
	
	\FR{Robustness of the WGB}\\
	As a test of robustness, gauge parameter dependence of the WGB has been studied in \cite{deBrito:2021pyi} and in \cite{Eichhorn:2021qet} and is weak in both cases. \cite{Eichhorn:2021qet} also demonstrates that it is important to include a full basis of generated interactions at leading order in canonical power counting. If not all interactions are included, the results strongly depend on the gauge on a qualitative level.\\
	In \cite{deBrito2022WIP} the WGB for a single scalar field has been investigated by considering a minimal coupling of gravity to a full function of the kinetic term for the scalar field. By studying the fixed-point structure of the pure-matter system upon expansion of the full function, it was concluded that only the free fixed point is a viable fixed point of the matter system, see also \cite{Laporte:2022ziz}. Accordingly, the WGB, which in this system arises as a collision between the (shifted) GFP and an interacting "fixed point", is interpreted as a truncation artifact. Furthermore, the WGB only appears in specific expansions of the full function of the kinetic term. In summary, it is not established whether the WGB is a strict boundary for asymptotically safe theories, or whether it merely indicates the transition to a more strongly-coupled regime, in which significantly larger truncations of the system are required to produce robust results.  In any case, a near-perturbative fixed point, which matches seamlessly to the perturbative SM at the Planck scale, is very likely incompatible with gravitational couplings beyond the WGB found in small truncations.

%% file: Input/Lightfermions.tex
\subsection{Light fermions}
\label{sec:lightfermions}
\emph{Synopsis: All fermions  in the SM are light compared to the Planck scale. This is a consequence of chiral symmetry, which prevents the generation of fermion masses, and which is only broken spontaneously at  the electroweak scale by the Higgs mechanism and below by QCD. In asymptotically safe quantum gravity, there are several conceivable mechanisms to break chiral symmetry, which would lead to inconsistencies with the observation of light fermions.  Avoidance of some of these mechanisms puts lower and upper bounds on the number of light fermions. For the number of fermions in the SM, chiral symmetry is not broken in asymptotically safe quantum gravity.} \\

The masses of the fermions in the SM range from several hundred keV to several $\mathrm{GeV}$, hence they are very light compared to the Planck scale. The reason for this is that chiral symmetry in the SM is only broken spontaneously at the electroweak scale. 
A chiral symmetry allows to rotate left and right-handed fermions $\psi_L$ and $\psi_R$\footnote{The left- and right-handed component of a Dirac fermion can be extracted through the projection operators $P_{R/L} = \frac{1}{2}\left(1\pm \gamma_5 \right)$ as $\psi_{R/L}= P_{R/L}\psi$.} independently. Depending on the number of fermions and their other symmetries, different symmetries can be chiral. For instance, at vanishing Yukawa couplings the quark sector of the Standard Model features an $SU(\Nferm)_L \times SU(\Nferm)_R$ symmetry, which rotates the $\Nferm$ quark flavors into each other separately for left- and right-handed components.
A mass term $m_{\psi}\,\bar{\psi}\psi= 1/2 m_{\psi} \left(\bar{\psi}_R \psi_L + \bar{\psi}_L \psi_R \right)$ breaks this symmetry explicitly, see also \autoref{sec:GlobalSymms}. If chiral symmetry is broken -- explicitly or spontaneously -- at an energy scale $k_{\chi\rm{SB}}$, quantum fluctuations automatically generate a mass-term for fermions, since it is allowed by the symmetries of the theory, see the discussion in \autoref{sec:GlobalSymms}.  Assuming that there is no fine-tuning, the generated, dimensionless mass is of order one at $k_{\chi \rm SB}$. Therefore, they decouple from the RG flow\footnote{Within the functional RG, this decoupling happens automatically due to the built-in threshold effects. Within perturbative RG schemes, the corresponding decoupling and matching at this scale has to be done by hand.} at $k_{\chi\rm SB}$ and their dimensionful masses are $\bar{m} = m\cdot k_{\chi\rm SB}$, where $m$ is of order one.

Therefore, light fermions in asymptotically safe quantum gravity can only be accommodated if gravity does not break chiral symmetry.
This breaking of chiral symmetry could in principle occur on different levels\footnote{ Not all of which are relevant for SM fermions, because some of them would be in conflict with SU(2) gauge symmetry.}: first,  interactions that break chiral symmetry explicitly (such as a mass term) could be induced by quantum fluctuations; second, the chirally symmetric subspace could feature no fixed point, necessitating that an asymptotically safe theory contains chiral-symmetry-breaking couplings; third, chiral symmetry might be broken spontaneously during the flow towards the IR; and fourth, the background spacetime might break chiral symmetry. In this last possibility, thermal fluctuations (as they are relevant, e.g., in the very early universe), could also play a role.
We will now discuss these four possibilities.

First, we consider the explicit breaking of chiral symmetry by induced interactions. In asymptotically safe gravity, there are no indications that this occurs. Technically, this is because in all studies so far, the coefficient $\beta_0$ in \eqref{eq:symmbreak} vanishes for interactions that break chiral symmetry explicitly \cite{Eichhorn:2016vvy, Daas:2020dyo}, unless a regulator is chosen that explicitly breaks chiral symmetry.
\\
Second, we focus on chirally symmetric, induced interactions, and ask whether they feature a fixed point. In all studies to date, the answer is positive \cite{Eichhorn:2011pc, Meibohm:2016mkp,  Eichhorn:2017eht, Eichhorn:2018nda, deBrito:2020dta}.
Just like for other matter fields, gravity induces self-interactions for fermions. Fermions are special with respect to the induced self-interactions, since the induced four-fermion interactions with the lowest mass dimension are of dimension six.  This is a lower dimension than  for induced interactions in the other sectors. One could therefore expect that gravity can more easily make the corresponding couplings relevant. 
These interactions are of the form\footnote{While both $\lambda_{\pm}$ are induced by gravitational fluctuations, there is a different basis, where only one interaction is induced \cite{Eichhorn:2011pc, Eichhorn:2017eht}, while the second linearly independent four-fermion interaction can be set to zero consistently. Hence, this four-fermion interaction is the only example discovered so far, where gravitational fluctuations do not induce an interaction that satisfies the symmetries of the kinetic term.}
\begin{equation}
S_{\mathrm{Ferm, int.}}=\frac{1}{2 k^{2}}\,\int\mathrm{d}^4x \sqrt{g}\,\left(\lambda_{-}(V-A)+\lambda_{+}(V+A)\right)\,,
\end{equation}
with
\begin{equation}
V=\left(\bar{\psi}^i\gamma_{\mu}\psi^i\right)\left(\bar{\psi}^j\gamma^{\mu}\psi^j\right)\,,\qquad A=-\left(\bar{\psi}^i\gamma_{\mu}\gamma_5\psi^i\right)\left(\bar{\psi}^j\gamma^{\mu}\gamma_5\psi^j\right)\,.
\end{equation}

Explicit studies of the above chirally symmetric four-fermion interactions confirm that gravitational fluctuations induce those \cite{Eichhorn:2011pc}, see also \cite{Meibohm:2016mkp, Eichhorn:2017eht}. However, induced four-fermion interactions do not feature an excluded strong gravity regime, where the chirally symmetric subsector would be UV-incomplete. This is independent of the strength of gravitational fluctuations. Thus, also the second possibility for chiral-symmetry-breaking is ruled out in the studies to date. This can be seen by inspecting the two beta functions $\beta_{\lambda_{\pm}}$, which feature four fixed points for any positive value of $G$ and any value of $\Lambda$:
\begin{equation}
\begin{aligned}
\label{eq: betaplusminus}
\beta_{\lambda_{\pm}}=&\,\,2\lambda_{\pm}+M_{\pm} 
-\frac{5 \lambda _{\pm} G}{8 \pi  (1-2 \Lambda )^2}\pm\frac{5 G^2}{8 (1-2 \Lambda )^3}\\
&+\frac{3 \lambda _{\pm} G}{4 \pi  (3-4 \Lambda )}+\frac{15 \lambda _{\pm} G}{8 \pi  (3-4 \Lambda )^2}\,,
\end{aligned}
\end{equation}
where the matter contributions $M_{\pm}$ are given by
\begin{equation}
M_+=\,\,\frac{8 \lambda _+ \left(\lambda _- \left(N_F+1\right) 
	\right)
}{32 \pi ^2}\,,\qquad
M_-=\,\,\frac{4 \lambda _-^2 \left(N_F-1\right)+4 \lambda _+^2 N_F
}{32 \pi ^2}\,.
\end{equation}
These beta functions admit four partial fixed points, one of which is the shifted Gaussian fixed point of interest, for all values of $G$ and $\Lambda$.

Next, we consider the spontaneous breaking of chiral symmetry by gravitational fluctuations. The spontaneous breaking of chiral symmetry is linked to the formation of bound states, as we will explain below. Because gravity is an attractive force, one may expect that it favors bound-state formation. However, explicit calculations show that this intuition, based on the classical nature of gravity, fails to correctly predict the effect of quantum gravitational fluctuations.\\

To understand the relation between the spontaneous breaking of chiral symmetry, the associated massless Goldstone boson and the induced four-fermion interactions $\lambda_{\pm}$, we first perform a Fierz transformation into a scalar-pseudoscalar basis. Focusing on $\lambda_+$, the transformation reads \cite{Gies:2001nw, Braun:2011pp}
\begin{equation}
\left(V+A\right)=-\frac{1}{2}\left[\left(\bar{\psi}^i\psi^j\right)^2-\left(\bar{\psi}^i\gamma_5\psi^j\right)^2\right]\,.
\end{equation}
In this basis, the four-fermion interactions can be rewritten in terms of auxiliary fields using a Hubbard-Stratonovich transformation, i.e., a change of fields in the path integral, see, e.g., \cite{Braun:2011pp}. Focusing on the case of a single flavor for illustration, the scalar part of the four-fermion interaction can be rewritten as
\begin{equation}
\label{eq:hubstrat}
-\frac{\lambda_{\psi}}{4}\left(\bar{\psi}\psi\right)^2=\left[h(\bar{\psi}\psi)\phi+m^2_{\phi}\phi^2\right]_{\mathrm{EoM}(\phi)}\,,
\end{equation}
where
\begin{equation}
\label{eq:ffrel}
m^2_{\phi}=\frac{h^2}{\lambda_{\psi}}\,,
\end{equation}
and where $h$ is some arbitrary, real, constant, which holds on the equation of motion for the scalar field $\phi$, i.e., when the auxiliary field $\phi$ is integrated out in the path integral. 

\begin{figure}[!t]
\includegraphics[width=\linewidth,clip=true, trim=3cm 16cm 8cm 2cm]{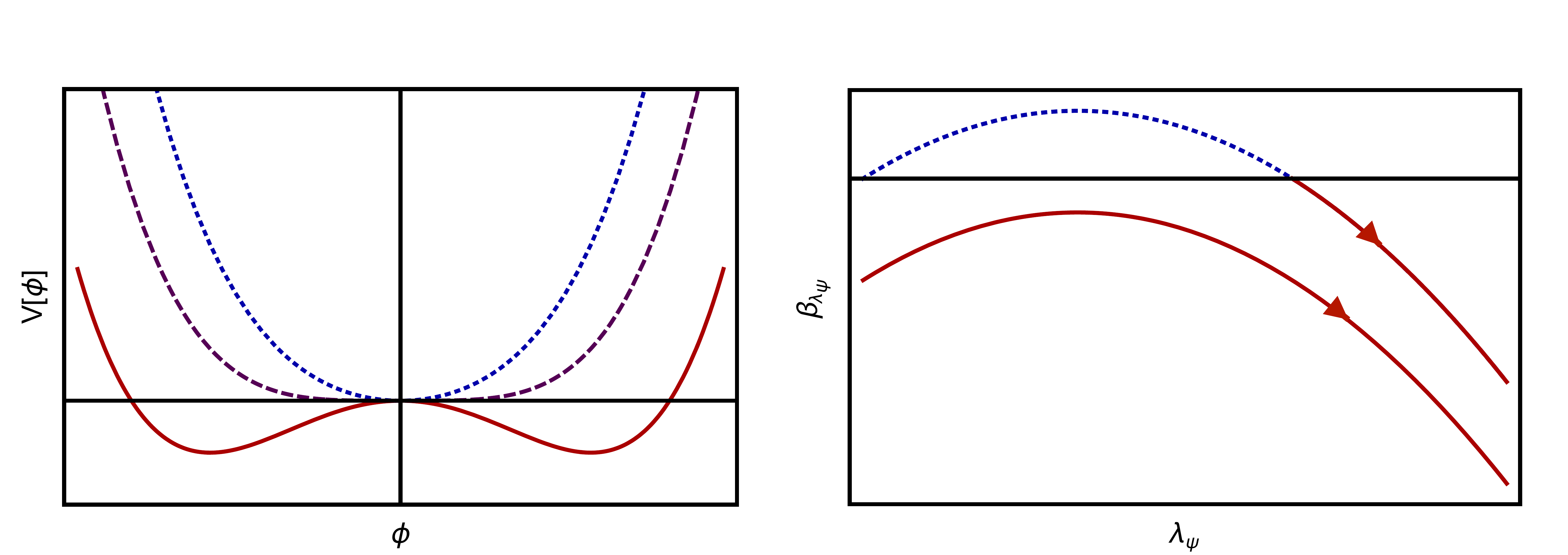}
\caption{\label{fig:chSB} We show the schematics of spontaneous chiral symmetry breaking: The effective potential for the bound-state field $\phi$ has a nontrivial minimum -- breaking chiral symmetry, because a vev for $\phi$ corresponds to a vev for the fermion bilinear $\bar{\psi}\psi$ -- after going through the point $m_{\phi}^2=0$ (purple dashed curve). In turn, the mass is inversely proportional  to $\lambda_{\psi}$. Thus, the beta function in red corresponds to a chirally broken regime, because $\lambda_{\psi}$ diverges. For the other beta function, the initial conditions can be chosen in the blue dashed region, corresponding to finite $\lambda_{\psi}$ and thus unbroken chiral symmetry. The left fixed point corresponds to the sGFP, while the other one is interacting, even in the absence of additional fields.}
\end{figure}

In terms of the scalar field $\phi$, chiral symmetry is spontaneously broken, when the mass term $m^2_{\phi}$ becomes negative. Since the mass term is related to the four-fermion interaction, see \eqref{eq:ffrel}, the onset of chiral-symmetry breaking is indicated by a divergence of the four-fermion interaction. This argument, which we exemplified on a single channel and for a single flavor, generalizes to the full system of $\Nferm$ flavors, see also \cite{Braun:2011pp} for a review. Hence, if the four-fermion interaction diverges, quantum gravity has broken chiral symmetry. This could happen (and does happen, e.g., in QCD \cite{Braun:2011pp}), when the beta function for the four-fermion interaction has no fixed points and is negative. Then, under the RG flow towards the infrared, the four-fermion interaction diverges.
However, since gravitational fluctuations do not induce a fixed-point collision of the SGFP, the four-fermion interaction cannot grow beyond a bound. Hence, chiral symmetry is not broken spontaneously by gravitational fluctuations.\\

This is very different to fluctuations of gauge fields, which induce the same four-fermion interactions and indeed give rise to a fixed-point collision during the flow towards the IR, which can be related with spontaneous chiral-symmetry-breaking, see, e.g., \cite{Gies:2001nw, Braun:2005uj, Braun:2006wu, Braun:2011pp}.\\
Combining the result from the gravitational and the gauge sector, one can obtain a lower bound on the number of fermions:
If one assumes that an interacting fixed point for the  Abelian gauge coupling is realized, see \autoref{sec:Gaugesector} for details, this can indeed give rise to broken chiral symmetry: gravitational fluctuations set a fixed-point value $\gyast$ for the gauge coupling, which, in analogy to the situation in QCD, can induce a fixed-point collision in $\lambda_{\pm}$. This can be seen from the additional contributions proportional to the gauge coupling that arise in Eq.~\eqref{eq: betaplusminus}; where a contribution $\sim \gy^4$ arises in $\beta_{\lambda_+}$ and a contribution $\sim -\gy^4$ in $\beta_{\lambda_-}$.\\
Since $\gyast$ decreases when increasing the number of fermions in the system, this mechanism gives a lower bound on the number of fermions such that chiral symmetry is unbroken, see \cite{deBrito:2020dta}.
Similarly, if the presence of topology-changing gravitational instantons are assumed, chiral symmetry can also become anomalous and be broken spontaneously \cite{Hamada:2020mug}. \\

The fourth possibility through which quantum gravity might break chiral symmetry is via the background geometry. This is a mechanism which is already active for fermions on classical spacetimes: an anti-deSitter background adds an ``effective" negative mass term for fermions and thus breaks chiral symmetry through gravitational catalysis. One can think of this mechanism as follows: the effective potential for the fermion bilinear (or the corresponding scalar field) depends on the background geometry. An anti-deSitter geometry of sufficiently negative curvature favors a non-trivial ground state, and thus chiral symmetry breaking, over the trivial ground state \cite{Inagaki:1997kz, Ebert:2008pc}.  The pertinent quadratic term of the effective potential for the scalar $\phi$ is given by \cite{Gies:2018jnv}
\be
U(\phi)= -N_f \phi^2 \left(\# \frac{|R|^{3/2}}{k_{\rm IR}} + \xi\, |R| \right),
\ee
with $R<0$ and where $\#$ is a positive number that depends on the details of the regularization. This gives rise to a curvature bound that parametrically depends on the nonminimal coupling $\xi$.\\
Of course, our universe is not an anti-deSitter spacetime. Nevertheless, gravitational catalysis can become relevant, because the small-scale geometry of quantum spacetime could have both negative- and positive-curvature regions and anti-deSitter is an effective description of simple negative-curvature regions.\\
Using this reasoning, in \cite{Gies:2018jnv, Gies:2021upb}, the curvature bounds were used to constrain asymptotically safe gravity: because an increasing number of fermions shifts the (background) cosmological constant to negative values, and thus shift the background curvature to large negative values, gravitational catalysis may become active at large enough fermion numbers. 
Thus,
this mechanism gives rise to an upper bound on the number of fermions which can be light \cite{Gies:2018jnv, Gies:2021upb}. The exact value of this upper bound depends on the assumed structure of the space-time geometry, and on the presence and strength of thermal fluctuations \cite{Gies:2018jnv, Gies:2021upb}.

%% file: Input/dgreater4.tex
\section{Gravity-matter systems in $d\neq4$ dimensions}
\label{sec:dgreater4}
\emph{Synopsis:  Experiments support the hypothesis that we live in a $3+1$-dimensional spacetime. We do not, however, know, why this is the case, or whether it could be different. Here, we review evidence that asymptotic safety of gravity with Standard Model matter may not be achievable in dimensions much beyond four; i.e., evidence that the predictive power of asymptotic safety may extend to free parameters of the geometry of spacetime.}\\

Current observations indicate that our universe is four dimensional (or rather, 3+1 dimensional), at least down to  length scales corresponding to an energy of $\sim 10$ TeV. Accordingly, our universe might be of higher dimension in the deep UV, if the additional dimensions are compact and therefore inaccessible at low energies.\footnote{Here, we refer to the topological dimension. There are in fact indications that other notions of dimensionality, most importantly the spectral dimension, instead exhibit a dynamical reduction in the UV, \cite{Lauscher:2005qz,Reuter:2011ah,Rechenberger:2012pm,Calcagni:2013vsa}. Such different notions of dimensionality are not related to each other and may thus exhibit differences in the UV.} Indeed, in string theory, such extra dimensions are necessary for the internal consistency of the theory. It is therefore interesting to understand what the status of extra dimensions is in other approaches to quantum gravity.\\
The compatibility of extra dimensions with the asymptotic-safety paradigm for quantum gravity and matter has been tested by studying i) the impact of matter fields on the gravitational fixed point ii) mechanisms for a UV complete matter sector, and iii) gravitational contributions to LHC scattering amplitudes to connect to observational constraints on extra dimensions. We will briefly discuss these results in the following.\\
\\
First, the coefficients $b_{\mathrm{grav}}$ and $a_{i}$ in \eqref{eq:betaGN} are dimension-dependent. All studies so far indicate that gravitational contributions remain anti-screening in $d>4$, i.e., $b_{\mathrm{grav}}>0$ in \eqref{eq:betaGN} \cite{Litim:2003vp, Fischer:2006fz, Ohta:2013uca}, such that pure gravity can become asymptotically safe in a larger number of dimensions. Since the gravitational contribution and the matter contributions scale differently with the dimensionality, see \cite{Dona:2013qba} for an explicit example, bounds on the number of matter fields may arise in $d>4$.  It may thus be the case that asymptotic safety of gravity with Standard-Model matter is not achievable in $d>4$, see \cite{Dona:2013qba} for a first study of this question. A systematic investigation of this question, which lifts the approximations made in  \cite{Dona:2013qba}, has not been completed yet. \\
\\
Second, if we assume that the higher-dimensional theory features a fundamental Abelian gauge coupling, demanding a UV completion constrains the number of dimensions.
This is because the triviality problem in the Abelian gauge sector becomes more severe in larger dimensions, where the Abelian gauge coupling has a negative canonical mass dimension. This acts akin a screening contribution to the scale dependence of the gauge coupling. Hence, to induce asymptotic freedom in $d>4$ in the gauge sector, the anti-screening gravitational contribution has to overcome this screening dimensional contribution, see also \autoref{sec:Gaugesector}. This requires that $f_g$ (cf.~Eq.~\eqref{eq:matterbetaschem}) increases with increasing dimensionality.
Explicit studies indicate that this only possible if gravitational fluctuations become stronger when we increase the dimensionality \cite{Eichhorn:2019yzm, Schiffer:2021gwl, Eichhorn:2021qet}. Hence, a UV-completion of the Abelian gauge sector is only possible for a more strongly-coupled gravity theory. However, such a strongly coupled regime might be excluded due to the WGB in the Abelian gauge sector, see \autoref{sec:WGB}. Indeed, according to the studies in \cite{Eichhorn:2019yzm, Schiffer:2021gwl, Eichhorn:2021qet}, it is not possible to reconcile the strong coupling required to solve the triviality problem with the weak coupling required to satisfy the weak-gravity bound, if $d\geq6$.\\
\\
Third, a more phenomenological approach to extra dimensions was taken, prior to the start of the LHC, with the hope of constraining asymptotic safety in large extra dimensions by experiment.
If the extra dimensions are large enough, the fundamental Planck scale is close to $TeV$ scales, see \cite{Arkani-Hamed:1998jmv}. Accordingly, scattering processes at the $TeV$ scale would be sensitive to the production or exchange of virtual Kaluza-Klein-gravitons. These would leave an imprint on scattering amplitudes of SM particles, for example by missing energy signatures. In \cite{Litim:2007iu, Litim:2007ee, Gerwick:2011jw} this scenario was investigated within asymptotically safe quantum gravity\footnote{We caution here that the scattering cross sections in \cite{Litim:2007iu, Litim:2007ee, Gerwick:2011jw} were computed within an approximation where the RG scale $k$ is identified with the physical momentum scale of a scattering process. A more careful investigation of gravity-mediated scattering amplitudes requires to encode the full momentum dependence of the vertices, e.g., by means of form factors, see \cite{Knorr:2019atm, Draper:2020bop}. Recent studies of gravity-mediated scattering processes indicate that scattering amplitudes can indeed be finite, despite the presence of trans-Planckian modes \cite{Draper:2020bop}. We refer the reader to the chapter on form factors and scattering amplitudes \cite{Knorr:2022dsx} for details.}. Specifically, it was found that the gravitational di-lepton production at LHC energies would be well above SM backgrounds, if the fundamental Planck scale was at the $TeV$ scale. Similarly, \cite{Dobrich:2012nv} found that gravity-mediated photon-photon-scattering can rise above the SM background (which in this case is a pure loop effect, with no tree-level contribution).
To date, no such signatures were discovered,  constraining the radii of large extra dimensions.\\

At a more formal level, it is also of interest to explore the gravitational dressing of matter theories with asymptotic safety in $d=3$ or $d=2$. Such matter theories encode universal critical behavior at continuous phase transitions; with the Wilson-Fisher fixed point a paradigmatic example. For such systems in statistical physics, gravitational fluctuations are not expected to be relevant, because the Planck length is much smaller than the atomic scale in those systems, where the quantum-field-theoretic description breaks down. Nevertheless, it is of interest to understand whether such universality classes persist and are gravitationally dressed. This sheds light on how two distinct, asymptotically safe theories can be brought together. It may even, within an AdS/CFT-type of correspondence, become of interest for quantum gravity in a more indirect way.\\
 Explicit studies of gravitationally dressed universality classes have started from the Wilson-Fisher fixed point \cite{Percacci:2015wwa} and considered universality classes in three-dimensional $O(N)$ models in \cite{Labus:2015ska}, finding evidence that these universality classes can be dressed gravitationally.

%% file: Input/SMUVcompletion.tex
\label{sec:SMUVcompletion}
\emph{Synopsis: The Standard Model is not UV complete on its own. Under the impact of asymptotically safe gravity, it may become UV complete. The current state of the art sees two distinct possibilities for this UV completion: either, all SM couplings vanish in the far UV; or some of them are nonzero. Except for the Higgs selfcoupling, all couplings that vanish in the far UV correspond to free parameters of the theory. Therefore, an asymptotically safe gravity-matter theory has at least one free parameter less in the SM sector than the SM on its own. Each coupling that is nonzero in the far UV changes its status from a free parameter to a calculable quantity. Thereby, quantities like the low-energy value of the Abelian gauge coupling or the top Yukawa coupling may be calculable from first principles. In turn, these calculations provide observational tests of the asymptotically safe theory.\\
In the future, when the systematic uncertainty of calculations is decreased further, only one or none of these two possibilities will remain phenomenologically viable, because the fixed points at nonzero values generate upper bounds on the values of the couplings. If these upper bounds are lower than the measured values, then neither the fixed point at zero value nor the fixed point at nonzero value remains accessible. There is thus the distinct possibility of ruling out an asymptotically safe theory of gravity and the SM. Given the lack of experimental guidance on quantum gravity, such a result would be important progress.}\\

The SM accurately describes all visible matter at low energies and has been experimentally tested up to T$eV$ scales. Nevertheless, the SM is only an effective description of matter, which breaks down at high energies.\footnote{This is due to Landau poles in the Higgs sector \cite{Cabibbo:1979ay} and Abelian gauge sector \cite{Gockeler:1997dn}. It remains a theoretical possibility that there is a nonperturbative UV completion of the SM as a whole, even though these separate sectors are not UV complete due to a triviality problem.
There is, however, no indication that this theoretical possibility is realized.} It is therefore crucial to investigate under which conditions asymptotically safe quantum gravity can UV complete the SM. On the one hand, this is an important consistency test for any theory that aims to describe nature on the fundamental level. On the other hand, such an asymptotic-safety UV completion may be more predictive than the SM, allowing to calculate some of the SM's free parameters from first principles. This might even provide answers to some long-standing questions in particle physics, such as the origin of different masses for fermions.\\

In the following we will first briefly summarize the Landau pole/triviality problems in sectors of the SM and review how asymptotically safe quantum gravity might cure these.\\
The Landau poles/triviality problems arise in some of the canonically marginal couplings of the SM which we write as $c$ here (later, the gauge couplings will be $g_{Y/2/3}$, the Yukawa couplings $y_f$ and the Higgs self-interaction will be $\lambda$). Their scale dependence takes the following schematic form
\begin{equation}
\label{eq:matterbetaschem}
\beta_{c}=- f_{c}\,c +\beta_{c,\, 1}\, c^{n} +\mathcal{O}(c^{n+1})\,,
\end{equation}
where $n=2$ for the quartic Higgs self-interaction, and $n=3$ for the gauge and Yukawa couplings. Here,  $\beta_{c, \, 1}$  is the one-loop matter contribution, and $f_{c}$ is the gravitational contribution to the scale dependence of the respective coupling. The signs of the gravitational and the matter contributions differ from sector to sector, giving rise to distinct phenomenological implications for each of them.  We will see that three of the four possibilities in Tab.~\ref{tab:UVoverview} may be realized in the SM with gravity.

\begin{table}[!t]
\begin{tabular}{|c|c|c|c|c|c|}
sign of 
 $\beta_{c, \, 1}$
& sign of $f_c$ & UV complete & $c$ predicted & $c_{\ast}\neq 0$ possible& $c$ predicted \\
 &  & & from $c_{\ast}=0$ fixed point & &  from $c_{\ast}\neq 0$ fixed point\\
&  & & at low energies & & at low energies\\ \hline
negative& positive &yes & no & no & -   \\ \hline
negative & negative & yes & yes & yes & no \\ \hline
positive & positive & yes & no & yes & yes \\ \hline
positive & negative & yes & yes & no & - \\ \hline
\end{tabular}
\caption{\label{tab:UVoverview} We show the four distinct possibilities for the signs of coefficients in Eq.~\eqref{eq:matterbetaschem}: In the first line, matter and gravity are both antiscreening, such that there is only an asymptotically free fixed point. This situation may be realized for non-Abelian gauge couplings, see \autoref{sec:Gaugesector}. In the second line, matter is antiscreening and gravity screening, such that there are two fixed points, with the one at vanishing coupling generating a prediction. In the third line, matter is screening and gravity is antiscreening, such that there are two fixed points, with the one at nonvanishing coupling generating a prediction. This situation may be realized for the Abelian gauge coupling and for Yukawa couplings, see \autoref{sec:Gaugesector} and \autoref{sec:Yukawas}. In the fourth line, matter and gravity are both screening, such that there is only the free fixed point which generates a prediction. This may be realized for the Higgs selfinteraction, see \autoref{sec:Higgs}.}
\end{table}

In the absence of gravity, $\beta_{c, \, 1}>0$
implies that the coupling shrinks under the RG flow to the IR. Starting with a finite value in the very far UV, $k \rightarrow \infty$, the IR value of the coupling therefore vanishes, i.e., the theory becomes non-interacting, or trivial. This is called the triviality problem. In perturbation theory, it is visible because the RG flow, followed against its natural direction from IR to UV, produces a Landau pole, i.e., a divergent coupling, at a finite UV scale. Beyond perturbation theory, non-perturbative studies have confirmed that there is a triviality problem in the quartic scalar and the Abelian gauge coupling, see \autoref{sec:Gaugesector} for a more detailed discussion.

The physical interpretation of the Landau pole/triviality problem is that the SM is an effective field theory with a finite range of validity. It breaks down if one attempts to extrapolate it towards the UV past the scale of the Landau poles. Because those Landau poles are transplanckian in the SM\footnote{This is true for the measured value of the Higgs mass and Abelian gauge coupling. If those values were different, the Landau poles could occur already below the Planck scale. The absence of Landau poles below the Planck scale is also an important guiding principle for physics beyond the SM.}, this problem is often ignored in particle physics. In the context of quantum gravity, one may no longer ignore the problem and must find a solution. Such a solution may lie in the coupling to gravity -- one can indeed interpret the transplanckian scale of the Landau pole as an indication that the missing new physics is nothing but gravity.

\subsection{Gauge couplings in the Standard Model}\label{sec:Gaugesector}
\emph{Synopsis: The gauge sector of the SM has a problem and three riddles. The problem is the Landau pole (or triviality problem) in the Abelian gauge coupling. The riddle is, what sets the values of the three gauge couplings at low energies, in particular, what sets the value of the finestructure constant $\alpha=1/137$?\\
There are compelling indications that asymptotically safe gravity could solve the problem. Whether or not it may also solve one of the three riddles and explain why $\alpha=1/137$ is less sure, because it depends on additional properties of asymptotically safe gravity.}

The gauge sector of the Standard Model is divided into an Abelian and a non-Abelian sector. The Abelian hypercharge group $U_Y(1)$ with coupling $\gy(k)$ is, together with the non-Abelian $SU(2)$, spontaneously broken to the $U_{\rm em}(1)$ electromagnetic gauge group with the electromagnetic coupling $e(k)$, related to the finestructure constant $\alpha(k) = e(k)^2/(4\pi)$. This sector features a problem and a riddle: the problem is the Landau-pole problem; the riddle is why the finestructure constant takes the value $\alpha(k \rightarrow 0) = 1/137$. It is a compelling scenario that asymptotically safe gravity could solve the problem and the riddle at the same time. We review the evidence for this scenario below.

The Landau pole or triviality problem is present in the absence of gravity, i.e., with $f_{\gy}=0$ in Eq.~\eqref{eq:matterbetaschem}. Quantum fluctuations of charged matter, i.e., the charged lepton fields and the quarks, turn the vacuum into a screening medium, such that for the Abelian hypercharge 
 $\beta_{\gy, \, 1}>0$
in \eqref{eq:matterbetaschem} \cite{Gell-Mann:1954yli} (with $\beta_{\gy,\, 1}
=41/(6\cdot 16 \pi^2)$). Thus, the coupling decreases when flowing towards lower energies.
As a consequence, a divergent value of $\gy(k = \Lambda_{\rm Landau})$ is mapped to the measured value of $\alpha$ at low energies. This can already be seen at one-loop order.
To this order, the solution to $\beta_{\gy}$ reads
\begin{equation}
\gy^2(k)=\frac{\gy^2(k_0)}{1-2\, \beta_{\gy,\, 1}\,\gy^2(k_0)\ln\left(\frac{k}{k_0}\right)}\,,
\end{equation}
where $k_0$ is a reference scale. Because ${\beta_{\gy,\, 1}}>0$, $\gy^2(k)$ diverges at a finite scale $\Lambda_{\rm Landau}$
\be
\Lambda_{\rm Landau} =\exp\left(\frac{1}{2{\beta_{\gy,\, 1}}\, \gy^2(k_0)} \right) k_0.\label{eq:Landaupole}
\ee
In theory, this divergence, the so-called Landau pole, can be avoided by setting $\gy^2(k_0)=0$. However, then $\gy^2(k)$ would be zero at all scales $k$, which is clearly in contradiction with the experimental observation of an interacting electromagnetic sector at low energies in our universe. 
Hence, the presence of a Landau pole in the Abelian gauge sector signals the breakdown of the Standard Model. From Eq.~\eqref{eq:Landaupole}, with $e(k_0 = 511\, \rm keV)$ and $\beta_{e,\, 1}=1/(12 \pi^2)$ (i.e., looking at QED), one can infer that $\Lambda_{\rm Landau} \approx 10^{286}\, \rm eV$ (which is shifted to $\sim 10^{34}\,\rm GeV$ for the matter content of the SM and in a two-loop calculation), which is highly transplanckian.
The computation giving rise to the Landau pole was performed within a perturbative one-loop approximation. Clearly, this approximation is not expected to be accurate when $\gy$ becomes large. However, non-perturbative studies using lattice \cite{Gockeler:1997dn, Gockeler:1997kt} and functional \cite{Gies:2004hy} methods find indications that the triviality problem in the Abelian gauge coupling persists beyond perturbation theory.\\

The triviality problem indicates that new physics must exist at very high energies. As the energy-scale of the Landau pole is beyond the Planck scale, quantum gravity is one candidate for such new physics. In fact, it is a compelling candidate, because it is not really ``new physics", given that we know for sure that the gravitational field exists.\\

Asymptotically safe gravitational fluctuations contribute through $f_{\gy}$ to the scale dependence of $\gy$, cf.~\eqref{eq:matterbetaschem}. $f_{\gy}$ depends on the gravitational couplings; explicit forms have been calculated in \cite{Daum:2009dn, Folkerts:2011jz, Harst:2011zx, Christiansen:2017gtg, Eichhorn:2017lry, Christiansen:2017cxa, DeBrito:2019gdd, deBrito:2022vbr, Pastor-Gutierrez:2022nki}.\\
 It is crucial that the gravitational contribution is linear in $\gy$ itself. This may be understood in two ways: first, it is a consequence of counting powers of $\gy$ in the loop diagrams that generate the gravity-contribution to $\beta_{\gy}$. Second, no lower-order contribution $\sim \gy^0$ exists, because it would break the global chiral symmetry of charged fermions, where left- and right-handed components transform under independent phase rotations. In \autoref{sec:GlobalSymms}, we explain why such global symmetries prevent the generation of such low-order terms in beta functions.
Therefore, for small values of the coupling, i.e., close to the Gaussian fixed point, the gravitational contribution dominates over the pure-matter contribution. \\
Explicit computations using the FRG indicate that $f_{\gy}\geq0$ \cite{Daum:2009dn, Folkerts:2011jz, Harst:2011zx, Christiansen:2017gtg, Eichhorn:2017lry, Christiansen:2017cxa, deBrito:2022vbr, Pastor-Gutierrez:2022nki}, such that gravitational fluctuations have an anti-screening effect on the gauge coupling. One can also argue that the gravitational contribution should have a negative sign, i.e., be antiscreening: gravity generates a self-coupling of the Abelian gauge field, such that effectively, the Abelian gauge field behaves like a non-Abelian one. In a non-Abelian gauge theory, gauge-field fluctuations antiscreen the vacuum. Thus, one may argue, the combined effect of gravity and Abelian gauge field should also be antiscreening, and thus, $f_{\gy}>0$ is the expected result.

The gravitational contribution $f_{\gy}$ is a function of the dimensionless Newton coupling $\GN$, and, in the absence of other gravitational couplings, given by 
\be
f_{\gy}= \frac{5\GN}{18\pi}.
\ee
Above the Planck scale, $\GN$ assumes its fixed-point value, such that $f_{\gy} = \rm const$.
Below the Planck scale, $\GN$ scales like $k^2$, i.e., decreases towards the IR.
Thus, gravitational fluctuations decouple very quickly below the Planck scale, and are completely negligible at experimentally accessible scales -- just as one expects.
Therefore, $f_{\gy}$ can be approximated as zero below the Planck scale, such that the scale dependence of the gauge coupling is only driven by SM fields.

In addition to $\GN$, $f_{\gy}$ depends on the other gravitational couplings, including the cosmological constant $\Lambda$ and higher-order couplings \cite{DeBrito:2019gdd}, such as the $R_{\mu\nu}R^{\mu\nu}$-coupling $b$. In terms of $\GN$, $\Lambda$ and $b$, $f_{\gy}$ reads
\be
f_{\gy} =\frac{\GN}{36\pi} \frac{10+7b- 40 \Lambda}{\left(1+b-2 \Lambda\right)^2}
\ee

In the gravitational fixed-point regime, different scenarios are realized, depending on the sign of $f_{\gy}$. If $f_{\gy}<0$, then gravitational fluctuations are screening and the triviality problem persists.
If $f_{\gy}>0$, then gravitational fluctuations are anti-screening. This is the scenario that appears o be realized, when using fixed-point values obtained in the literature.
Thus, they compete with the screening fluctuations of charged matter fields. At small $\gy$, the gravitational contribution dominates and the gauge coupling becomes asymptotically free. At large $\gy$, the matter contribution dominates and the triviality problem persists. Inbetween, at $\gy=\gyast$, with
\be
\gyast = 4\pi\, \sqrt{\frac{6\,f_{\gy}}{41}}
\ee
the screening and antiscreening effects cancel out exactly and generate an interacting fixed point.

\begin{figure}[t]
\includegraphics[width=\linewidth, clip=true, trim=1cm 15cm 34cm 7cm]{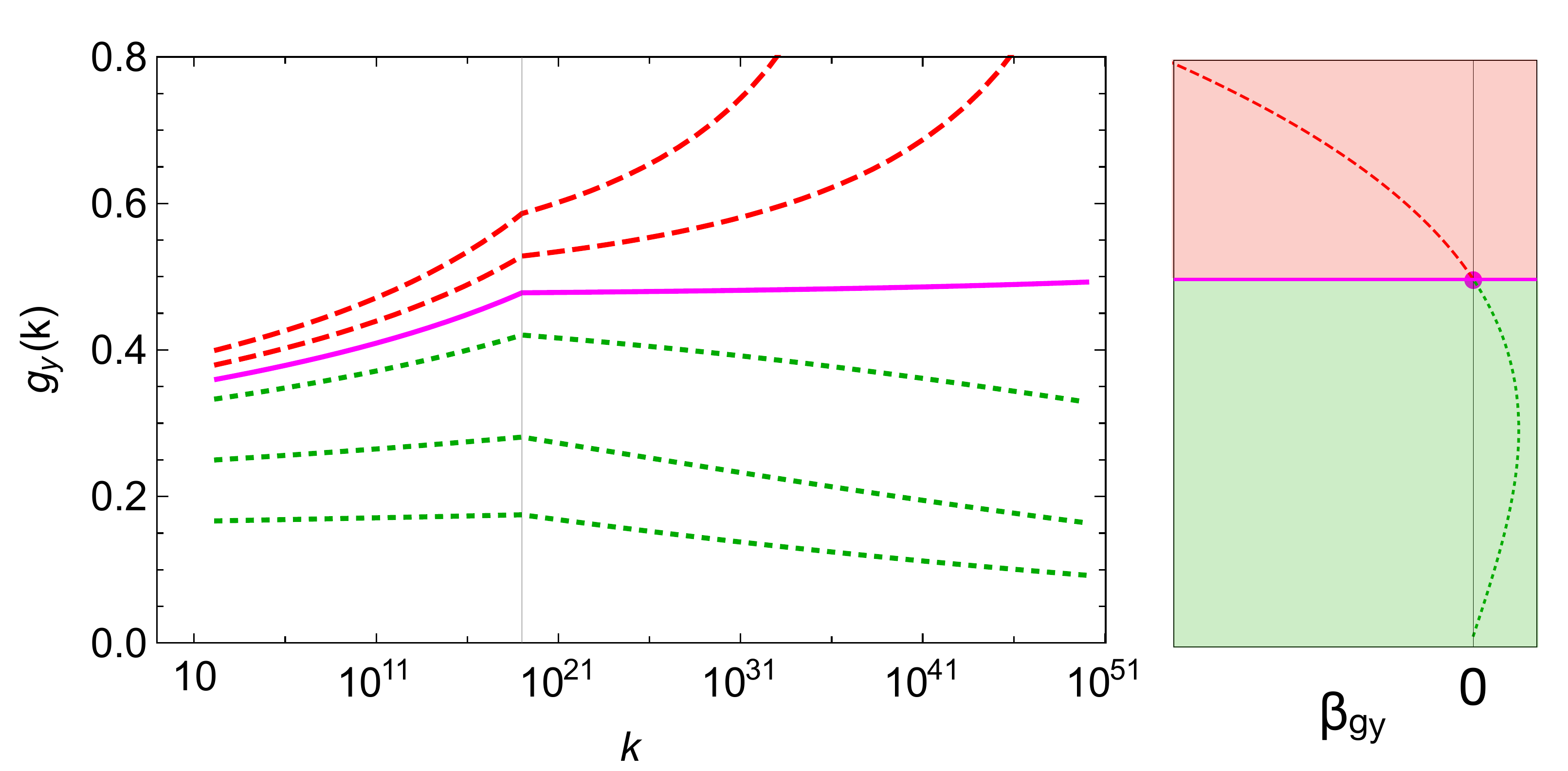}
	\caption{\label{fig:gyschem} We show the beta function for $\gy$ (right panel, rotated), such that the gravity-dominated regime is the lower, green-shaded range of coupling-values. The matter-dominated regime is the upper, red-shaded range of coupling values. Gravity and matter fluctuations balance out at an interacting fixed point (magenta). The resulting trajectories are shown in the left panel: in the matter-dominated regime, the triviality problem persists; in the gravity-dominated regime, trajectories emanate from an asymptotically free fixed point in the very far UV. A single trajectory is the asymptotically safe one, on which quantum scale symmetry holds at transplanckian scales. The fixed-point value is mapped to a unique value in the IR. For $f_{\gy} \approx 9.7\cdot 10^{-3}$, the IR value corresponds to the measured IR value of the Abelian hypercharge coupling.}
\end{figure}

This fixed point has predictive power: because gravity fluctuations antiscreen the coupling at $\gy<\gyast$, they drive the coupling towards the fixed-point value from below. Conversely, because matter fluctuations screen the coupling at $\gy>\gyast$, they drive the coupling towards the fixed-point value from above. Thus, the coupling stays fixed at $\gy=\gyast$ all the way down to the Planck scale. Asymptotic safety thereby produces a unique value of the coupling at the Planck scale. Below the Planck scale, the SM RG flow maps this unique value to a unique value in the IR.\\
Rephrased in a more technical manner, at this fixed point, $\gy$ is an irrelevant direction with $\theta = - \partial \beta_{\gy}/\partial \gy \vert_{\gy = \gyast}= -2 f_{\gy}<0$, and therefore does not contribute a free parameter. Hence, the IR value of the Abelian hypercharge following from this fixed point is a prediction of the UV completion. 

At the same time, the interacting fixed point generates an upper bound, above which the triviality problem cannot be avoided: trajectories for which $\gy(k=M_{\rm Planck})<\gyast$ are those that come from an asymptotically free fixed point and are therefore UV complete. They reach IR values that lie below the value from the interacting fixed point. However, trajectories for which $\gy(k=\rm Planck)> \gyast$ diverge if followed further into the UV. They reach IR values that lie above the value from the interacting fixed point. In summary, the IR value of the gauge coupling is bounded from above by the prediction from asymptotic safety, if one wants to avoid the triviality problem.

In consequence, asymptotic safety becomes testable: if the prediction for the coupling from the interacting fixed point (assuming this fixed point persists beyond the approximations it was seen in do date) is below the measured value, the triviality problem persists, and a UV completion of the SM with gravity is ruled out.

Intriguingly, the predicted value of the Abelian hypercharge comes out above the experimentally measured value, and, within  estimates of the systematic uncertainties of FRG computations, might even be in agreement with experimental observations \cite{Harst:2011zx, Eichhorn:2017lry}. Therefore, asymptotically safe quantum gravity might not only provide a UV completion for the Abelian gauge sector, but this UV completion might even be predictive, solving the long-standing riddle, why the finestructure constant is $1/137$ in the IR, see also \cite{Eichhorn:2018whv, Eichhorn:2017muy}.

The non-Abelian gauge sector of the SM with and without gravity has a simpler structure:
for non-Abelian gauge couplings, the matter contribution $\beta_{g_i, \, 1}$
is negative in the SM, giving rise to asymptotic freedom. The minimal coupling of gravity to matter is not sensitive to internal symmetries of the matter sector. Hence, the gravitational contribution $f_{\gy}$ is also the gravitational contribution to the scale dependence of non-Abelian gauge couplings. Hence, asymptotic freedom for the strong coupling remains intact in the presence of asymptotically safe quantum gravity \cite{Folkerts:2011jz, Pastor-Gutierrez:2022nki}. The IR values of the two non-Abelian gauge couplings remain free parameters.\\

\FRT{Further reading:}\\

\FR{Chiral-symmetry breaking with the Abelian gauge coupling:}\\
Assuming the realization of the interacting fixed point for $\gy$, a fixed-point collision in induced chirally symmetric four-fermion interactions, see \autoref{sec:lightfermions}, may occur, see \cite{deBrito:2020dta}. In this scenario the non-vanishing fixed point value for $\gy$ triggers a fixed-point collision in four-fermion interactions. Such a fixed point collision is well studied in QCD, where it is associated with spontaneous breaking of chiral symmetry.  If this effect were to occur in the asymptotically safe system, it would prevent the existence of light fermions, i.e., fermions with masses below the Planck scale.
To prevent this, the value of the interacting fixed-point has to be small enough to allow for a UV complete and chirally symmetric theory. This is only possible if the number of fermions in the system exceeds a critical number, see \cite{deBrito:2020dta}. Hence, the interplay of quantum gravity and matter might put lower bounds on the number of fermions, in addition to the upper bounds discussed in \autoref{sec:lightfermions}.\\

\FR{Subtleties in $f_{\gy}$:}\\
The interpretation of $f_{\gy}$ has some subtleties which we did not discuss above. For this discussion, it is useful to isolate the $\GN$-dependence in $f_{\gy}$ and write $\tilde{f}_{\gy} = \GN\, f_{\gy}$.
Individual terms in beta functions are not necessarily physical, and may therefore depend on the scheme.
Indeed, in perturbative studies, it depends on the scheme, whether $\tilde{f}_{\rm gy}$ vanishes or not, see, e.g., \cite{Robinson:2005fj,Pietrykowski:2006xy, Toms:2007sk, Ebert:2007gf, Tang:2008ah,Toms:2010vy, Anber:2010uj}. From this, it was concluded that there is no gravitational contribution to the scale dependence of the Abelian gauge coupling.
However, there is an important hidden assumption in these perturbative studies: they treat the gravitational coupling $\GN$ as a constant which is finite. The gravitational contribution $f_{\gy} = \GN\, \tilde{f}_{\gy}$ vanishes, when $\tilde{f}_{\gy}$ and $\GN$ is finite. However, when $\GN$ diverges, one must be careful when evaluating $f_{\gy}$. In \cite{deBrito:2022vbr}, it was shown that there are FRG regulators \cite{Baldazzi:2020vxk, Baldazzi:2021guw}, for which $\tilde{f}_{\gy}$ also vanishes. In contrast to perturbative studies using dimensional regularization, the FRG regulator features a smooth limit in which $\tilde{f}_{\gy}$ goes to zero as a function of a control parameter. As a function of the same control parameter, the fixed-point value $\GNast$ diverges. We therefore find that $f_{\gy}$ may indeed vanish for some schemes, when $\GN$ is held fixed. However, in at least one of those schemes, $\GNast$ diverges such that $f_{\gy}$ remains finite while $\tilde{f}_{\gy}$ goes to zero.\\
Further, $f_{\gy}$ evaluated at $\GN = \GNast$ is a universal quantity, because it is a critical exponent at $\gyast=0$. As a universal quantity, it may not depend on the scheme (in practise, in approximations, it still does). It is therefore reassuring that even in schemes, in which $f_{\gy} (\GN) \rightarrow 0$, for the universal quantity, it holds that $f_{\gy} (\GNast) \neq 0$.\\
We therefore conclude that in perturbation theory, when gravity does not assume a fixed point, it may be the case that gravitational contributions to beta functions vanish in some schemes. However, when gravity assumes a fixed point, and $f_{\gy}$ is evaluated with the appropriate care, there is a nonzero gravitational contribution and $f_{\gy}$ is a universal quantity.

\subsection{Yukawa couplings in the Standard Model}\label{sec:Yukawas}
\emph{Synopsis: Gravity can either screen or antiscreen a Yukawa coupling, depending on the gravitational fixed-point values. In the antiscreening case, the Yukawa coupling can become asymptotically free or safe, with an upper bound on its IR value. In the screening case, the Yukawa coupling is not UV complete.\\
Based on this result, there is a mechanism that ties the quark masses to their charges: If gravity is antiscreening, and the Abelian gauge coupling is at its interacting fixed point, there is an interacting fixed point in the Yukawa sector, for which the up-type quarks and down-type quarks have different fixed-point values, because they are charged differently under the Abelian hypercharge.\\
For the third generation, this mechanism gives rise to a SM-like IR phenomenology, with bottom quark mass and top quark mass predicted at or in the vicinity, of their measured values.\\
For the full quark sector of the SM, mixing between flavors becomes important at highly transplanckian scales, and fixed points with nonzero Yukawa couplings no longer produce SM-like IR phenomenology. Instead, a fixed point which is made asymptotically free under the impact of gravity is available for all quark Yukawa couplings and CKM matrix elements, rendering the SM quark Yukawa sector UV complete.}

In the SM, quark masses are generated by two mechanisms: first, by electroweak symmetry breaking in the Higgs-Yukawa sector -- called the current mass, and second by chiral symmetry breaking in the strongly-interacting phase of QCD -- called the constituent mass.
Lepton masses are generated only through electroweak symmetry breaking. The ratio of lepton masses and current quark masses to the Higgs vacuum expectation value is determined by Yukawa couplings -- one for each quark flavor and lepton species.
Schematically, this is the same as for a simple Yukawa system built out of a Dirac fermion and a real scalar; although the SM is based on Weyl fermions and a complex Higgs scalar that is an SU(2) doublet. The Yukawa coupling $y\,\phi\bar{\psi}\psi$ between the Dirac fermion $\psi$, the corresponding antifermion $\bar{\psi}$ and the real scalar $\phi$ gives rise to a mass term, when the scalar develops a vacuum-expectation value $\langle \phi \rangle = v$ in the symmetry-broken phase. There, one can express the scalar field as excitations $\varphi$ around its expectation value, leading to $y\,\phi\bar{\psi}\psi \rightarrow m\, \bar{\psi}\psi + y\varphi \bar{\psi}\psi$, where $m = y\, v$. In the SM, the IR values of the Yukawa couplings are therefore known, because the masses of all fermions have been measured. In addition, both ATLAS and CMS have measured the Yukawa couplings of the heaviest quarks \cite{CMS:2018uxb,ATLAS:2018mme,CMS:2018nsn,ATLAS:2018kot} and the heaviest lepton \cite{ATLAS:2015xst,CMS:2017zyp}. This motivates a study of the Yukawa sector coupled to asymptotically safe gravity, to find out, whether (i) the Yukawa sector is UV complete when gravity is present and (ii) whether the measured IR values can either be accommodated or even ``retrodicted".

\subsubsection{Simple Yukawa system}
The structure of the gravity-Yukawa-system for the SM follows from the basic structure of a single Yukawa coupling and the gravitational effect on it: That structure is the same as for gauge couplings. Out of a competition between a screening matter contribution and an antiscreening gravity contribution, an asymptotically free fixed point arises. Trajectories that start from it, reach a range of values in the IR, which is bounded from above by the prediction from an asymptotically safe fixed point. The difference between gauge and Yukawa sector is that this mechanism is only at work in a part of the gravitational parameter space, cf.~\autoref{fig:mattermattersinterplay}.

For a simple Yukawa system as introduced above, the matter contribution  in Eq.~\eqref{eq:matterbetaschem} is positive, $\beta_{y,\, 1}>0$, such that a simple Yukawa system features a Landau pole.
The sign of the gravitational contribution $f_y$ depends on the fixed-point value of the cosmological constant \cite{Oda:2015sma, Hamada:2017rvn, Eichhorn:2016esv, Eichhorn:2017eht}, 
see also \autoref{fig:mattermattersinterplay}: for fixed-point values below a critical value $\Lambda_{\mathrm{crit}}$, $f_y>0$ holds, such that the Gaussian fixed point $y_*=0$ is IR-repulsive. 
Starting from this fixed point, finite values for the Yukawa couplings in the IR can be reached and the IR value is a free parameter. Just like in the Abelian gauge sector, $f_y>0$ gives rise to a second, interacting fixed point, where the Yukawa coupling corresponds to an irrelevant, i.e., IR repulsive direction. This fixed point hence is connected to a single predictive trajectory, where the IR value of the coupling is a prediction. This predictive trajectory produces an upper bound of the IR value of the Yukawa coupling, which can be reached from the Gaussian fixed point $y_*=0$. \\
For $\Lambda>\Lambda_{\mathrm{crit}}$, the gravitational contribution is screening, i.e., $f_y<0$ \cite{Oda:2015sma, Hamada:2017rvn, Eichhorn:2016esv, Eichhorn:2017eht}. This makes the Landau-pole problem worse. Accordingly, if $\Lambda_*>\Lambda_{\mathrm{crit}}$, the only possibility to achieve a UV-complete Yukawa sector is to set $y=0$ at the Planck scale. Once $y$ is set to zero, there is an additional global symmetry, namely a chiral rotation for the fermion, $\psi \rightarrow e^{i\gamma_5\, \alpha}\psi$. This symmetry protects the Yukawa coupling, such that it cannot be regenerated below the Planck scale. Accordingly, the case $f_y<0$ results in a prediction, namely of a vanishing Yukawa coupling.
Thus, fixed-point values $\Lambda_{\ast}>\Lambda_{\mathrm{crit}}$ are excluded from the viable parameterspace for the gravitational fixed-point values, because vanishing Yukawa couplings are in contradiction with observations.\\

Calculations of the gravitational fixed-point values in the presence of a single Dirac fermion and real scalar (i.e., the fields that make up a simple Yukawa system) yield the result $\Lambda_{\ast}> \Lambda_{\rm crit}$ \cite{Dona:2013qba, Meibohm:2015twa, Eichhorn:2016vvy, Eichhorn:2018ydy}. At larger number of fields, in particular, in the presence of all SM fields, different studies find differing results; however, e.g., \cite{Dona:2013qba, Eichhorn:2016vvy}  (which rely on the background field approximation\footnote{In \cite{Pastor-Gutierrez:2022nki} $\Lambda_{\ast}< \Lambda_{\rm crit}$ is achieved in fluctuation computations by integrating out gravitational and matter fluctuations at slightly different scales.}) find that, once a third generation of SM fermions is present, the gravitational fixed-point value has moved to $\Lambda_{\ast}< \Lambda_{\rm crit}$. \\

\FRT{Further reading:}\\

\FR{Retrodicting the top mass}\\
In \cite{Eichhorn:2017ylw}, the gravity-generated interacting fixed point for the top Yukawa coupling allows to calculate the top quark mass from first principles, yielding a value of about 171 GeV, which is, within the systematic uncertainties of the calculation, very well compatible with the experimental value of  172.8 GeV.

\subsubsection{Top-bottom-system}\label{sec:tby}
In the SM, the Yukawa couplings couple the right-handed $SU(2)$-singlets to left-handed $SU(2)$-doublets and the Higgs field. Nevertheless, the gravitational contribution is the same as for a real scalar coupling to a Dirac fermion and its antifermion. The underlying reason is that gravity is ``blind" to internal symmetries (in this case, the $SU(2)$). Therefore, the gravitational contribution to the top-quark Yukawa coupling $y_t$ and the bottom-quark Yukawa coupling $y_b$ is the same. Thus, one can, if $\Lambda_{\ast}< \Lambda_{\rm crit}$, achieve asymptotic freedom for the two Yukawa couplings. Asymptotic safety for both Yukawa couplings is ruled out, because the fixed-point value for the top and bottom Yukawa is equal to each other. Thus, their IR values are also close to each other (they are not equal, because gauge field contributions to the two scale-dependences differ), and this contradicts experiment: the top quark is about forty times as heavy as the bottom quark.

\begin{figure}[!t]
\includegraphics[width=0.45\linewidth]{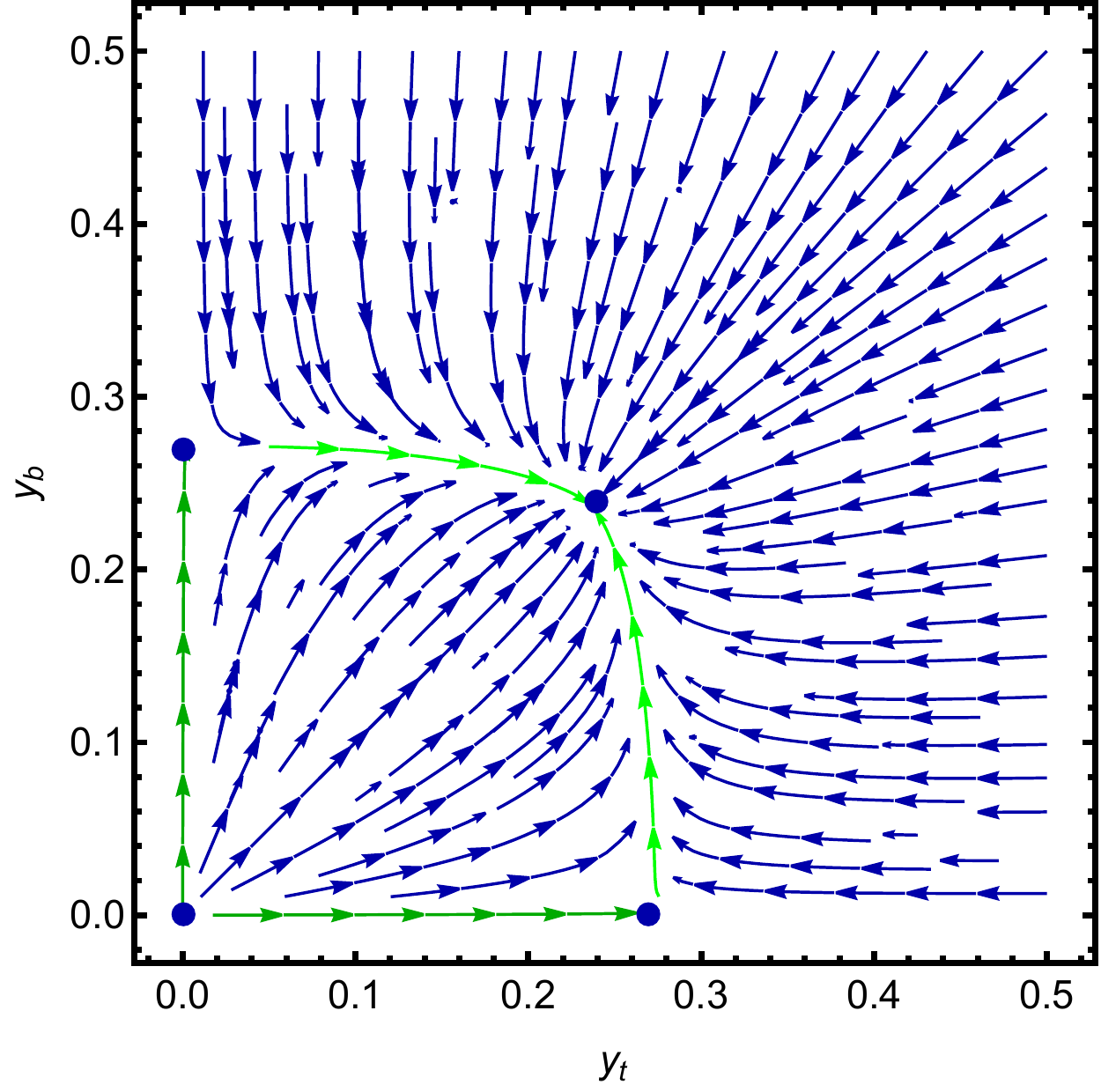}\quad \includegraphics[width=0.45\linewidth]{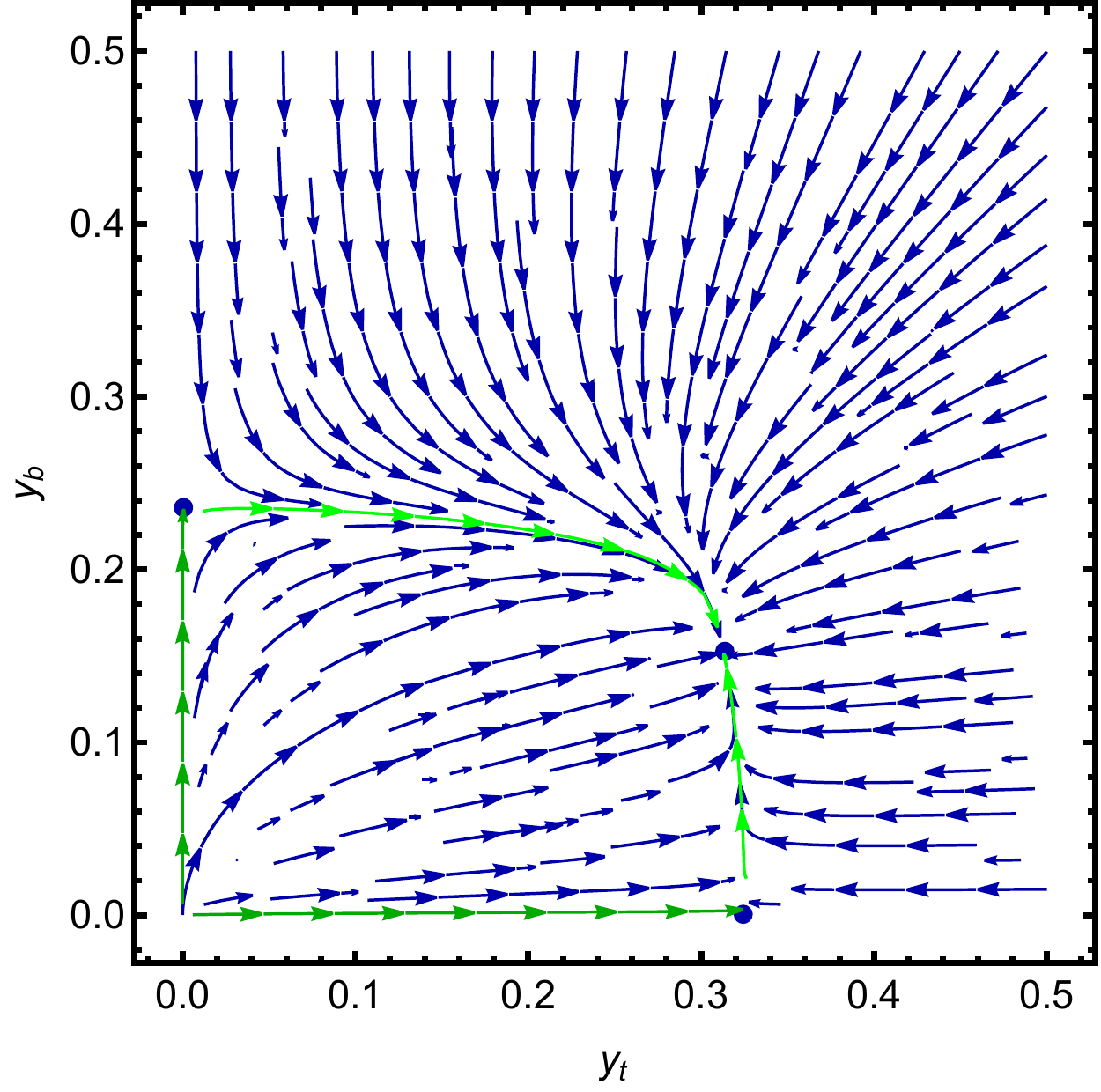}
\caption{\label{fig:yukawastreams} We show the RG flow towards the IR in the plane spanned by $y_t$ and $y_b$. In the absence of the Abelian gauge field (left panel) there is complete symmetry between $y_t$ and $y_b$. In particular, the fully interacting fixed point, which attracts all trajectories, lies at $y_{t\, \ast} = y_{b\, \ast}$, resulting in a prediction of $y_t(M_{\rm Planck})= y_b(M_{\rm Planck})$, which cannot result in viable IR phenomenology. In the presence of an Abelian gauge coupling (right panel) the symmetry is broken. For purposes of illustration, we have chosen a large $f_y$. For $f_y = 1.188 \cdot 10^{-4}$, as in \cite{Eichhorn:2018whv}, the fully interacting fixed point lies at $y_{b\, \ast}\ll y_{t\, \ast}$, very close to the fixed point at $y_{t\, \ast}\neq 0, y_{b\, \ast}=0$.}
\end{figure}

\begin{figure}[!t]
\begin{center}
\includegraphics[width=0.8\linewidth]{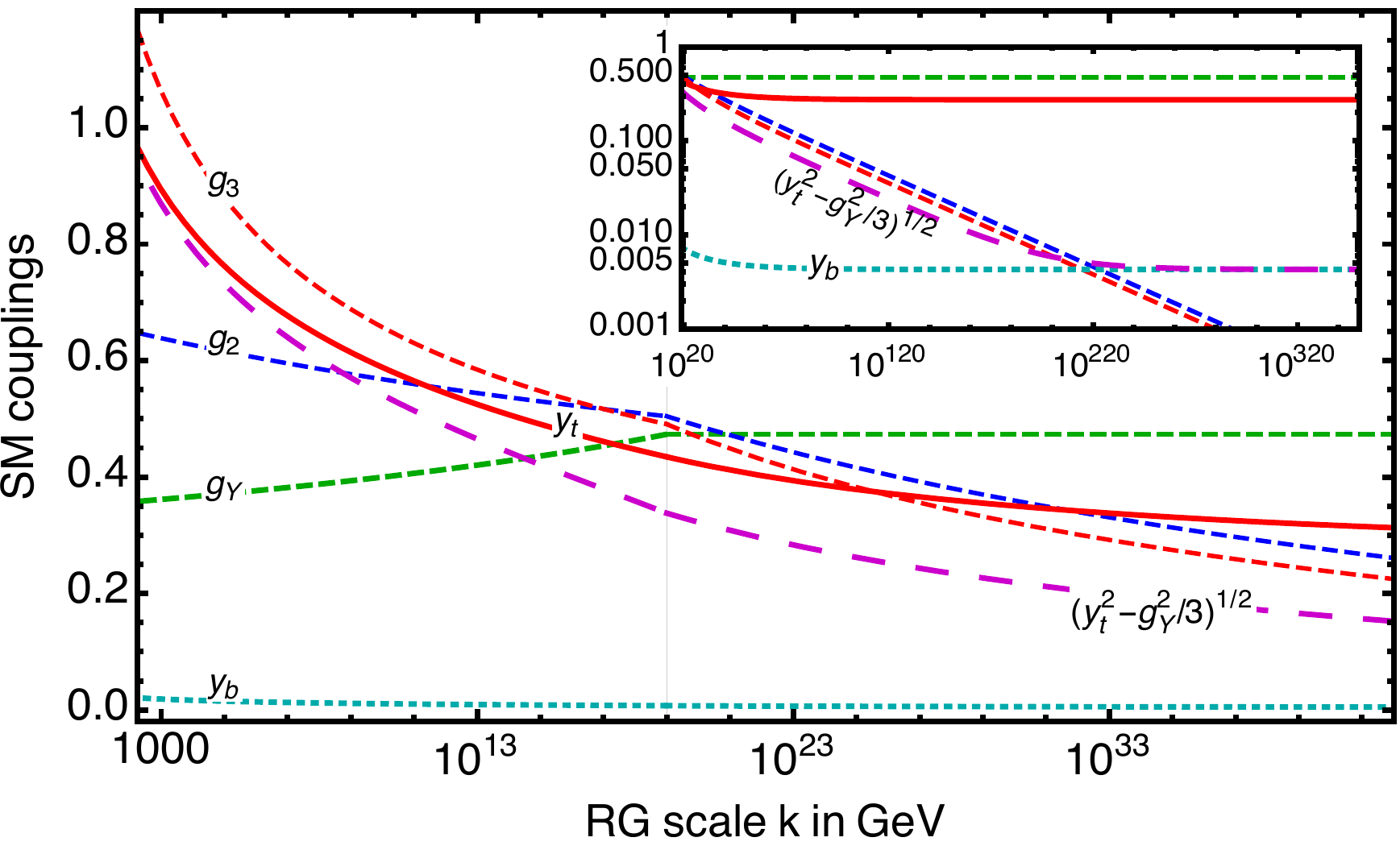}
\end{center}
\caption{\label{fig:tbyflow} We show the scale-dependence of the top-bottom-gauge-system, with all three gauge couplings and the two Yukawa couplings, cf.~\cite{Eichhorn:2018whv}. Above the Planck scale, $y_t$, $y_b$ and $g_Y$ start out at an interacting fixed point that satisfies relation Eq.~\eqref{eq:FPrelationtby}. $f_g=9.7\cdot 10^{-3}$ and $f_y = 1.188\cdot 10^{-4}$ are chosen such that the bottom quark Yukawa and Abelian gauge coupling are predicted to agree with their measured value. The resulting top quark Yukawa coupling is somewhat higher than in the SM.}
\end{figure}

However, besides the gravitational contribution, there can also be a contribution from the Abelian gauge field. If $\gy$ starts out at the asymptotically safe fixed point in \autoref{sec:Gaugesector}, then $y_{t\, \ast}\neq y_{b\, \ast}$ follows. This is because the top quark and the bottom quark do not have the same electric charge, and thus also not the same $U(1)$ hypercharge. This can be seen from their beta functions, together with that for the Abelian hypercharge\footnote{ The non-Abelian gauge couplings have vanishing fixed-point values and are therefore not included in the beta function when one searches for fixed points. They are included, when one follows the RG flow from the fixed points down to the IR, as in \autoref{fig:tbyflow}.}  and in a truncation without higher-order couplings  and to leading order in the couplings, read: 
\bea
\beta_{y_t}&=&\frac{y_t}{16\pi^2}\left( \frac{9}{2}y_t^2 + \frac{3}{2}y_b^2 -\frac{17}{12} g_Y^2\right)-f_y\, y_t,\label{eq:betayt}\\
\beta_{y_b}&=&\frac{y_b}{16\pi^2}\left( \frac{9}{2}y_b^2 + \frac{3}{2}y_t^2 -\frac{5}{12} g_Y^2\right)-f_y\, y_b,\label{eq:betayb}\\
\beta_{g_Y}&=&\frac{g_{Y}^3}{16\pi^2}\frac{41}{6}-f_{g}\, g_Y \label{eq:betagy}.
\eea
The system features an interacting fixed point, for which the fixed-point relation
\be
y_{t\, \ast}^2 - y_{b\, \ast}^2 = \frac{1}{3}g_{Y\, \ast}^2\label{eq:FPrelationtby}
\ee
holds. This fixed-point relation distinguishes the fixed-point values for top and bottom Yukawa, as soon as a non-zero fixed-point value for the Abelian gauge coupling is realized. Then, the top quark Yukawa coupling is also automatically much larger than the bottom-quark Yukawa coupling.  The resulting beta functions for top and bottom are illustrated in \autoref{fig:yukawastreams}.
Taking into account the flow of all three gauge couplings of the SM, and choosing $f_g$ and $f_y$ appropriately, produces an IR phenomenology that is rather close to that of the SM, cf.~\autoref{fig:tbyflow}.\\

This mechanism is remarkable, because it links the charge ratio of the two quarks to its mass ratio. In fact, other charge ratios are incompatible with the measured masses of top and bottom quark, even if arbitrary values of $f_y$ and $f_g$ are considered.

 In calculations based on Eq.~\eqref{eq:betayt}-\eqref{eq:betagy}, full agreement with the measured values cannot be reached and the top quark is $\sim 5-10$ GeV too heavy (depending on which approximation is used), cf.~\cite{Eichhorn:2018whv,Alkofer:2020vtb}. However, these calculations come with significant systematic uncertainties, e.g., by neglecting further, higher-order interactions. 
 
Further, the Yukawa couplings of the other generations cannot be neglected at very high scales, contrary to what one may first think. This is because, although the other Yukawa couplings themselves are very small compared to $y_t$, the three-generation quark system features CKM-mixing. At very high energies, the CKM matrix elements are scale dependent, triggering a deviation of the fixed-point structure from Eq.~\eqref{eq:FPrelationtby}. Therefore, we consider the full quark sector of the SM next.

\subsubsection{Quark sector of the SM}
\begin{figure}
 \includegraphics[width=\linewidth]{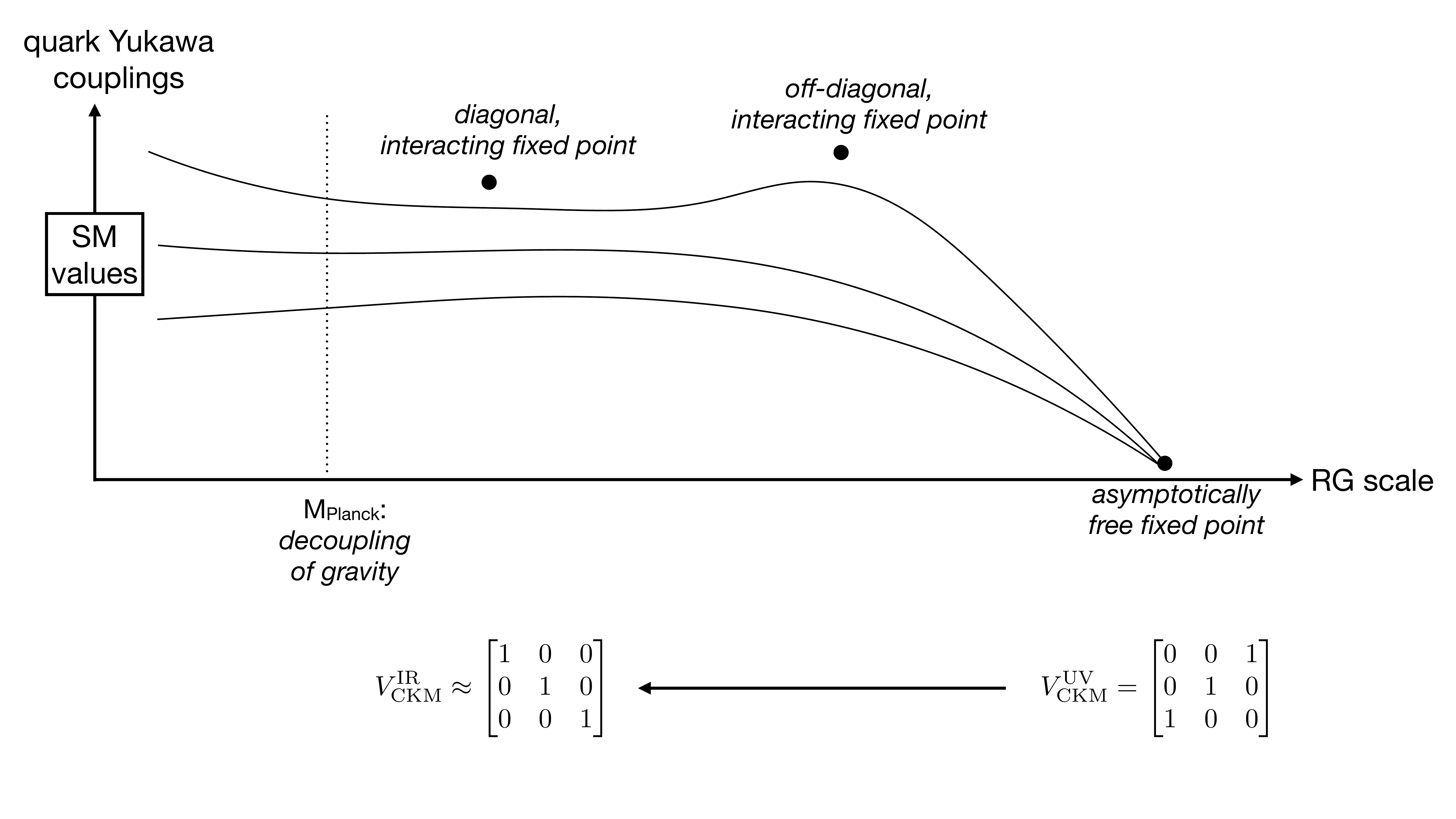}
 \caption{\label{fig:CKMschematic} Under the impact of asymptotically safe gravity, the quark Yukawa couplings start out at an asymptotically free fixed point in the very deep UV, at which the CKM matrix is off-diagonal. They are attracted towards an interacting fixed point, due to its irrelevant directions. When the CKM matrix elements transition towards a near-diagonal configuration, the fixed-point values at the interacting fixed point for the Yukawa couplings change. Over a large range of scales, that fixed point determines the properties of the Yukawa system. At the Planck scale, gravity decouples dynamically, and the flow of the quark Yukawa sector is exactly that of the SM, cf.~\cite{Alkofer:2020vtb} for more details.}
 \end{figure}

In the SM, the quark Yukawa sector contains ten beta functions: Six for the Yukawa couplings and four for the CKM matrix elements. The CKM matrix describes mixing in the quark sector, i.e., the electroweak interaction can change the flavors. 
Because it is unitary, the CKM matrix contains four physical parameters, with gravity-independent beta functions.\footnote{The independence of the CKM matrix from gravity contributions can be shown from the flavor-universality of gravity, i.e., the ``blindness" of gravity to internal symmetries.}\\ 
In \cite{Alkofer:2020vtb}, see also \cite{Kowalska:2022ypk}, the resulting complexity of the analysis was dealt with by making an assumption about the fixed-point structure, namely that the CKM-matrix elements assume fixed-point values which are independent of the Yukawa fixed-point values. The fixed-point conditions then factorize, and a fixed-point search for the CKM matrix elements can be conducted first.\\
Thereby, one obtains several simple fixed-point configurations for the CKM matrix, which have zeros or ones as the only entries. Two of those fixed-point solutions are phenomenologically important:
It was already observed in \cite{Pendleton:1980as} that a diagonal CKM matrix (which is close to the actual measured values) is an IR fixed point, because it has three IR attractive directions. This fixed point is approached in the IR, starting from another fixed point, namely an off-diagonal CKM matrix with four IR repulsive directions. The corresponding flow is very slow  (even on logarithmic scales, the CKM matrix elements are essentially constant); therefore the transition from the off-diagonal to the near-diagonal configuration occurs at highly transplanckian scales.\\
In a second step, one can analyze the consequences for the Yukawa system. Because the CKM-matrix elements enter the beta functions for the Yukawa couplings, those beta functions change, when the CKM matrix changes from an off-diagonal to a near-diagonal configuration.
Therefore, in the very far UV, where the CKM matrix is off-diagonal, the fixed-point values for the Yukawa couplings are modified compared to an analysis without flavor mixing, as in \cite{Eichhorn:2018whv}. 
It turns out that among the many fixed points that the beta functions have, only the asymptotically free one is phenomenologically relevant. It can be achieved if $f_y>-2.2\cdot 10^{-4}$. Interacting fixed points, most importantly one at which $y_{t\ast} \neq 0$, remain important, because, starting from the asymptotically free fixed point in the very far UV, the system approaches such an interacting fixed point at intermediate scales, see the schematic illustration in \autoref{fig:CKMschematic}.

In summary, asymptotically safe gravity may UV complete the quark Yukawa sector of the SM. It may even have predictive power: if one assumes that asymptotic safety holds above the Planck scale, but not to arbitrarily high scales (e.g., because of a more fundamental UV completion), the mechanism discussed in \autoref{sec:tby} may link quark masses to their charges. Only if one insists that the RG flow should continue to make sense at scales as high as $k\approx 10^{1000}\, \rm GeV$, does the flow of the CKM matrix matter, triggering an approach towards the asymptotically free fixed point (if one follows the RG flow in the reverse direction, i.e., towards the UV).
 
Several open questions currently remain, including (i) a study including the full lepton sector of the SM, (ii) a search for fixed points, for which the factorization hypothesis between CKM matrix elements and Yukawa couplings is given up and (iii) the inclusion of higher-order effects in the beta functions, coming from additional interactions beyond the truncations considered to date.\\

\FRT{Further reading:}\\

\FR{Learning about dark sectors from predictions for the SM:}\\
All matter gravitates. This is also true for dark matter, such that the scale-dependence of the gravitational couplings are affected by all visible and dark matter. Hence, dark matter influences the fixed-point values of the Newton coupling and the cosmological constant. These in turn enter the gravitational contribution to the scale-dependence of SM couplings, and hence determine the interacting fixed-point value for the Abelian gauge coupling and the Yukawa couplings. On the one hand, this changes the prediction for the low-energy values arising from the interacting fixed point. At the same time, the presence of dark matter lowers the upper bound on the low-energy couplings, which can be reached by the free fixed points. Hence, if too many dark-matter fields are present, the lower bound might drop below the experimentally observed value, thereby ruling such dark-matter models out. Thus, the predicted low-energy value of SM couplings might put constraints on the number of dark matter fields. In \cite{Eichhorn:2017ylw} a proof-of-principle for this idea was given, because it was shown that the upper bound on the top quark mass depends on the presence of additional matter fields. According to that analysis, if three right-handed neutrinos and an axion are added to the SM, the upper bound increases, which results in viable IR phenomenology (and implies an asymptotically free fixed point for the top Yukawa).

\subsection{Higgs quartic coupling in the Standard Model}
\label{sec:Higgs}
\emph{Synopsis: Gravity screens the Higgs quartic interaction, such that the ratio of the Higgs mass to the Higgs vacuum expectation value is a prediction from asymptotically safe quantum gravity, dating back to before the measurement of the Higgs mass at the LHC.\\
The value for this prediction depends on the gauge and Yukawa couplings in the SM, in particular the Abelian gauge coupling and the top Yukawa coupling. If both are asymptotically free, the Higgs mass is predicted to  be only a few GeV above the experimental value (at least if the central value for the top quark mass is assumed).
If both are asymptotically safe, the Higgs mass comes out larger than that, but can be lowered by a BSM Higgs portal coupling to a dark sector.}

The beta function for the Higgs quartic coupling $\lambda_H$ is given by
\bea
  \label{eqn:beta_lambda}
 \beta_{\lambda_\text{H}}= - f_s \lambda_\text{H} 
   &+& \frac{1}{16\pi^2} \left( - 6 y_t^4 + \frac{3}{8}\left(2 g_2^4 + (g_2^2 + \frac{5}{3} g_Y^2)^2\right) \right)  \nonumber \\
     &+& \frac{1}{16\pi^2} \lambda_{\rm H} \left(12 y_t^2-9 g_2^2-5 g_Y^2 \right) + \frac{3}{2\pi^2}\lambda_{\rm H}^2\,
,\eea
with the top Yukawa coupling $y_t$, the Abelian hypercharge coupling $g_Y$, the ${\rm SU}(2)_L$ gauge coupling $g_2$ and the gravitational contribution $f_s$. All other Yukawa couplings also contribute in principle, but are in practise negligible, because all other fermions are much lighter than the top quark. In \cite{Shaposhnikov:2009pv}, the following idea was developed: if the gauge couplings and the top Yukawa coupling are asymptotically free under the impact of quantum gravity and $f_s<0$, then Eq.~\eqref{eqn:beta_lambda} has the fixed-point solution $\lambda_{H\, \ast}=0$. This fixed point is IR-attractive, i.e., quantum-gravity fluctuations ensure that $\lambda_H(k=M_{\rm Planck})=0$. It is one of the astonishing results of the LHC, that $\lambda_H(k=M_{\rm Planck})\approx0$ is what needs to be realized to obtain the measured Higgs mass. Explaining this special value is indeed one of the key challenges for particle physics at the moment.

Below the Planck scale, the Higgs quartic interaction is regenerated by gauge and top quark fluctuations, which enter through the terms $y_t^4$ and $g_i^4$ in Eq.~\eqref{eqn:beta_lambda}. The vanishing Planck-scale-value of $\lambda_\text{H}$ is thereby mapped onto a unique value at the electroweak scale, where it determines the ratio of Higgs mass $M_{\rm Higgs}$ and Higgs vacuum expectation value $v_{\rm Higgs} = 246\, \rm GeV$:
\be
\lambda_H(k_{\rm IR})=\frac{1}{2} \left(\frac{M_{\rm Higgs}}{v_{\rm Higgs}}\right)^2.
\ee
The map depends sensitively on the top Yukawa coupling \cite{Bezrukov:2014ina}, which is only known with a significant systematic uncertainty. Thereby, a Higgs mass of $M_{\rm Higgs} = 129 \, \rm GeV$ comes out for a top mass of $M_t = 172.9\, \rm GeV$, but a top mass of $M_t= 170.9\, \rm GeV$ is not ruled out, which means that the measured value $M_{\rm Higgs} = 126 \, \rm GeV$ could be compatible with $\lambda_H(k=M_{\rm Planck})=0$.\\
Thus, the idea in \cite{Shaposhnikov:2009pv} led to a successful prediction of the Higgs mass in the vicinity of the measured value, several years prior to the experimental discovery of the Higgs particle at the LHC \cite{ATLAS:2012yve, CMS:2012qbp}. This distinguishes the Higgs sector from the other sectors of the SM, where asymptotic safety may also allow to calculate masses and couplings from first principles, however, only after they have been measured, i.e., as ``postdictions", not genuine predictions.\\
There is a second difference to the prediction in the gauge and Yukawa sector: there, the IR values of the couplings depend sensitively on the values of $f_g$ and $f_y$. In contrast, the prediction for the Higgs mass only depends on the sign of $f_s$. This is because the fixed point for $\lambda_H$ is always the non-interacting one, $\lambda_{H\, \ast}=0$, independent of the value of $f_s$. As long as $f_s<0$, this fixed point is IR attractive and a prediction follows.\\
Following \cite{Percacci:2015wwa, Labus:2015ska, Oda:2015sma, Hamada:2017rvn, Eichhorn:2017als,Eichhorn:2017ylw, Pawlowski:2018ixd, Wetterich:2019rsn, Eichhorn:2020sbo} and even predating \cite{Narain:2009fy} the idea for the prediction of the Higgs mass \cite{Shaposhnikov:2009pv}, evidence for $f_s<0$ has accumulated (with many papers working in the opposite sign convention with $\beta_{\lambda_H}|_{\mathrm{gravity}} = f_s \, \lambda_H$).\\

As a second possibility for a UV-complete Higgs sector, the Abelian hypercharge and top Yukawa coupling may assume an interacting fixed point, in turn inducing an interacting fixed point for the Higgs quartic coupling
\be
  \label{eq:lambda_interacting_fp}
 \lambda_{\text{H}\, \ast}= \frac{5}{48}\gyast^2-\frac{1}{4} y_{t \ast}^2 + \frac{\pi^2}{3} f_s + \frac{1}{48}\sqrt{\left(12 y_{t\ast}^2-5\gyast^2 - 16 \pi^2 f_s \right)^2+576 y_{t \ast}^4 - 100 \gyast^4}.
\ee
The fixed-point value now depends on the value of $f_s$, not just its sign, and also depends on gauge and Yukawa coupling. If they assume fixed-point values which result in the measured values in the IR, the Higgs mass comes out larger than $129 \, \rm GeV$. This rules out such an interacting fixed point in the SM with gravity.  New physics is thus required in this scenario, see below.\\

\FRT{Further reading:}\\

\FR{Higgs mass prediction with dark sector}\\
In \cite{Eichhorn:2021tsx}, it was proposed that adding a Higgs portal, i.e. a coupling between the Higgs-field and a dark scalar field, to an interacting dark sector with scalar and fermions could simultaneously solve two problems: first, it could provide a particle candidate to explain the observed dark-matter relic density. Second, it could shift the predicted Higgs mass towards lower values, starting from the fixed point at which top Yukawa and Abelian gauge coupling are nonzero. This would place a truly asymptotically safe UV completion of the SM with gravity and dark matter, with a high predictive power, within reach. The study in \cite{Eichhorn:2021tsx} is done within a toy model of the full SM (containing a real scalar as the Higgs and a Dirac fermion as the top quark, but no gauge fields), with the extension to the full SM an obvious and important open question.

Similarly, in \cite{Kwapisz:2019wrl} it was found that a new massive $Z'$ boson, corresponding to a gauged $U(1)_{B-L}$ symmetry lowers the predicted value of the Higgs mass, while a sterile quark axion only has little impact on the prediction.

\FR{Resurgence mechanism and scalar mass parameter}\\
In \cite{Shaposhnikov:2009pv}, the ratio of Higgs mass to electroweak scale is predicted, but not the electroweak scale itself. This is because the Higgs mass parameter, i.e., the quadratic term in an expansion of the potential about vanishing field value, is a free parameter, i.e., assumed to be RG relevant. In \cite{Wetterich:2016uxm}, it was suggested that this could change, consistent with results in \cite{}, which confirm that quantum gravity contributes negatively to the corresponding critical exponent. Thus, if gravitational fluctuations are large enough, the Higgs mass parameter becomes irrelevant. If it does so at vanishing fixed-point value, the resulting low-energy prediction is a vanishing electroweak scale. However, if the asymptotically safe gravity-matter fixed point has nonvanishing gauge and/or Yukawa couplings, these shift this fixed-point value away from zero. Whether or not this may lead to a scenario in which the electroweak scale is predicted at the right value is currently an open question. It should be stressed, though, that the required strength of gravitational fluctuations is large and not compatible with weak-gravity bounds, see \autoref{sec:WGB}.\\

\FR{Higgs inflation}\\
Higgs inflation \cite{Bezrukov:2007ep} is based on a nonminimal coupling $\xi$ between the Higgs scalar and the curvature scalar. It could explain inflation without the need for extra fields beyond the SM.  A change to the Einstein-frame, i.e., a conformal transformation of the metric that removes the nonminimal coupling, produces a potential that is appropriate for inflation for suitable values of the couplings. This model is attractive due to its predictive power, because it does not require any BSM fields and contains only one free parameter, namely the nonminimal coupling. In \cite{Eichhorn:2020kca}, it was found that the nonminimal coupling is predicted in asymptotic safety, at least for those values of $G$ and $\Lambda$, for which Yukawa couplings can be nonzero. It turns out that the predicted ratio $\lambda_4/\xi^2$ between the Higgs quartic coupling and the nonminimal coupling is much too large to be compatible with CMB data. This result, if confirmed in extended truncations, rules out Higgs inflation in asymptotically safe gravity. The same conclusion was achieved already in \cite{Wetterich:2019rsn}, based on a study of the Higgs potential at large field values.

%% file: Input/DarkMatter.tex
\label{sec:DarkBSM}
There are indications that a complete description of nature requires particle physics beyond the SM.
On the one hand, observations of galactic rotation curves, the cosmic microwave background and gravitational lensing show that we do not fully understand the sources of gravity in the universe. A modification of the gravitational law itself remains a potential explanation, but missing matter is widely considered a more likely explanation. Primordial black holes may be one candidate, but are only produced in sufficient abundance under appropriate conditions in the early universe.\footnote{It remains an interesting challenge to explore whether or not these are realizable within the asymptotic-safety paradigm.} Alternatively, elementary particles beyond the SM are a candidate. These broadly fall into two categories, namely weakly-interacting massive particles, with masses roughly in the GeV range and couplings of the order of the weak interaction; and light particles with very small interaction strengths, such as axions or axion-like particles (ALPs). Both options have been investigated within the asymptotic-safety paradigm, including their interaction with gravity, and we will review the status in the following section.

On the other hand, there are  also indications that the SM is not a complete description of visible matter. These indications come, for example, from the observation of neutrino oscillations, which prove that the neutrinos of the SM are massive, or from  a tension between the measured and predicted value of the anomalous magnetic dipole moment of the muon. Adding new physics at the electroweak scale or above might solve these shortcomings of the SM. Within asymptotically safe quantum gravity, the gravitational impact on the anomalous magnetic dipole moment of the muon, the generation of neutrino masses, and grand unified theories have been studied explicitly. We will review their status in the following section. 

\subsection{Higgs portal to dark sectors}
\emph{Synopsis: The portal coupling between the SM Higgs and a dark scalar, which is a popular, but increasingly tightly constrained coupling between the SM and a WIMP, is predicted to vanish in asymptotic safety. In contrast, a portal coupling to a dark sector with additional fields beyond the dark scalar may be generated either above or below the Planck scale. In contrast to phenomenological models of dark matter, such asymptotically safe portal models have a high predictive power.}

A (massive) dark scalar $d$ may couple to the Higgs scalar $H$ of the SM through a portal coupling
\be
S = \lambda_{p}\, \int d^4x\, H^{\dagger} H\, d^2.
\ee
If the coupling is of similar size to SM couplings, the dark scalar is in thermal equilibrium and thus produced through a standard freeze-out mechanism in the early universe. Whether or not a single scalar is a viable dark-matter candidate therefore depends on the size of the coupling.
The coupling is also key to observational constraints, e.g., through the non-observation of scattering off SM particles in dedicated dark-matter experiments and non-observation of production at the LHC.\\
Similar to the quartic Higgs self-interaction, the gravitational contribution to $\lambda_p$ is towards irrelevance at the free fixed point, which therefore is IR-attractive, and no other fixed point is generated:
\be
\beta_{\lambda_p}\vert_{\rm grav} = - f_{s} \lambda_p.
\ee

Furthermore, since no contributions from gauge or Yukawa couplings contribute to the beta function of $\lambda_p$, it is not regenerated below the Planck scale.
Thus, $\lambda_p=0$ is a prediction from asymptotic safety that holds at all scales \cite{Eichhorn:2017als}.\\
Intriguingly, experiments continuously improve the strength of bounds on $\lambda_p$, but have not led to a detection, making the prediction from asymptotic safety compatible with the current experimental situation.\\

This result may be circumvented in a more complex dark sector which contains more than one field. For instance, the portal coupling is regenerated below the Planck scale, if a new gauge field couples to the Higgs and the dark scalar. Gauge field fluctuations generate the portal coupling below the Planck scale, leading to a prediction for the portal coupling as a function of the new gauge coupling \cite{Reichert:2019car,Hamada:2020vnf}. As a second example, an additional dark fermion with a Yukawa coupling to the dark scalar, can also generate the Higgs portal coupling \cite{Reichert:2019car, Eichhorn:2020kca}. Dark fermions may even generate a portal coupling at transplanckian scales, i.e., in the fixed-point regime: in a two-step mechanism, a fixed point with a finite dark and visible Yukawa coupling necessarily features a dark and visible non-minimal coupling. Together, the two non-minimal couplings generate a portal coupling. This model, although a toy model with an incomplete SM sector, is a striking example of the predictive power asymptotic safety may have: when viewed as an effective, phenomenological model, it has 9 free parameters (two scalar masses, two quartic scalar couplings, one portal coupling, two non-minimal couplings, two Yukawa couplings). At an asymptotically safe fixed point, only the two mass parameters remain as free parameters. This reduces the parameter space of the model dramatically, see \cite{Eichhorn:2020kca,Eichhorn:2020sbo}. \\
Similarly, \cite{Grabowski:2018fjj} derives specific predictions for the masses of dark-matter particles, if the \emph{conformal Standard Model} \cite{Meissner:2006zh} becomes asymptotically safe. A similar predictive power was observed in a study \cite{Kowalska:2020zve} of BSM physics which provide a dark matter candidate simultaneously with explaining the value of the muon $g-2$ measurement. There, many phenomenological models were ruled out when the coupling to gravity was included in a parameterized way, by including appropriate terms $\sim f_g, \, f_y$ into the beta functions, see also \autoref{sec:SMUVcompletion}. \\

\FRT{Further Reading}\\

\FR{Higgs portal couplings and gauged $B-L$ symmetry}\\
A Higgs portal to a dark sector could also become relevant for models involving a gauged $B-L$ symmetry and thus a new gauge boson beyond the SM. In this case, the dark scalar field spontaneously breaks the $B-L$ symmetry. Imposing that these models become asymptotically safe under the inclusion of quantum gravity restricts the parameterspace of new physics significantly. In particular, the kinetic mixing between the new gauge boson and the SM gauge bosons is fixed, and in some cases the branching fraction of the new gauge boson to SM particles can be predicted, see \cite{Boos:2022jvc}. This model can also be extended to accommodate fermionic dark matter, which is consistent with the observed relic density \cite{Boos:2022pyq}. Imposing asymptotic safety allows constraining the mass of the dark-matter particles as a function of the mass of the new gauge boson.

\subsection{Axion-like particles in the asymptotically safe landscape}
\emph{Synopsis: ALPs couple to the electromagnetic field through a dimension-five operator, which allows a conversion between photons and ALPs that enables experimental searches for ALPs. Within asymptotically safe gravity, there are indications that the ALP-photon coupling is driven to zero, unless gravity is strongly coupled. At strong coupling, there may be a tension with the weak-gravity bound (if it indeed exists), such that there may be a prediction from asymptotic safety, that the ALP-photon coupling vanishes.}\\
The axion is a conjectured BSM particle which solves the strong CP problem. That ``problem" consists in the observation that the coupling of the term $F_{\mu\nu}\tilde{F}^{\mu\nu}$ is very small or potentially zero in QCD.\footnote{For an Abelian gauge theory, $F_{\mu\nu}\tilde{F}^{\mu\nu}$ is a total derivative, but not for a non-Abelian one.} \footnote{Here, we have put quotation marks to highlight that the CP problem is not actually a consistency problem, but a finetuning problem. Such finetuning problems start from the assumption that it is ``natural" for dimensionless numbers to be close to one. Therefore, a small number is said to require an explanation. We disagree with the expectation that a small number requires an explanation more than a number of order one does. Ultimately, a theory in which all free parameters become calculable from first principles would be most satisfying. In the absence of such a theory, any value can be chosen for a free parameter and a particular deviation from 1 does not require less explanation than a particular deviation from 0.} It can be solved by introducing an additional degree of freedom, namely the axion, \cite{Peccei:1977hh}, which takes the place of the coupling of that term. A dynamical mechanism \cite{Peccei:1977ur} drives the expectation value of the axion, i.e., the coupling of that term, to zero.\footnote{It remains an intriguing open question whether the work in \cite{Peccei:1977ur} can be extended to a gravitational setting, where gravitational contributions to the anomalous dimension of the gauge field generate a flow for the coupling. This may solve the strong CP "problem" without the need for new degrees of freedom.} At the same time, a coupling between the axion field $a$ and the electromagnetic field strength is generated, which takes the form
\be
S_{\rm axion-photon} = \int d^4x\, \bar{g}_a\, a\,F_{\mu\nu}\tilde{F}^{\mu\nu}.
\ee 
ALPs are pseudoscalars, like the axion, which have the same coupling to photons. Axions and ALPs are very weakly-coupled dark-matter candidates which can be generated out of equilibrium in the early universe, see \cite{Ferreira:2020fam} for a review.\\
In string theory, the axion and ALPs are expected to exists \cite{Ringwald:2012cu}. If, therefore, they do not exist (or do not couple to photons) in asymptotic safety, that would be a discriminator between the two candidates for quantum gravity. In view of numerous searches for axion-photon and ALP-photon couplings, this discriminator is highly relevant.

In fact, the gravitational contribution to the flow of the ALP-photon coupling $g_a = \bar{g}_a \, k$ is towards relevance at the fixed point $g_{a\, \ast}=0$. However, the gravitational contribution needs to overwhelm the canonical scaling dimension. Otherwise, the fixed point at $g_{a\, \ast}=0$ cannot be connected to a nonzero ALP-photon coupling in the IR. The gravitational contribution can overwhelm the canonical scaling dimension, if gravitational fluctuations are strong enough. This is in conflict with the weak-gravity bound, if the latter indeed exists. In \cite{deBrito:2021akp}, it is therefore concluded that the ALP-photon coupling may be predicted to vanish in asymptotically safe gravity.\\
Should the weak-gravity bound not persist, the ALP-photon coupling satisfies an upper bound in asymptotic safety, because schematically, the beta function is of the form
\be
\beta_{g_a} = g_a +\beta_{1}\, g_a\, G_N + \beta_2 g_a^3,
\ee
where $\beta_1$ is a function of the cosmological constant that is typically negative and $\beta_2$ is positive. The associated fixed-point structure is therefore the same as for the Abelian gauge coupling or some of the Yukawa couplings. The fixed point at $g_{a\, \ast}>0$ hence imposes an upper bound on IR values of the ALP-photon coupling. In contrast to the Abelian gauge coupling and some of the Yukawa couplings, this fixed point only becomes available if $\beta_1\, G_N <-1$, because of the canonical dimension of $g_a$.

In summary, if the weak-gravity bound persists, ALP-photon couplings are likely driven to zero in asymptotic safety, implying that experiments will continue to place tighter constraints without a discovery. If the weak-gravity bound turns out to be a truncation artefact, an upper bound on the ALP-photon coupling exists. Both scenarios are experimentally testable and are in contrast to string theory.

%% file: Input/nondarkBSM.tex
\subsection{Grand unified theories}
\emph{Synopsis: If the matter content of a grand unified theory is compatible with an asymptotically safe fixed point in gravity, then asymptotically safe GUTs are much more predictive than GUTs without gravity. The scalar potential for the many scalars that are required to spontaneously break the large gauge group to the SM gauge groups is completely fixed, except for a single parameter for each scalar field. Thereby, many breaking chains that are considered in gravity-free GUTs are no longer available in asymptotically safe GUTs.}

GUTs are attractive, because they explain charge quantization and explain why the charge of proton and electron are exactly equal in absolute value. Further they can, upon spontaneous symmetry breaking to the SM gauge group, automatically give rise to a right-handed neutrino with the required SM charges. Besides this motivation, the near-crossing of the values of SM gauge couplings at energy scales of about $10^{16}\, \rm GeV$ may be interpreted as an indication for a unification of the SM gauge groups to one larger group.

However, GUTs are unattractive, because they come with a plethora of free parameters. These are linked to the scalar potential: in order to break the GUT gauge group to the SM gauge groups, several scalars are needed, which typically introduce numerous quartic couplings. There are multiple quartic invariants, because there are typically several scalar fields transforming in different representations of the gauge group. Depending on the representation, several different quartic invariants exist. In addition, quartic interactions can be built from quadratic interactions of two different scalars. 
In a typical GUT setting, these are free parameters. In turn, starting from a grand unified symmetry, many different chains of spontaneous symmetry breaking are available, depending on the values one chooses for these free parameters. It is therefore unexplained, why the SM, instead of a theory with a different symmetry, should come out as the low-energy limit of the GUT.

In \cite{Eichhorn:2019dhg}, it was proposed that, if a GUT can become asymptotically safe under the impact of quantum gravity\footnote{It is not conclusively established whether this is the case; the numerous matter fields of a GUT may even destroy the fixed point in the gravitational sector. Different studies come to different conclusions on this point, see, e.g., \cite{Dona:2013qba, Wetterich:2019zdo}. Because these studies differ mainly in their choice of regulator function, the differing results may be interpreted as indicating the need to include further interactions in the studies.}, then scalar potentials may largely be fixed. Specifically, by the same mechanism as in \autoref{sec:Higgs} for the Higgs sector of the SM, all quartic couplings of the scalar potential may be predicted. The quadratic couplings, linked to the mass parameters, are expected to remain free parameters, but these add only a single free parameter for each scalar. In addition, the unified gauge coupling may also be predicted, by the same mechanism as the Abelian gauge coupling in the SM, if the matter content of the GUT is such that the gauge coupling is no longer asymptotically free \cite{Eichhorn:2017muy}.\\
In turn, fixing the values of the quartic couplings removes much of the freedom in choosing scalar potentials to accommodate different breaking chains. As a consequence, one may expect that multiple breaking chains may be excluded and the asymptotically safe GUT setting may combine the attractive attributes of a GUT with the high predictive power of asymptotic safety, achieving explanations of many of the properties of the SM.

This general idea was put to the test in \cite{Held:2022hnw}, where it was indeed shown that for SO(10) GUTs with a 16- and 45-dimensional scalar representation, particular breaking chains can be excluded.
Such a result motivates further studies to determine whether (i) asymptotic safety can be achieved in gravity-GUT-theories and (ii) whether breaking chains to the SM are available or not.

\subsection{Neutrino masses}
\emph{Synopsis: Neutrinos masses may be generated by adding a right-handed Weyl fermion to each generation in the SM, together with a Yukawa coupling to the Higgs field. At an asymptotically safe fixed point, such Yukawa couplings are automatically driven to zero. A cross-over trajectory may therefore exist, which spends many scales close to such a fixed point and thereby drives the neutrino Yukawa coupling to tiny values, thus providing an explanation for the smallness of neutrino masses.\\
Alternatively, the Seesaw mechanism may provide naturally small neutrino masses through heavy right-handed neutrinos. Asymptotic safety may constrain models based on the seesaw mechanism.}

There are several possible ways to generate neutrino masses, two of which we will discuss, namely the inclusion of right-handed Weyl neutrinos, and the addition of Majorana masses for right-handed neutrinos.\\

First, by adding a right-handed Weyl fermion to each generation of the SM, one can introduce a Yukawa coupling $y_{\nu}$ for neutrinos. To generate neutrino masses in the meV range, this Yukawa coupling has to be as small as $y_{\nu}\sim 10^{-13}$. In  
\cite{Held:2019vmi,Kowalska:2022ypk,Eichhorn:2022vgp}, it was shown that asymptotic safety may provide an explanation for such a small value: Extending the studies of the Yukawa sector of the SM in \cite{Eichhorn:2017ylw,Eichhorn:2018whv}, see \autoref{sec:Yukawas}, by the neutrino Yukawa coupling, one finds two possible fixed points: one, at which all Yukawa couplings are asymptotically free, and a second one, at which top and bottom Yukawa coupling are nonzero. The first fixed point can accommodate the tiny IR-value of the neutrino-Yukawa coupling by choosing a suitable trajectory, but cannot explain it. The second fixed point, where top and bottom Yukawa couplings are non-vanishing in the UV predicts a vanishing neutrino Yukawa coupling, and is therefore phenomenologically not viable. 
In combination, however, these two fixed points dynamically generate a tiny neutrino Yukawa coupling: starting from the asymptotically free fixed point, all Yukawa couplings grow, until the top and bottom Yukawa coupling reach the vicinity of the interacting fixed point. There, the critical exponent of the neutrino Yukawa coupling switches sign from positive (relevant) to negative (irrelevant), driving the neutrino Yukawa coupling back down to tiny values. 
Hence, the neutrino generically ends up much lighter than the other fermions.

Because, without an asymptotically safe fixed point of the above type, no mechanism appears to exist to explain a tiny neutrino Yukawa coupling, this form of neutrino mass generation was long considered unappealing, because it is not ``natural".\footnote{There is clearly some arbitrariness in the notion of naturalness: the ratio between electron mass and top quark mass is already $10^{-6}$, but is usually not viewed as a motivation to think about alternative mass generation mechanisms for the electron. Instead, the line between ``natural" and ``unnatural" is in this case drawn somewhere below $10^{-6}$, so that the ratio of about $10^{-9}$ between an meV-neutrino-mass-scale and the electron mass scale is considered ``unnatural". This arbitrariness already suggests that naturalness may at best be a slight motivation to search for alternative explanations, but not a strong and unequivocal reason to rule an ``unnatural" setting out.}

As an alternative, one can introduce heavy (with masses around the GUT scale) right-handed neutrinos with a Majorana mass term. The neutrino mass matrix, upon diagonalization, produces neutrino masses which are inversely proportional to the heavy Majorana mass scale, making neutrinos ``naturally" light.\\
Majorana masses were investigated in the context of asymptotically safe systems in \cite{DeBrito:2019rrh},  where they were found to remain relevant under the impact of quantum gravity fluctuations. Accordingly, the corresponding mass scale can be chosen freely, providing a basis for the seesaw-mechanism with heavy right-handed neutrinos. In \cite{Domenech:2020yjf}, the seesaw mechanism for a specific choice of heavy fields was investigated and constrained. In particular, the additional fields have an impact on the prediction of the Higgs mass and top quark mass, because both depend on physics at the new, heavy scale. \cite{Domenech:2020yjf} therefore also constitutes an example for how an embedding into an asymptotically safe UV completion constrains not just the deep IR (around the electroweak scale), but also constrains physics at intermediate scales which are beyond the reach of current experiments.

\subsection{$g-2$ and flavor anomalies}
\emph{Synopsis: There might be a possibility for new physics at the electroweak scale, in order to resolve tensions between SM predictions and experimental data on the muon magnetic moment and on lepton-flavor non-universality in rare B meson decays. Among the phenomenological models that have been proposed, asymptotic safety could act as a discriminator, because its predictive power may rule out values of couplings which are required to resolve the tensions.}

Current experimental data on parameters of the SM indicate several anomalies, i.e., discrepancies between measurement and theoretical prediction. Since these discrepancies are below a statistical significance of $5\sigma$, they are not discoveries, but merely anomalies. Future experiments and updated theoretical methods will either resolve the tension, or increase the significance, possibly beyond $5\sigma$. The most commonly anomalies concern the anomalous magnetic moment of the muon, $(g-2)_{\mu}$, and flavor anomalies in the $b\to s$ and the $b\to c$ transitions. While these anomalies do not provide sufficient evidence for a significant deviation from SM predictions (yet), many models involving particle physics beyond the SM were developed to explain the anomalies.\\
Some of these models have been investigated in the context of asymptotically safe quantum gravity, see \cite{Kowalska:2020gie, Kowalska:2020zve, Chikkaballi:2022urc}. In particular, it was investigated whether quantum gravity might turn some parameters of the extensions into irrelevant directions. In this case, the predictive power of asymptotically safe quantum gravity would extend to physics beyond the SM and predict, for example, the mass of dark-matter particles that are required to resolve the anomaly. Confronting these predictions with existing bounds from searches for physics beyond the SM can either rule out such solutions or provide strong constraints on the parameter-space. These constraints might guide experimental searches, and allow insights in the most promising next-generation particle colliders.
For instance, in \cite{Kowalska:2020gie}, the leptoquark solution to flavor anomalies was investigated and it was found that asymptotic safety limits the mass range of the leptoquark to $4-7\, \rm TeV$, where it is within reach of future colliders.\\

On the technical level, these studies proceed within a \emph{parameterized} framework, first introduced in \cite{Eichhorn:2018whv}, in which the gravity contributions are parameterized by $f_{c}$'s and resulting fixed points in matter beta functions are investigated. These studies therefore provide experimentally testable consequences of asymptotic safety under the assumption that asymptotic safety is realized in the full system. Checking this assumption would require (i) to account for the impact of the new matter fields on the gravitational fixed point to check whether it exists, (ii) to calculate the resulting values of $f_{c}$ to compare with those required on a phenomenological level and (iii) to check that higher-order interactions in the matter sector as well as non-minimal interactions between matter and gravity are subleading and do not change the conclusions much.

%% file: Input/nearperturbative.tex
\section{On the near-perturbative nature of gravity-matter systems}
\label{sec:ner-pert}
\emph{Synopsis: An asymptotically safe fixed point can be near-perturbative, i.e., be close to canonical power counting. For such a fixed point, calculations are easier and systematic uncertainties are simpler to control. The SM coupled to gravity may have such a near-perturbative fixed point  which is easy to connect to the perturbative RG flow of the SM at and below the Planck scale.  More generally, phenomenological studies of asymptotically safe gravity-matter systems typically rely on the assumption that the system is near-perturbative.}

In the previous sections, we relied on an implicit assumption about the nature of the systems we investigated: this assumption is implicit in the beta functions we used, which are all limited to leading order terms (i.e., low orders in the couplings) and to the canonically least irrelevant interactions of the system.\\
This assumption is that the matter-gravity systems we investigated are sufficiently weakly coupled to be near-perturbative, despite being asymptotically safe. Technically, this makes robust calculations possible. Physically, this goes hand in hand with the idea that to control the (trans)planckian regime in a quantum field theory of gravity, a mechanism of dynamical weakening has to apply to quantum gravity.

To provide the technical and conceptual underpinning of the discussion in the previous sections, we therefore define in more detail what we mean by near-perturbative, and discuss the evidence for the near-perturbative nature of gravity-matter systems.

In general, interacting systems can be strongly coupled and governed by non-perturbative effects, or can be weakly coupled and governed by perturbative effects, or anything inbetween. This has phenomenological implications -- for instance, in strongly-coupled systems, the fundamental degrees of freedom can bind together and form new, stable or unstable states. It also has formal implications -- most importantly for the set of tools that is best to analyze the system and also for the type of approximations that can be made.

For gravity in the UV, one may first expect that it is strongly coupled and non-perturbative. This expectation arises, because the dimensionless Newton coupling, $\GN$, grows when one goes from low to high momenta. If one simply extrapolates from the classical regime, where $\GN(k) \sim k^2$, $\GN$ becomes of order one at the Planck scale. This is typically interpreted as a sign of strong coupling. We rush to caution that values of couplings are not a good measure of perturbativity, because, e.g., the fixed-point value of a coupling can be changed arbitrarily by rescalings of the coupling.

However,
several sets of results indicate that asymptotically safe gravity-matter systems are near-perturbative at high energies. With near-perturbative we refer to a situation where the theory is interacting, i.e., not strictly perturbative, but at the same time lacks non-perturbative phenomena such as the formation of stable bound states.
Such near-perturbative behavior of asymptotically safe gravity matter systems is indicated by i) the critical exponents of higher-order interactions which remain near-canonical, ii) the contributions of quantum gravity to beta functions in the matter sector, which are small, and iii) symmetry identities between gravity-matter interactions which are near-trivial, as they are in perturbative settings. We will discuss each of these indications in the following.\\

\emph{Canonically irrelevant couplings remain irrelevant}\\
At a non-interacting fixed point, the critical exponents of all couplings correspond to their canonical mass dimension. At an interacting fixed point, the critical exponents still contain a dimensional contribution, but also an additional contribution $\delta_i$ which is induced by quantum fluctuations, i.e,
\begin{equation}
\Theta_i=d_{\bar{g}_i}+\delta_i\,.
\end{equation}
If the quantum contributions $\delta_i$ grow very large, they can turn canonically irrelevant couplings into relevant directions at an interacting fixed point. This would decrease the predictivity of the system, compared to the non-interacing fixed point, since more relevant directions indicate more free parameters that need to be fixed by experiments.
More importantly, this would indicate that the system is very non-perturbative, since quantum fluctuations drastically change qualitative features of the system. Conversely, an interacting fixed point where all canonically irrelevant couplings remain irrelevant, is near-perturbative and quantum fluctuations only change quantitative features, such as the value of couplings at low energies.\\
In all studies so far, the critical exponents of interactions involving matter fields follow their canonical mass dimension.
In particular, canonically irrelevant matter interactions \cite{Eichhorn:2011pc, Eichhorn:2012va, Eichhorn:2016esv, Christiansen:2017gtg, Eichhorn:2017eht, Eichhorn:2017als, Eichhorn:2020sbo} and non-minimal gravity-matter interactions \cite{Eichhorn:2016vvy, Eichhorn:2017sok, Eichhorn:2018nda, Eichhorn:2020sbo, Daas:2020dyo, Daas:2021abx} remain irrelevant at the asymptotically safe fixed point. This justifies truncations based on canonical power counting a-posteriori.
\\

\emph{The impact of gravity on the matter sector}\\
As mentioned previously, the strength with which gravitational fluctuations impact the matter sector is encoded in effective gravitational couplings
\begin{equation}
G_{\rm eff}^{(n)} = \frac{\GN}{(1-2\Lambda)^n}\,.
\end{equation}
Accordingly, the impact of gravity on the matter sector becomes weaker, the more negative the fixed-point value of the cosmological constant, and the smaller the fixed-point value of the Newton coupling gets. Independent of the details of the setup, fermionic matter was found to decrease the effective gravitational coupling \cite{Eichhorn:2018nda}. Similarly, gauge fields decrease  ${\GN}_{,\,\mathrm{eff}\, n}$, since the fixed-point value for \GN{} approaches zero for $\Nvec\to\infty$. While scalar fields might increase ${\GN}_{,\,\mathrm{eff}\, n}$, the effective gravitational couplings for the field content of the SM remains small.\\
As a consequence, the gravitational contribution to the scale dependence of matter couplings is subleading compared to matter contributions, such that the matter sector might remain near-perturbative at the UV-fixed point. This can also be seen by explicitly evaluating $f_g$ or $f_y$, e.g., at the fixed point found in \cite{Dona:2013qba}, as was done in \cite{Eichhorn:2017lry}, finding $f_g=0.048$. In particular, the observation of non-vanishing fermion masses requires a small-enough impact of gravity on the matter sector, as discussed in Fig.~\ref{fig:mattermattersinterplay}, see also \cite{Pastor-Gutierrez:2022nki}.\\
These results imply that the asymptotically safe fixed point might provide a straightforward UV-completion of the SM, which is perturbative at the Planck scale. Conversely, one may interpret the fact that the SM couplings are perturbative at the Planck scale as an indication, that a UV completion with gravity must be near-perturbative.
\\

\emph{Non-trivial symmetry identities}\\
Just like in Abelian and non-Abelian gauge theories, one breaks the gauge symmetry of gravity when computing gravitational fluctuations with the FRG. As a consequence, the scale dependence of different gravity-matter vertices differs from one another. In other words: the scale dependence of the Newton coupling, when read-off from different vertices, differs. This however does not mean that diffeomorphism invariance is manifestly broken. Instead, non-trivial symmetry identities, the Slavnov-Taylor identities, encode how diffeomorphism invariance is restored. If these identities are solved together with the scale-dependence of the effective action, diffeomorphism invariance is retained along the RG-trajectory.\\
Naively these identities are trivial in the perturbative regime, where a single gauge coupling can be defined. In Yang-Mills theories, the one-loop universality of beta-functions ensures exactly this feature: the scale dependence for the gauge coupling is identical, when extracted from pure gauge, or gauge-ghost vertices. In the non-perturbative regime however, the scale-dependence of different vertices disagrees, and can even feature opposite signs, see \cite{Cyrol:2016tym}.\\
In asymptotically safe gravity, different gravity-matter vertices are found to agree on a semi-quantitative level at the fixed point \cite{Christiansen:2017cxa, Eichhorn:2018akn, Eichhorn:2018ydy, Eichhorn:2018nda}, but can disagree significantly away from the fixed point. 

The semi-quantitative agreement is defined in \cite{Eichhorn:2018akn, Eichhorn:2018ydy} by comparing the scale dependences of different gravity-matter vertices, and setting all versions of the Newton coupling equal to one another. If these differences between scale dependences are zero, one unique Newton coupling can be defined. If these differences are large, all different vertices have to be treated independently to fully capture the UV-behavior of asymptotically safe gravity-matter systems. In analogy to QCD, a semi-quantitative agreement between different vertices at the fixed-point indicates that the theory might be near-perturbative. In particular, it might imply that the underlying Slavnov-Taylor identities are trivial at the fixed point.\\
 This provides another piece of evidence that asymptotically safe quantum gravity might be near-perturbative: it is non-perturbative enough to induce scale invariance at high energies, but remains as perturbative as possible.

%% file: Input/summary.tex
What are the key challenges in high-energy physics today? The first key challenge is to test the quantum nature of gravity, i.e., predict observable and testable consequences from candidate quantum theories of gravity. In this field, predictions from fundamental theory are rare, and most results are based on phenomenological models.
The second key challenge is to solve open problems in particle physics, such as the nature of dark matter, the origin of the free parameters of the Standard Model, the origin of matter-antimatter asymmetry, the mechanism for neutrino-mass generation. In this field, there is a vast collection of more or less ad-hoc phenomenological models.\\
In this section, we have summarized research that aims at addressing both challenges at once. The key assumption underlying this line of research posits that only by addressing both challenges at once can one find a meaningful solution: predictions from fundamental theory require a quantum theory of gravity and matter as their starting point; such a theory in turn is expected to have high predictive power and thus select among the ad-hoc phenomenological models often proposed to solve specific problems in particle physics.\\
The paradigm in which such a quantum theory of gravity and matter is being developed is the asymptotic-safety paradigm. It requires quantum scale symmetry at UV scales. In turn, the presence of this symmetry constrains the UV and IR properties of the theory. Constraints on the IR, where the symmetry is no longer realized, arise because a departure from scale symmetry is only possible for the relevant parameters of the theory, which are very few. To use a literary analogy, quantum scale symmetry is like the Cheshire cat: even when it is no longer present, its smile remains behind -- the smile being the relations between couplings in the theory that hold in the IR.\\
The current state-of-the-art suggests that asymptotic safety of gravity with matter may indeed be realized for Standard-Model-like theories, i.e., theories with the degrees of freedom of the Standard Model, and possibly a few additional degrees of freedom; and with coupling values close to or exactly those of the Standard Model. There is evidence that, under the impact of the matter fields of the Standard Model, gravity is asymptotically safe. In turn, under the impact of asymptotically safe gravity, the Standard Model couplings become asymptotically free; and the Higgs quartic coupling emerges as a calculable quantity in the IR. There is even evidence of higher predictive power, where the Abelian gauge coupling also emerges as a calculable quantity in the IR; and potentially some of the Yukawa couplings do as well. The status of these predictions is as follows: the existence of predictive (partial) fixed points for the Abelian gauge coupling and the Yukawa couplings is a robust result\footnote{For the Yukawa couplings, this constrains the gravitational fixed-point values.}; however, the resulting IR values of the couplings are only known within significant systematic uncertainties. Therefore, it is currently not known whether or not the predictions match observations or not.\\
The current state-of-the-art also includes theories beyond the Standard Model, which are significantly constrained by the predictive power of asymptotic safety. Among a growing number of theories that have been explored, most noteworthy is the high predictive power for theories including dark-matter candidates. Among popular proposals, such as a Higgs portal to a dark scalar, an axion-like-particle, or a dark photon, many possibilities are, according to the most advanced calculations in these settings, ruled out.\\
These successes of the asymptotic-safety paradigm for gravity with matter have by now given rise to a phenomenological approach to such theories, which one may call a \emph{principled-parameterized-approach}, first put forward in \cite{Eichhorn:2018whv}: the approach consists in adding contributions to the beta functions of a matter theory. These contributions come with free parameters (thus ``parameterized" approach); but their dependence on the couplings of the theory and the scale at which they are important are that of quantum gravity (thus ``principled" approach). This approach is based on the observation that quantum-gravity contributions to matter couplings are ``blind" to internal symmetries; thus, for instance, the quantum-gravity contribution to the beta function of a gauge coupling does not depend on the gauge group. Therefore, in the principled-parameterized approach, one assumes that a fixed point is present in the gravity theory under the impact of the matter fields of the theory. One then parameterizes the effect of gravity on the matter couplings, searches for fixed points and analyzes their predictive power. Due to its relative technical simplicity (compared to fully-fledged calculations of the gravitational contributions), this principled-parameterized approach holds significant promise to solve the two key challenges in high-energy physics: first, it allows to derive observable consequences of quantum gravity. Second, it allows to search for fundamental theories which solve outstanding problems in particle physics.\newline\\

There are of course open questions in asymptotically safe settings with gravity and matter. First, there are several open questions that apply to asymptotically safe quantum gravity in general. These include, for example, unitarity, the Lorentzian signature, and convergence of results, see \cite{Bonanno:2020bil} for a thorough discussion. Second, there are open questions specifically within the interplay of gravity and matter. We will comment on some of them in the following.\\

\emph{The nature of dark matter}\\
While dark matter particles remain a compelling theoretical explanation of numerous astrophysical and cosmological observations, no such elementary particles have been detected yet. Therefore, there is a lot of freedom in constructing phenomenological models that can explain experimental observations. Typically, these models come with several free parameters and therefore have high-dimensional parameter spaces, which can only in part be probed experimentally. Embedding these dark-matter models within asymptotic safety can reduce the amount of free parameters significantly, as we have reviewed in this chapter. Furthermore, some models could even be ruled out by not admitting an asymptotically safe fixed point. Therefore, the \emph{principled-parametrized-approach} to dark matter in asymptotic safety  may be a successful new paradigm for the  nature of dark matter. It might educate model-building efforts and in the future even guide experimental searches for dark matter.\\

\emph{Explaining the baryon asymmetry of our universe}\\
According to current models for the dynamics in the early universe, matter and anti-matter are produced at the same rate. Furthermore, neither the SM nor General Relativity include a mechanism that generates a large enough asymmetry between baryons and anti-baryons. Thus, the origin of the observed asymmetry between the amount of matter and antimatter in the universe is still unknown.  Just like for dark matter, there are phenomenological models which satisfy the Sakharov-conditions \cite{Sakharov:1967dj} and may thus explain the asymmetry.
So far, within asymptotically safe quantum gravity, matter-antimatter-asymmetry is mostly unexplored. Again, the \emph{principled-parametrized-approach} and the strong predictive power of scale symmetry might provide insights into which form of physics beyond the SM is viable.\\

\emph{The origin of neutrino masses}\\
To account for neutrino oscillations, neutrinos have to be massive. Various mechanisms for mass generation have been proposed and led to phenomenological models. Among this large number of models, some of which can be probed experimentally, the \emph{principled-parameterized-approach} is expected to select a smaller subset, providing us with theoretical guidance on the origin of neutrino masses.\\

These three open questions -- the nature of dark matter, the origin of matter-antimatter-asymmetry and the origin of neutrino masses -- are just three examples of physics beyond the SM which is unexplained, and described by a large number of phenomenological models, among which the \emph{principled-parameterized-approach} may select a smaller subset. More generally, we expect that the embedding into asymptotic safety, and the description of this embedding through the \emph{principled-parameterized-approach} can be a powerful discriminator for model-building. This can be useful in developing models to describe new physics and guide experiment. At the same time, it provides a way to derive potential phenomenological consequences of quantum gravity \emph{at energies much below the Planck scale}. There is therefore the genuine possibility of ruling out asymptotically safe gravity-matter theories experimentally.

%% file: draft.bbl
\begin{thebibliography}{215}
\providecommand{\natexlab}[1]{#1}
\providecommand{\url}[1]{{#1}}
\providecommand{\urlprefix}{URL }
\expandafter\ifx\csname urlstyle\endcsname\relax
  \providecommand{\doi}[1]{DOI~\discretionary{}{}{}#1}\else
  \providecommand{\doi}{DOI~\discretionary{}{}{}\begingroup
  \urlstyle{rm}\Url}\fi
\providecommand{\eprint}[2][]{\url{#2}}

\bibitem[{Addazi et~al(2022)}]{Addazi:2021xuf}
Addazi A, et~al (2022) {Quantum gravity phenomenology at the dawn of the
  multi-messenger era\textemdash{}A review}. Prog Part Nucl Phys 125:103,948,
  \doi{10.1016/j.ppnp.2022.103948}, \eprint{2111.05659}

\bibitem[{Berry and Gair(2011)}]{Berry:2011pb}
Berry CPL, Gair JR (2011) {Linearized f(R) Gravity: Gravitational Radiation and
  Solar System Tests}. Phys Rev D 83:104,022, \doi{10.1103/PhysRevD.83.104022},
  [Erratum: Phys.Rev.D 85, 089906 (2012)], \eprint{1104.0819}

\bibitem[{Niedermaier(2009)}]{Niedermaier:2009zz}
Niedermaier MR (2009) {Gravitational Fixed Points from Perturbation Theory}.
  Phys Rev Lett 103:101,303, \doi{10.1103/PhysRevLett.103.101303}

\bibitem[{Niedermaier(2010)}]{Niedermaier:2010zz}
Niedermaier M (2010) {Gravitational fixed points and asymptotic safety from
  perturbation theory}. Nucl Phys B 833:226--270,
  \doi{10.1016/j.nuclphysb.2010.01.016}

\bibitem[{Eichhorn(2019)}]{Eichhorn:2018yfc}
Eichhorn A (2019) {An asymptotically safe guide to quantum gravity and matter}.
  Front Astron Space Sci 5:47, \doi{10.3389/fspas.2018.00047},
  \eprint{1810.07615}

\bibitem[{Gastmans et~al(1978)Gastmans, Kallosh, and Truffin}]{Gastmans:1977ad}
Gastmans R, Kallosh R, Truffin C (1978) {Quantum Gravity Near Two-Dimensions}.
  Nucl Phys B 133:417--434, \doi{10.1016/0550-3213(78)90234-1}

\bibitem[{Christensen and Duff(1978)}]{Christensen:1978sc}
Christensen SM, Duff MJ (1978) {Quantum Gravity in Two + $\epsilon$
  Dimensions}. Phys Lett B 79:213--216, \doi{10.1016/0370-2693(78)90225-3}

\bibitem[{Kawai et~al(1993{\natexlab{a}})Kawai, Kitazawa, and
  Ninomiya}]{Kawai:1992np}
Kawai H, Kitazawa Y, Ninomiya M (1993{\natexlab{a}}) {Scaling exponents in
  quantum gravity near two-dimensions}. Nucl Phys B 393:280--300,
  \doi{10.1016/0550-3213(93)90246-L}, \eprint{hep-th/9206081}

\bibitem[{Kawai et~al(1993{\natexlab{b}})Kawai, Kitazawa, and
  Ninomiya}]{Kawai:1993mb}
Kawai H, Kitazawa Y, Ninomiya M (1993{\natexlab{b}}) {Ultraviolet stable fixed
  point and scaling relations in (2+epsilon)-dimensional quantum gravity}. Nucl
  Phys B 404:684--716, \doi{10.1016/0550-3213(93)90594-F},
  \eprint{hep-th/9303123}

\bibitem[{Aida and Kitazawa(1997)}]{Aida:1996zn}
Aida T, Kitazawa Y (1997) {Two loop prediction for scaling exponents in
  (2+epsilon)-dimensional quantum gravity}. Nucl Phys B 491:427--460,
  \doi{10.1016/S0550-3213(97)00091-6}, \eprint{hep-th/9609077}

\bibitem[{Loll(2020)}]{Loll:2019rdj}
Loll R (2020) {Quantum Gravity from Causal Dynamical Triangulations: A Review}.
  Class Quant Grav 37(1):013,002, \doi{10.1088/1361-6382/ab57c7},
  \eprint{1905.08669}

\bibitem[{Catterall et~al(2018)Catterall, Laiho, and
  Unmuth-Yockey}]{Catterall:2018dns}
Catterall S, Laiho J, Unmuth-Yockey J (2018) {K\"ahler-Dirac fermions on
  Euclidean dynamical triangulations}. Phys Rev D 98(11):114,503,
  \doi{10.1103/PhysRevD.98.114503}, \eprint{1810.10626}

\bibitem[{Dai et~al(2021)Dai, Laiho, Schiffer, and Unmuth-Yockey}]{Dai:2021fqb}
Dai M, Laiho J, Schiffer M, Unmuth-Yockey J (2021) {Newtonian binding from
  lattice quantum gravity}. Phys Rev D 103(11):114,511,
  \doi{10.1103/PhysRevD.103.114511}, \eprint{2102.04492}

\bibitem[{Bassler et~al(2021)Bassler, Laiho, Schiffer, and
  Unmuth-Yockey}]{Bassler:2021pzt}
Bassler S, Laiho J, Schiffer M, Unmuth-Yockey J (2021) {The de Sitter Instanton
  from Euclidean Dynamical Triangulations}. Phys Rev D 103:114,504,
  \doi{10.1103/PhysRevD.103.114504}, \eprint{2103.06973}

\bibitem[{Asaduzzaman and Catterall(2022)}]{Asaduzzaman:2022kxz}
Asaduzzaman M, Catterall S (2022) {Euclidean Dynamical Triangulations
  Revisited} \eprint{2207.12642}

\bibitem[{Hamber(2009)}]{Hamber:2009mt}
Hamber HW (2009) {Quantum Gravity on the Lattice}. Gen Rel Grav 41:817--876,
  \doi{10.1007/s10714-009-0769-y}, \eprint{0901.0964}

\bibitem[{Kelly and Trugenberger(2019)}]{Kelly:2018diy}
Kelly C, Trugenberger CA (2019) {Combinatorial Quantum Gravity: Emergence of
  Geometric Space from Random Graphs}. J Phys Conf Ser 1275(1):012,016,
  \doi{10.1088/1742-6596/1275/1/012016}, \eprint{1811.12905}

\bibitem[{Kelly et~al(2019)Kelly, Trugenberger, and Biancalana}]{Kelly:2019rpx}
Kelly C, Trugenberger CA, Biancalana F (2019) {Self-Assembly of Geometric Space
  from Random Graphs}. Class Quant Grav 36(12):125,012,
  \doi{10.1088/1361-6382/ab1c7d}, \eprint{1901.09870}

\bibitem[{Eichhorn et~al(2020)Eichhorn, Lumma, Pereira, and
  Sikandar}]{Eichhorn:2019hsa}
Eichhorn A, Lumma J, Pereira AD, Sikandar A (2020) {Universal critical behavior
  in tensor models for four-dimensional quantum gravity}. JHEP 02:110,
  \doi{10.1007/JHEP02(2020)110}, \eprint{1912.05314}

\bibitem[{Eichhorn(2018)}]{Eichhorn:2017bwe}
Eichhorn A (2018) {Towards coarse graining of discrete Lorentzian quantum
  gravity}. Class Quant Grav 35(4):044,001, \doi{10.1088/1361-6382/aaa0a3},
  \eprint{1709.10419}

\bibitem[{Eichhorn(2019)}]{Eichhorn:2019xav}
Eichhorn A (2019) {Steps towards Lorentzian quantum gravity with causal sets}.
  J Phys Conf Ser 1275(1):012,010, \doi{10.1088/1742-6596/1275/1/012010},
  \eprint{1902.00391}

\bibitem[{Manrique and Reuter(2011)}]{Manrique:2009tj}
Manrique E, Reuter M (2011) {Bare versus Effective Fixed Point Action in
  Asymptotic Safety: The Reconstruction Problem}. PoS CLAQG08:001,
  \doi{10.22323/1.079.0001}, \eprint{0905.4220}

\bibitem[{Morris and Slade(2015)}]{Morris:2015oca}
Morris TR, Slade ZH (2015) {Solutions to the reconstruction problem in
  asymptotic safety}. JHEP 11:094, \doi{10.1007/JHEP11(2015)094},
  \eprint{1507.08657}

\bibitem[{Fraaije et~al(2022)Fraaije, Platania, and
  Saueressig}]{Fraaije:2022uhg}
Fraaije M, Platania A, Saueressig F (2022) {On the reconstruction problem in
  quantum gravity}. Phys Lett B 834:137,399,
  \doi{10.1016/j.physletb.2022.137399}, \eprint{2206.10626}

\bibitem[{Wetterich(1993)}]{Wetterich:1992yh}
Wetterich C (1993) {Exact evolution equation for the effective potential}. Phys
  Lett B 301:90--94, \doi{10.1016/0370-2693(93)90726-X}, \eprint{1710.05815}

\bibitem[{Morris(1994)}]{Morris:1993qb}
Morris TR (1994) {The Exact renormalization group and approximate solutions}.
  Int J Mod Phys A 9:2411--2450, \doi{10.1142/S0217751X94000972},
  \eprint{hep-ph/9308265}

\bibitem[{Ellwanger(1994)}]{Ellwanger:1993mw}
Ellwanger U (1994) {FLow equations for N point functions and bound states}. Z
  Phys C 62:503--510, \doi{10.1007/BF01555911}, \eprint{hep-ph/9308260}

\bibitem[{Reuter(1998)}]{Reuter:1996cp}
Reuter M (1998) {Nonperturbative evolution equation for quantum gravity}. Phys
  Rev D 57:971--985, \doi{10.1103/PhysRevD.57.971}, \eprint{hep-th/9605030}

\bibitem[{Manrique et~al(2011)Manrique, Rechenberger, and
  Saueressig}]{Manrique:2011jc}
Manrique E, Rechenberger S, Saueressig F (2011) {Asymptotically Safe Lorentzian
  Gravity}. Phys Rev Lett 106:251,302, \doi{10.1103/PhysRevLett.106.251302},
  \eprint{1102.5012}

\bibitem[{Bonanno et~al(2022)Bonanno, Denz, Pawlowski, and
  Reichert}]{Bonanno:2021squ}
Bonanno A, Denz T, Pawlowski JM, Reichert M (2022) {Reconstructing the
  graviton}. SciPost Phys 12(1):001, \doi{10.21468/SciPostPhys.12.1.001},
  \eprint{2102.02217}

\bibitem[{Fehre et~al(2021)Fehre, Litim, Pawlowski, and
  Reichert}]{Fehre:2021eob}
Fehre J, Litim DF, Pawlowski JM, Reichert M (2021) {Lorentzian quantum gravity
  and the graviton spectral function} \eprint{2111.13232}

\bibitem[{Dupuis et~al(2021)Dupuis, Canet, Eichhorn, Metzner, Pawlowski,
  Tissier, and Wschebor}]{Dupuis:2020fhh}
Dupuis N, Canet L, Eichhorn A, Metzner W, Pawlowski JM, Tissier M, Wschebor N
  (2021) {The nonperturbative functional renormalization group and its
  applications}. Phys Rept 910:1--114, \doi{10.1016/j.physrep.2021.01.001},
  \eprint{2006.04853}

\bibitem[{Balog et~al(2019)Balog, Chat\'e, Delamotte, Marohnic, and
  Wschebor}]{Balog:2019rrg}
Balog I, Chat\'e H, Delamotte B, Marohnic M, Wschebor N (2019) {Convergence of
  Nonperturbative Approximations to the Renormalization Group}. Phys Rev Lett
  123(24):240,604, \doi{10.1103/PhysRevLett.123.240604}, \eprint{1907.01829}

\bibitem[{Martini et~al(2022)Martini, Vacca, and Zanusso}]{Martini:2022sll}
Martini R, Vacca GP, Zanusso O (2022) {Perturbative approaches to
  non-perturbative quantum gravity} \eprint{2210.13910}

\bibitem[{Pawlowski and Reichert(2021)}]{Pawlowski:2020qer}
Pawlowski JM, Reichert M (2021) {Quantum Gravity: A Fluctuating Point of View}.
  Front in Phys 8:551,848, \doi{10.3389/fphy.2020.551848}, \eprint{2007.10353}

\bibitem[{Manrique and Reuter(2010)}]{Manrique:2009uh}
Manrique E, Reuter M (2010) {Bimetric Truncations for Quantum Einstein Gravity
  and Asymptotic Safety}. Annals Phys 325:785--815,
  \doi{10.1016/j.aop.2009.11.009}, \eprint{0907.2617}

\bibitem[{Manrique et~al(2011)Manrique, Reuter, and
  Saueressig}]{Manrique:2010mq}
Manrique E, Reuter M, Saueressig F (2011) {Matter Induced Bimetric Actions for
  Gravity}. Annals Phys 326:440--462, \doi{10.1016/j.aop.2010.11.003},
  \eprint{1003.5129}

\bibitem[{Falls et~al(2013)Falls, Litim, Nikolakopoulos, and
  Rahmede}]{Falls:2013bv}
Falls K, Litim DF, Nikolakopoulos K, Rahmede C (2013) {A bootstrap towards
  asymptotic safety} \eprint{1301.4191}

\bibitem[{Falls et~al(2018)Falls, King, Litim, Nikolakopoulos, and
  Rahmede}]{Falls:2017lst}
Falls K, King CR, Litim DF, Nikolakopoulos K, Rahmede C (2018) {Asymptotic
  safety of quantum gravity beyond Ricci scalars}. Phys Rev D 97(8):086,006,
  \doi{10.1103/PhysRevD.97.086006}, \eprint{1801.00162}

\bibitem[{Kluth and Litim(2020)}]{Kluth:2020bdv}
Kluth Y, Litim DF (2020) {Fixed Points of Quantum Gravity and the
  Dimensionality of the UV Critical Surface} \eprint{2008.09181}

\bibitem[{Knorr et~al(2019)Knorr, Ripken, and Saueressig}]{Knorr:2019atm}
Knorr B, Ripken C, Saueressig F (2019) {Form Factors in Asymptotic Safety:
  conceptual ideas and computational toolbox}. Class Quant Grav 36(23):234,001,
  \doi{10.1088/1361-6382/ab4a53}, \eprint{1907.02903}

\bibitem[{Knorr et~al(2022)Knorr, Ripken, and Saueressig}]{Knorr:2022dsx}
Knorr B, Ripken C, Saueressig F (2022) {Form Factors in Asymptotically Safe
  Quantum Gravity} \eprint{2210.16072}

\bibitem[{Benedetti et~al(2011)Benedetti, Groh, Machado, and
  Saueressig}]{Benedetti:2010nr}
Benedetti D, Groh K, Machado PF, Saueressig F (2011) {The Universal RG
  Machine}. JHEP 06:079, \doi{10.1007/JHEP06(2011)079}, \eprint{1012.3081}

\bibitem[{Falls et~al(2020)Falls, Ohta, and Percacci}]{Falls:2020qhj}
Falls K, Ohta N, Percacci R (2020) {Towards the determination of the dimension
  of the critical surface in asymptotically safe gravity}. Phys Lett B
  810:135,773, \doi{10.1016/j.physletb.2020.135773}, \eprint{2004.04126}

\bibitem[{Sen et~al(2022)Sen, Wetterich, and Yamada}]{Sen:2021ffc}
Sen S, Wetterich C, Yamada M (2022) {Asymptotic freedom and safety in quantum
  gravity}. JHEP 03:130, \doi{10.1007/JHEP03(2022)130}, \eprint{2111.04696}

\bibitem[{Wetterich(2021)}]{Wetterich:2020cxq}
Wetterich C (2021) {Fundamental scale invariance}. Nucl Phys B 964:115,326,
  \doi{10.1016/j.nuclphysb.2021.115326}, \eprint{2007.08805}

\bibitem[{Reuter and Saueressig(2002)}]{Reuter:2001ag}
Reuter M, Saueressig F (2002) {Renormalization group flow of quantum gravity in
  the Einstein-Hilbert truncation}. Phys Rev D 65:065,016,
  \doi{10.1103/PhysRevD.65.065016}, \eprint{hep-th/0110054}

\bibitem[{Lauscher and Reuter(2002{\natexlab{a}})}]{Lauscher:2001ya}
Lauscher O, Reuter M (2002{\natexlab{a}}) {Ultraviolet fixed point and
  generalized flow equation of quantum gravity}. Phys Rev D 65:025,013,
  \doi{10.1103/PhysRevD.65.025013}, \eprint{hep-th/0108040}

\bibitem[{Lauscher and Reuter(2002{\natexlab{b}})}]{Lauscher:2002sq}
Lauscher O, Reuter M (2002{\natexlab{b}}) {Flow equation of quantum Einstein
  gravity in a higher derivative truncation}. Phys Rev D 66:025,026,
  \doi{10.1103/PhysRevD.66.025026}, \eprint{hep-th/0205062}

\bibitem[{Litim(2004)}]{Litim:2003vp}
Litim DF (2004) {Fixed points of quantum gravity}. Phys Rev Lett 92:201,301,
  \doi{10.1103/PhysRevLett.92.201301}, \eprint{hep-th/0312114}

\bibitem[{Niedermaier and Reuter(2006)}]{Niedermaier:2006wt}
Niedermaier M, Reuter M (2006) {The Asymptotic Safety Scenario in Quantum
  Gravity}. Living Rev Rel 9:5--173, \doi{10.12942/lrr-2006-5}

\bibitem[{Codello et~al(2009)Codello, Percacci, and Rahmede}]{Codello:2008vh}
Codello A, Percacci R, Rahmede C (2009) {Investigating the Ultraviolet
  Properties of Gravity with a Wilsonian Renormalization Group Equation}.
  Annals Phys 324:414--469, \doi{10.1016/j.aop.2008.08.008}, \eprint{0805.2909}

\bibitem[{Benedetti et~al(2009{\natexlab{a}})Benedetti, Machado, and
  Saueressig}]{Benedetti:2009rx}
Benedetti D, Machado PF, Saueressig F (2009{\natexlab{a}}) {Asymptotic safety
  in higher-derivative gravity}. Mod Phys Lett A 24:2233--2241,
  \doi{10.1142/S0217732309031521}, \eprint{0901.2984}

\bibitem[{Benedetti et~al(2009{\natexlab{b}})Benedetti, Machado, and
  Saueressig}]{Benedetti:2009iq}
Benedetti D, Machado PF, Saueressig F (2009{\natexlab{b}}) {Four-derivative
  interactions in asymptotically safe gravity}. AIP Conf Proc 1196(1):44,
  \doi{10.1063/1.3284399}, \eprint{0909.3265}

\bibitem[{Manrique et~al(2011)Manrique, Reuter, and
  Saueressig}]{Manrique:2010am}
Manrique E, Reuter M, Saueressig F (2011) {Bimetric Renormalization Group Flows
  in Quantum Einstein Gravity}. Annals Phys 326:463--485,
  \doi{10.1016/j.aop.2010.11.006}, \eprint{1006.0099}

\bibitem[{Groh et~al(2011)Groh, Rechenberger, Saueressig, and
  Zanusso}]{Groh:2011vn}
Groh K, Rechenberger S, Saueressig F, Zanusso O (2011) {Higher Derivative
  Gravity from the Universal Renormalization Group Machine}. PoS
  EPS-HEP2011:124, \doi{10.22323/1.134.0124}, \eprint{1111.1743}

\bibitem[{Rechenberger and Saueressig(2012)}]{Rechenberger:2012pm}
Rechenberger S, Saueressig F (2012) {The $R^2$ phase-diagram of QEG and its
  spectral dimension}. Phys Rev D 86:024,018, \doi{10.1103/PhysRevD.86.024018},
  \eprint{1206.0657}

\bibitem[{Donkin and Pawlowski(2012)}]{Donkin:2012ud}
Donkin I, Pawlowski JM (2012) {The phase diagram of quantum gravity from
  diffeomorphism-invariant RG-flows} \eprint{1203.4207}

\bibitem[{Christiansen et~al(2014)Christiansen, Litim, Pawlowski, and
  Rodigast}]{Christiansen:2012rx}
Christiansen N, Litim DF, Pawlowski JM, Rodigast A (2014) {Fixed points and
  infrared completion of quantum gravity}. Phys Lett B 728:114--117,
  \doi{10.1016/j.physletb.2013.11.025}, \eprint{1209.4038}

\bibitem[{Benedetti and Caravelli(2012)}]{Benedetti:2012dx}
Benedetti D, Caravelli F (2012) {The Local potential approximation in quantum
  gravity}. JHEP 06:017, \doi{10.1007/JHEP06(2012)017}, [Erratum: JHEP 10, 157
  (2012)], \eprint{1204.3541}

\bibitem[{Dietz and Morris(2013)}]{Dietz:2012ic}
Dietz JA, Morris TR (2013) {Asymptotic safety in the f(R) approximation}. JHEP
  01:108, \doi{10.1007/JHEP01(2013)108}, \eprint{1211.0955}

\bibitem[{Christiansen et~al(2016)Christiansen, Knorr, Pawlowski, and
  Rodigast}]{Christiansen:2014raa}
Christiansen N, Knorr B, Pawlowski JM, Rodigast A (2016) {Global Flows in
  Quantum Gravity}. Phys Rev D 93(4):044,036, \doi{10.1103/PhysRevD.93.044036},
  \eprint{1403.1232}

\bibitem[{Becker and Reuter(2014)}]{Becker:2014qya}
Becker D, Reuter M (2014) {En route to Background Independence: Broken
  split-symmetry, and how to restore it with bi-metric average actions}. Annals
  Phys 350:225--301, \doi{10.1016/j.aop.2014.07.023}, \eprint{1404.4537}

\bibitem[{Falls et~al(2016)Falls, Litim, Nikolakopoulos, and
  Rahmede}]{Falls:2014tra}
Falls K, Litim DF, Nikolakopoulos K, Rahmede C (2016) {Further evidence for
  asymptotic safety of quantum gravity}. Phys Rev D 93(10):104,022,
  \doi{10.1103/PhysRevD.93.104022}, \eprint{1410.4815}

\bibitem[{Gies et~al(2015)Gies, Knorr, and Lippoldt}]{Gies:2015tca}
Gies H, Knorr B, Lippoldt S (2015) {Generalized Parametrization Dependence in
  Quantum Gravity}. Phys Rev D 92(8):084,020, \doi{10.1103/PhysRevD.92.084020},
  \eprint{1507.08859}

\bibitem[{Christiansen et~al(2015)Christiansen, Knorr, Meibohm, Pawlowski, and
  Reichert}]{Christiansen:2015rva}
Christiansen N, Knorr B, Meibohm J, Pawlowski JM, Reichert M (2015) {Local
  Quantum Gravity}. Phys Rev D 92(12):121,501,
  \doi{10.1103/PhysRevD.92.121501}, \eprint{1506.07016}

\bibitem[{Demmel et~al(2015)Demmel, Saueressig, and Zanusso}]{Demmel:2015oqa}
Demmel M, Saueressig F, Zanusso O (2015) {A proper fixed functional for
  four-dimensional Quantum Einstein Gravity}. JHEP 08:113,
  \doi{10.1007/JHEP08(2015)113}, \eprint{1504.07656}

\bibitem[{Ohta et~al(2016)Ohta, Percacci, and Vacca}]{Ohta:2015fcu}
Ohta N, Percacci R, Vacca GP (2016) {Renormalization Group Equation and scaling
  solutions for f(R) gravity in exponential parametrization}. Eur Phys J C
  76(2):46, \doi{10.1140/epjc/s10052-016-3895-1}, \eprint{1511.09393}

\bibitem[{Gies et~al(2016)Gies, Knorr, Lippoldt, and Saueressig}]{Gies:2016con}
Gies H, Knorr B, Lippoldt S, Saueressig F (2016) {Gravitational Two-Loop
  Counterterm Is Asymptotically Safe}. Phys Rev Lett 116(21):211,302,
  \doi{10.1103/PhysRevLett.116.211302}, \eprint{1601.01800}

\bibitem[{Denz et~al(2018)Denz, Pawlowski, and Reichert}]{Denz:2016qks}
Denz T, Pawlowski JM, Reichert M (2018) {Towards apparent convergence in
  asymptotically safe quantum gravity}. Eur Phys J C 78(4):336,
  \doi{10.1140/epjc/s10052-018-5806-0}, \eprint{1612.07315}

\bibitem[{Christiansen et~al(2018)Christiansen, Falls, Pawlowski, and
  Reichert}]{Christiansen:2017bsy}
Christiansen N, Falls K, Pawlowski JM, Reichert M (2018) {Curvature dependence
  of quantum gravity}. Phys Rev D 97(4):046,007,
  \doi{10.1103/PhysRevD.97.046007}, \eprint{1711.09259}

\bibitem[{Knorr and Lippoldt(2017)}]{Knorr:2017fus}
Knorr B, Lippoldt S (2017) {Correlation functions on a curved background}. Phys
  Rev D 96(6):065,020, \doi{10.1103/PhysRevD.96.065020}, \eprint{1707.01397}

\bibitem[{Gonzalez-Martin et~al(2017)Gonzalez-Martin, Morris, and
  Slade}]{Gonzalez-Martin:2017gza}
Gonzalez-Martin S, Morris TR, Slade ZH (2017) {Asymptotic solutions in
  asymptotic safety}. Phys Rev D 95(10):106,010,
  \doi{10.1103/PhysRevD.95.106010}, \eprint{1704.08873}

\bibitem[{Falls et~al(2019)Falls, Litim, and Schr\"oder}]{Falls:2018ylp}
Falls KG, Litim DF, Schr\"oder J (2019) {Aspects of asymptotic safety for
  quantum gravity}. Phys Rev D 99(12):126,015,
  \doi{10.1103/PhysRevD.99.126015}, \eprint{1810.08550}

\bibitem[{De~Brito et~al(2018)De~Brito, Ohta, Pereira, Tomaz, and
  Yamada}]{DeBrito:2018hur}
De~Brito GP, Ohta N, Pereira AD, Tomaz AA, Yamada M (2018) {Asymptotic safety
  and field parametrization dependence in the $f(R)$ truncation}. Phys Rev D
  98(2):026,027, \doi{10.1103/PhysRevD.98.026027}, \eprint{1805.09656}

\bibitem[{Knorr(2021)}]{Knorr:2021slg}
Knorr B (2021) {The derivative expansion in asymptotically safe quantum
  gravity: general setup and quartic order}. SciPost Phys Core 4:020,
  \doi{10.21468/SciPostPhysCore.4.3.020}, \eprint{2104.11336}

\bibitem[{Bonanno et~al(2020)Bonanno, Eichhorn, Gies, Pawlowski, Percacci,
  Reuter, Saueressig, and Vacca}]{Bonanno:2020bil}
Bonanno A, Eichhorn A, Gies H, Pawlowski JM, Percacci R, Reuter M, Saueressig
  F, Vacca GP (2020) {Critical reflections on asymptotically safe gravity}.
  Front in Phys 8:269, \doi{10.3389/fphy.2020.00269}, \eprint{2004.06810}

\bibitem[{Narain and Percacci(2010)}]{Narain:2009fy}
Narain G, Percacci R (2010) {Renormalization Group Flow in Scalar-Tensor
  Theories. I}. Class Quant Grav 27:075,001,
  \doi{10.1088/0264-9381/27/7/075001}, \eprint{0911.0386}

\bibitem[{Dona and Percacci(2013)}]{Dona:2012am}
Dona P, Percacci R (2013) {Functional renormalization with fermions and
  tetrads}. Phys Rev D 87(4):045,002, \doi{10.1103/PhysRevD.87.045002},
  \eprint{1209.3649}

\bibitem[{Don\`a et~al(2014)Don\`a, Eichhorn, and Percacci}]{Dona:2013qba}
Don\`a P, Eichhorn A, Percacci R (2014) {Matter matters in asymptotically safe
  quantum gravity}. Phys Rev D 89(8):084,035, \doi{10.1103/PhysRevD.89.084035},
  \eprint{1311.2898}

\bibitem[{Percacci and Vacca(2015)}]{Percacci:2015wwa}
Percacci R, Vacca GP (2015) {Search of scaling solutions in scalar-tensor
  gravity}. Eur Phys J C 75(5):188, \doi{10.1140/epjc/s10052-015-3410-0},
  \eprint{1501.00888}

\bibitem[{Meibohm et~al(2016)Meibohm, Pawlowski, and
  Reichert}]{Meibohm:2015twa}
Meibohm J, Pawlowski JM, Reichert M (2016) {Asymptotic safety of gravity-matter
  systems}. Phys Rev D 93(8):084,035, \doi{10.1103/PhysRevD.93.084035},
  \eprint{1510.07018}

\bibitem[{Labus et~al(2016)Labus, Percacci, and Vacca}]{Labus:2015ska}
Labus P, Percacci R, Vacca GP (2016) {Asymptotic safety in $O(N)$ scalar models
  coupled to gravity}. Phys Lett B 753:274--281,
  \doi{10.1016/j.physletb.2015.12.022}, \eprint{1505.05393}

\bibitem[{Don\`a et~al(2016)Don\`a, Eichhorn, Labus, and
  Percacci}]{Dona:2015tnf}
Don\`a P, Eichhorn A, Labus P, Percacci R (2016) {Asymptotic safety in an
  interacting system of gravity and scalar matter}. Phys Rev D 93(4):044,049,
  \doi{10.1103/PhysRevD.93.129904}, [Erratum: Phys.Rev.D 93, 129904 (2016)],
  \eprint{1512.01589}

\bibitem[{Meibohm and Pawlowski(2016)}]{Meibohm:2016mkp}
Meibohm J, Pawlowski JM (2016) {Chiral fermions in asymptotically safe quantum
  gravity}. Eur Phys J C 76(5):285, \doi{10.1140/epjc/s10052-016-4132-7},
  \eprint{1601.04597}

\bibitem[{Biemans et~al(2017)Biemans, Platania, and
  Saueressig}]{Biemans:2017zca}
Biemans J, Platania A, Saueressig F (2017) {Renormalization group fixed points
  of foliated gravity-matter systems}. JHEP 05:093,
  \doi{10.1007/JHEP05(2017)093}, \eprint{1702.06539}

\bibitem[{Christiansen et~al(2018)Christiansen, Litim, Pawlowski, and
  Reichert}]{Christiansen:2017cxa}
Christiansen N, Litim DF, Pawlowski JM, Reichert M (2018) {Asymptotic safety of
  gravity with matter}. Phys Rev D 97(10):106,012,
  \doi{10.1103/PhysRevD.97.106012}, \eprint{1710.04669}

\bibitem[{Alkofer and Saueressig(2018)}]{Alkofer:2018fxj}
Alkofer N, Saueressig F (2018) {Asymptotically safe $f(R)$-gravity coupled to
  matter I: the polynomial case}. Annals Phys 396:173--201,
  \doi{10.1016/j.aop.2018.07.017}, \eprint{1802.00498}

\bibitem[{Eichhorn et~al(2018)Eichhorn, Labus, Pawlowski, and
  Reichert}]{Eichhorn:2018akn}
Eichhorn A, Labus P, Pawlowski JM, Reichert M (2018) {Effective universality in
  quantum gravity}. SciPost Phys 5(4):031, \doi{10.21468/SciPostPhys.5.4.031},
  \eprint{1804.00012}

\bibitem[{Eichhorn et~al(2019{\natexlab{a}})Eichhorn, Lippoldt, Pawlowski,
  Reichert, and Schiffer}]{Eichhorn:2018ydy}
Eichhorn A, Lippoldt S, Pawlowski JM, Reichert M, Schiffer M
  (2019{\natexlab{a}}) {How perturbative is quantum gravity?} Phys Lett B
  792:310--314, \doi{10.1016/j.physletb.2019.01.071}, \eprint{1810.02828}

\bibitem[{Eichhorn et~al(2019{\natexlab{b}})Eichhorn, Lippoldt, and
  Schiffer}]{Eichhorn:2018nda}
Eichhorn A, Lippoldt S, Schiffer M (2019{\natexlab{b}}) {Zooming in on fermions
  and quantum gravity}. Phys Rev D 99(8):086,002,
  \doi{10.1103/PhysRevD.99.086002}, \eprint{1812.08782}

\bibitem[{B\"urger et~al(2019)B\"urger, Pawlowski, Reichert, and
  Schaefer}]{Burger:2019upn}
B\"urger B, Pawlowski JM, Reichert M, Schaefer BJ (2019) {Curvature dependence
  of quantum gravity with scalars} \eprint{1912.01624}

\bibitem[{Daas et~al(2020)Daas, Oosters, Saueressig, and Wang}]{Daas:2020dyo}
Daas J, Oosters W, Saueressig F, Wang J (2020) {Asymptotically safe gravity
  with fermions}. Phys Lett B 809:135,775,
  \doi{10.1016/j.physletb.2020.135775}, \eprint{2005.12356}

\bibitem[{Daas et~al(2021)Daas, Oosters, Saueressig, and Wang}]{Daas:2021abx}
Daas J, Oosters W, Saueressig F, Wang J (2021) {Asymptotically Safe
  Gravity-Fermion Systems on Curved Backgrounds}. Universe 7(8):306,
  \doi{10.3390/universe7080306}, \eprint{2107.01071}

\bibitem[{Oda and Yamada(2016)}]{Oda:2015sma}
Oda Ky, Yamada M (2016) {Non-minimal coupling in Higgs\textendash{}Yukawa model
  with asymptotically safe gravity}. Class Quant Grav 33(12):125,011,
  \doi{10.1088/0264-9381/33/12/125011}, \eprint{1510.03734}

\bibitem[{Eichhorn and Lippoldt(2017)}]{Eichhorn:2016vvy}
Eichhorn A, Lippoldt S (2017) {Quantum gravity and Standard-Model-like
  fermions}. Phys Lett B 767:142--146, \doi{10.1016/j.physletb.2017.01.064},
  \eprint{1611.05878}

\bibitem[{Hamada and Yamada(2017)}]{Hamada:2017rvn}
Hamada Y, Yamada M (2017) {Asymptotic safety of higher derivative quantum
  gravity non-minimally coupled with a matter system}. JHEP 08:070,
  \doi{10.1007/JHEP08(2017)070}, \eprint{1703.09033}

\bibitem[{Eichhorn et~al(2018)Eichhorn, Lippoldt, and
  Skrinjar}]{Eichhorn:2017sok}
Eichhorn A, Lippoldt S, Skrinjar V (2018) {Nonminimal hints for asymptotic
  safety}. Phys Rev D 97(2):026,002, \doi{10.1103/PhysRevD.97.026002},
  \eprint{1710.03005}

\bibitem[{Laporte et~al(2021)Laporte, Pereira, Saueressig, and
  Wang}]{Laporte:2021kyp}
Laporte C, Pereira AD, Saueressig F, Wang J (2021) {Scalar-tensor theories
  within Asymptotic Safety}. JHEP 12:001, \doi{10.1007/JHEP12(2021)001},
  \eprint{2110.09566}

\bibitem[{Knorr(2022)}]{Knorr:2022ilz}
Knorr B (2022) {Safe essential scalar-tensor theories} \eprint{2204.08564}

\bibitem[{Wetterich and Yamada(2019)}]{Wetterich:2019zdo}
Wetterich C, Yamada M (2019) {Variable Planck mass from the gauge invariant
  flow equation}. Phys Rev D 100(6):066,017, \doi{10.1103/PhysRevD.100.066017},
  \eprint{1906.01721}

\bibitem[{Kabat(1995)}]{Kabat:1995eq}
Kabat DN (1995) {Black hole entropy and entropy of entanglement}. Nucl Phys B
  453:281--299, \doi{10.1016/0550-3213(95)00443-V}, \eprint{hep-th/9503016}

\bibitem[{Larsen and Wilczek(1996)}]{Larsen:1995ax}
Larsen F, Wilczek F (1996) {Renormalization of black hole entropy and of the
  gravitational coupling constant}. Nucl Phys B 458:249--266,
  \doi{10.1016/0550-3213(95)00548-X}, \eprint{hep-th/9506066}

\bibitem[{Narain and Rahmede(2010)}]{Narain:2009gb}
Narain G, Rahmede C (2010) {Renormalization Group Flow in Scalar-Tensor
  Theories. II}. Class Quant Grav 27:075,002,
  \doi{10.1088/0264-9381/27/7/075002}, \eprint{0911.0394}

\bibitem[{Eichhorn and Pauly(2021)}]{Eichhorn:2020sbo}
Eichhorn A, Pauly M (2021) {Constraining power of asymptotic safety for scalar
  fields}. Phys Rev D 103(2):026,006, \doi{10.1103/PhysRevD.103.026006},
  \eprint{2009.13543}

\bibitem[{Pastor-Guti\'errez et~al(2022)Pastor-Guti\'errez, Pawlowski, and
  Reichert}]{Pastor-Gutierrez:2022nki}
Pastor-Guti\'errez A, Pawlowski JM, Reichert M (2022) {The Asymptotically Safe
  Standard Model: From quantum gravity to dynamical chiral symmetry breaking}
  \eprint{2207.09817}

\bibitem[{Don\`a et~al(2015)Don\`a, Eichhorn, and Percacci}]{Dona:2014pla}
Don\`a P, Eichhorn A, Percacci R (2015) {Consistency of matter models with
  asymptotically safe quantum gravity}. Can J Phys 93(9):988--994,
  \doi{10.1139/cjp-2014-0574}, \eprint{1410.4411}

\bibitem[{Eichhorn and Held(2018)}]{Eichhorn:2017ylw}
Eichhorn A, Held A (2018) {Top mass from asymptotic safety}. Phys Lett B
  777:217--221, \doi{10.1016/j.physletb.2017.12.040}, \eprint{1707.01107}

\bibitem[{Eichhorn et~al(2016)Eichhorn, Held, and Pawlowski}]{Eichhorn:2016esv}
Eichhorn A, Held A, Pawlowski JM (2016) {Quantum-gravity effects on a
  Higgs-Yukawa model}. Phys Rev D 94(10):104,027,
  \doi{10.1103/PhysRevD.94.104027}, \eprint{1604.02041}

\bibitem[{Eichhorn and Held(2017)}]{Eichhorn:2017eht}
Eichhorn A, Held A (2017) {Viability of quantum-gravity induced ultraviolet
  completions for matter}. Phys Rev D 96(8):086,025,
  \doi{10.1103/PhysRevD.96.086025}, \eprint{1705.02342}

\bibitem[{Sirunyan et~al(2018)}]{CMS:2018uxb}
Sirunyan AM, et~al (2018) {Observation of $\mathrm{t\overline{t}}$H
  production}. Phys Rev Lett 120(23):231,801,
  \doi{10.1103/PhysRevLett.120.231801}, \eprint{1804.02610}

\bibitem[{Aaboud et~al(2018)}]{ATLAS:2018mme}
Aaboud M, et~al (2018) {Observation of Higgs boson production in association
  with a top quark pair at the LHC with the ATLAS detector}. Phys Lett B
  784:173--191, \doi{10.1016/j.physletb.2018.07.035}, \eprint{1806.00425}

\bibitem[{Sirunyan et~al(2018)}]{CMS:2018nsn}
Sirunyan AM, et~al (2018) {Observation of Higgs boson decay to bottom quarks}.
  Phys Rev Lett 121(12):121,801, \doi{10.1103/PhysRevLett.121.121801},
  \eprint{1808.08242}

\bibitem[{Aaboud et~al(2018)}]{ATLAS:2018kot}
Aaboud M, et~al (2018) {Observation of $H \rightarrow b\bar{b}$ decays and $VH$
  production with the ATLAS detector}. Phys Lett B 786:59--86,
  \doi{10.1016/j.physletb.2018.09.013}, \eprint{1808.08238}

\bibitem[{Aad et~al(2015)}]{ATLAS:2015xst}
Aad G, et~al (2015) {Evidence for the Higgs-boson Yukawa coupling to tau
  leptons with the ATLAS detector}. JHEP 04:117, \doi{10.1007/JHEP04(2015)117},
  \eprint{1501.04943}

\bibitem[{Sirunyan et~al(2018)}]{CMS:2017zyp}
Sirunyan AM, et~al (2018) {Observation of the Higgs boson decay to a pair of
  $\tau$ leptons with the CMS detector}. Phys Lett B 779:283--316,
  \doi{10.1016/j.physletb.2018.02.004}, \eprint{1708.00373}

\bibitem[{Banks and Dixon(1988)}]{Banks:1988yz}
Banks T, Dixon LJ (1988) {Constraints on String Vacua with Space-Time
  Supersymmetry}. Nucl Phys B 307:93--108, \doi{10.1016/0550-3213(88)90523-8}

\bibitem[{Banks and Seiberg(2011)}]{Banks:2010zn}
Banks T, Seiberg N (2011) {Symmetries and Strings in Field Theory and Gravity}.
  Phys Rev D 83:084,019, \doi{10.1103/PhysRevD.83.084019}, \eprint{1011.5120}

\bibitem[{Daus et~al(2020)Daus, Hebecker, Leonhardt, and
  March-Russell}]{Daus:2020vtf}
Daus T, Hebecker A, Leonhardt S, March-Russell J (2020) {Towards a Swampland
  Global Symmetry Conjecture using weak gravity}. Nucl Phys B 960:115,167,
  \doi{10.1016/j.nuclphysb.2020.115167}, \eprint{2002.02456}

\bibitem[{Borissova and Eichhorn(2021)}]{Borissova:2020knn}
Borissova JN, Eichhorn A (2021) {Towards black-hole singularity-resolution in
  the Lorentzian gravitational path integral}. Universe 7(3):48,
  \doi{10.3390/universe7030048}, \eprint{2012.08570}

\bibitem[{Bonanno and Reuter(2006)}]{Bonanno:2006eu}
Bonanno A, Reuter M (2006) {Spacetime structure of an evaporating black hole in
  quantum gravity}. Phys Rev D 73:083,005, \doi{10.1103/PhysRevD.73.083005},
  \eprint{hep-th/0602159}

\bibitem[{Falls and Litim(2014)}]{Falls:2012nd}
Falls K, Litim DF (2014) {Black hole thermodynamics under the microscope}. Phys
  Rev D 89:084,002, \doi{10.1103/PhysRevD.89.084002}, \eprint{1212.1821}

\bibitem[{Susskind(1995)}]{Susskind:1995da}
Susskind L (1995) {Trouble for remnants} \eprint{hep-th/9501106}

\bibitem[{de~Brito et~al(2021)de~Brito, Eichhorn, and Santos}]{deBrito:2021pyi}
de~Brito GP, Eichhorn A, Santos RRLd (2021) {The weak-gravity bound and the
  need for spin in asymptotically safe matter-gravity models}. JHEP 11:110,
  \doi{10.1007/JHEP11(2021)110}, \eprint{2107.03839}

\bibitem[{Eichhorn et~al(2018)Eichhorn, Hamada, Lumma, and
  Yamada}]{Eichhorn:2017als}
Eichhorn A, Hamada Y, Lumma J, Yamada M (2018) {Quantum gravity fluctuations
  flatten the Planck-scale Higgs potential}. Phys Rev D 97(8):086,004,
  \doi{10.1103/PhysRevD.97.086004}, \eprint{1712.00319}

\bibitem[{Eichhorn(2012)}]{Eichhorn:2012va}
Eichhorn A (2012) {Quantum-gravity-induced matter self-interactions in the
  asymptotic-safety scenario}. Phys Rev D 86:105,021,
  \doi{10.1103/PhysRevD.86.105021}, \eprint{1204.0965}

\bibitem[{Eichhorn(2013)}]{Eichhorn:2013ug}
Eichhorn A (2013) {Faddeev-Popov ghosts in quantum gravity beyond perturbation
  theory}. Phys Rev D 87(12):124,016, \doi{10.1103/PhysRevD.87.124016},
  \eprint{1301.0632}

\bibitem[{Ali et~al(2021)Ali, Eichhorn, Pauly, and Scherer}]{Ali:2020znq}
Ali P, Eichhorn A, Pauly M, Scherer MM (2021) {Constraints on discrete global
  symmetries in quantum gravity}. JHEP 05:036, \doi{10.1007/JHEP05(2021)036},
  \eprint{2012.07868}

\bibitem[{Eichhorn and Gies(2011)}]{Eichhorn:2011pc}
Eichhorn A, Gies H (2011) {Light fermions in quantum gravity}. New J Phys
  13:125,012, \doi{10.1088/1367-2630/13/12/125012}, \eprint{1104.5366}

\bibitem[{de~Brito et~al(2021)de~Brito, Eichhorn, and
  Schiffer}]{deBrito:2020dta}
de~Brito GP, Eichhorn A, Schiffer M (2021) {Light charged fermions in quantum
  gravity}. Phys Lett B 815:136,128, \doi{10.1016/j.physletb.2021.136128},
  \eprint{2010.00605}

\bibitem[{Eichhorn et~al(2022)Eichhorn, Kwapisz, and
  Schiffer}]{Eichhorn:2021qet}
Eichhorn A, Kwapisz JH, Schiffer M (2022) {Weak-gravity bound in asymptotically
  safe gravity-gauge systems}. Phys Rev D 105(10):106,022,
  \doi{10.1103/PhysRevD.105.106022}, \eprint{2112.09772}

\bibitem[{Christiansen and Eichhorn(2017)}]{Christiansen:2017gtg}
Christiansen N, Eichhorn A (2017) {An asymptotically safe solution to the U(1)
  triviality problem}. Phys Lett B 770:154--160,
  \doi{10.1016/j.physletb.2017.04.047}, \eprint{1702.07724}

\bibitem[{Eichhorn and Schiffer(2019)}]{Eichhorn:2019yzm}
Eichhorn A, Schiffer M (2019) {$d=4$ as the critical dimensionality of
  asymptotically safe interactions}. Phys Lett B 793:383--389,
  \doi{10.1016/j.physletb.2019.05.005}, \eprint{1902.06479}

\bibitem[{Eichhorn(2020)}]{Eichhorn:2020mte}
Eichhorn A (2020) {Asymptotically safe gravity}. In: {57th International School
  of Subnuclear Physics}: {In Search for the Unexpected}, \eprint{2003.00044}

\bibitem[{de~Brito et~al(to appear)de~Brito, Knorr, and
  Schiffer}]{deBrito2022WIP}
de~Brito G, Knorr B, Schiffer M (to appear)

\bibitem[{Laporte et~al(2022)Laporte, Locht, Pereira, and
  Saueressig}]{Laporte:2022ziz}
Laporte C, Locht N, Pereira AD, Saueressig F (2022) {Evidence for a novel
  shift-symmetric universality class from the functional renormalization group}
  \eprint{2207.06749}

\bibitem[{Gies and Wetterich(2002)}]{Gies:2001nw}
Gies H, Wetterich C (2002) {Renormalization flow of bound states}. Phys Rev D
  65:065,001, \doi{10.1103/PhysRevD.65.065001}, \eprint{hep-th/0107221}

\bibitem[{Braun(2012)}]{Braun:2011pp}
Braun J (2012) {Fermion Interactions and Universal Behavior in Strongly
  Interacting Theories}. J Phys G 39:033,001,
  \doi{10.1088/0954-3899/39/3/033001}, \eprint{1108.4449}

\bibitem[{Braun and Gies(2007)}]{Braun:2005uj}
Braun J, Gies H (2007) {Running coupling at finite temperature and chiral
  symmetry restoration in QCD}. Phys Lett B 645:53--58,
  \doi{10.1016/j.physletb.2006.11.059}, \eprint{hep-ph/0512085}

\bibitem[{Braun(2006)}]{Braun:2006wu}
Braun J (2006) {Chiral Phase Boundary of QCD from the Functional
  Renormalization Group}. In: {ECT* School on Renormalization Group and
  Effective Field Theory Approaches to Many-Body Systems},
  \eprint{hep-ph/0611145}

\bibitem[{Hamada et~al(2021)Hamada, Pawlowski, and Yamada}]{Hamada:2020mug}
Hamada Y, Pawlowski JM, Yamada M (2021) {Gravitational instantons and anomalous
  chiral symmetry breaking}. Phys Rev D 103(10):106,016,
  \doi{10.1103/PhysRevD.103.106016}, \eprint{2009.08728}

\bibitem[{Inagaki et~al(1997)Inagaki, Muta, and Odintsov}]{Inagaki:1997kz}
Inagaki T, Muta T, Odintsov SD (1997) {Dynamical symmetry breaking in curved
  space-time: Four fermion interactions}. Prog Theor Phys Suppl 127:93,
  \doi{10.1143/PTPS.127.93}, \eprint{hep-th/9711084}

\bibitem[{Ebert et~al(2009)Ebert, Tyukov, and Zhukovsky}]{Ebert:2008pc}
Ebert D, Tyukov AV, Zhukovsky VC (2009) {Gravitational catalysis of chiral and
  color symmetry breaking of quark matter in hyperbolic space}. Phys Rev D
  80:085,019, \doi{10.1103/PhysRevD.80.085019}, \eprint{0808.2961}

\bibitem[{Gies and Martini(2018)}]{Gies:2018jnv}
Gies H, Martini R (2018) {Curvature bound from gravitational catalysis}. Phys
  Rev D 97(8):085,017, \doi{10.1103/PhysRevD.97.085017}, \eprint{1802.02865}

\bibitem[{Gies and Salek(2021)}]{Gies:2021upb}
Gies H, Salek AS (2021) {Curvature bound from gravitational catalysis in
  thermal backgrounds}. Phys Rev D 103(12):125,027,
  \doi{10.1103/PhysRevD.103.125027}, \eprint{2103.05542}

\bibitem[{Lauscher and Reuter(2005)}]{Lauscher:2005qz}
Lauscher O, Reuter M (2005) {Fractal spacetime structure in asymptotically safe
  gravity}. JHEP 10:050, \doi{10.1088/1126-6708/2005/10/050},
  \eprint{hep-th/0508202}

\bibitem[{Reuter and Saueressig(2011)}]{Reuter:2011ah}
Reuter M, Saueressig F (2011) {Fractal space-times under the microscope: A
  Renormalization Group view on Monte Carlo data}. JHEP 12:012,
  \doi{10.1007/JHEP12(2011)012}, \eprint{1110.5224}

\bibitem[{Calcagni et~al(2013)Calcagni, Eichhorn, and
  Saueressig}]{Calcagni:2013vsa}
Calcagni G, Eichhorn A, Saueressig F (2013) {Probing the quantum nature of
  spacetime by diffusion}. Phys Rev D 87(12):124,028,
  \doi{10.1103/PhysRevD.87.124028}, \eprint{1304.7247}

\bibitem[{Fischer and Litim(2006)}]{Fischer:2006fz}
Fischer P, Litim DF (2006) {Fixed points of quantum gravity in extra
  dimensions}. Phys Lett B 638:497--502, \doi{10.1016/j.physletb.2006.05.073},
  \eprint{hep-th/0602203}

\bibitem[{Ohta and Percacci(2014)}]{Ohta:2013uca}
Ohta N, Percacci R (2014) {Higher Derivative Gravity and Asymptotic Safety in
  Diverse Dimensions}. Class Quant Grav 31:015,024,
  \doi{10.1088/0264-9381/31/1/015024}, \eprint{1308.3398}

\bibitem[{Schiffer(2021)}]{Schiffer:2021gwl}
Schiffer M (2021) {Probing Quantum Gravity: Theoretical and phenomenological
  consistency tests of asymptotically safe quantum gravity}. PhD thesis, U.
  Heidelberg (main), \doi{10.11588/heidok.00030645}

\bibitem[{Arkani-Hamed et~al(1998)Arkani-Hamed, Dimopoulos, and
  Dvali}]{Arkani-Hamed:1998jmv}
Arkani-Hamed N, Dimopoulos S, Dvali GR (1998) {The Hierarchy problem and new
  dimensions at a millimeter}. Phys Lett B 429:263--272,
  \doi{10.1016/S0370-2693(98)00466-3}, \eprint{hep-ph/9803315}

\bibitem[{Litim and Plehn(2008)}]{Litim:2007iu}
Litim DF, Plehn T (2008) {Signatures of gravitational fixed points at the LHC}.
  Phys Rev Lett 100:131,301, \doi{10.1103/PhysRevLett.100.131301},
  \eprint{0707.3983}

\bibitem[{Litim and Plehn(2007)}]{Litim:2007ee}
Litim DF, Plehn T (2007) {Virtual gravitons at the LHC}. In: {15th
  International Conference on Supersymmetry and the Unification of Fundamental
  Interactions (SUSY07)}, pp 628--631, \eprint{0710.3096}

\bibitem[{Gerwick et~al(2011)Gerwick, Litim, and Plehn}]{Gerwick:2011jw}
Gerwick E, Litim D, Plehn T (2011) {Asymptotic safety and Kaluza-Klein
  gravitons at the LHC}. Phys Rev D 83:084,048,
  \doi{10.1103/PhysRevD.83.084048}, \eprint{1101.5548}

\bibitem[{Draper et~al(2020)Draper, Knorr, Ripken, and
  Saueressig}]{Draper:2020bop}
Draper T, Knorr B, Ripken C, Saueressig F (2020) {Finite Quantum Gravity
  Amplitudes: No Strings Attached}. Phys Rev Lett 125(18):181,301,
  \doi{10.1103/PhysRevLett.125.181301}, \eprint{2007.00733}

\bibitem[{Dobrich and Eichhorn(2012)}]{Dobrich:2012nv}
Dobrich B, Eichhorn A (2012) {Can we see quantum gravity? Photons in the
  asymptotic-safety scenario}. JHEP 06:156, \doi{10.1007/JHEP06(2012)156},
  \eprint{1203.6366}

\bibitem[{Cabibbo et~al(1979)Cabibbo, Maiani, Parisi, and
  Petronzio}]{Cabibbo:1979ay}
Cabibbo N, Maiani L, Parisi G, Petronzio R (1979) {Bounds on the Fermions and
  Higgs Boson Masses in Grand Unified Theories}. Nucl Phys B 158:295--305,
  \doi{10.1016/0550-3213(79)90167-6}

\bibitem[{Gockeler et~al(1998)Gockeler, Horsley, Linke, Rakow, Schierholz, and
  Stuben}]{Gockeler:1997dn}
Gockeler M, Horsley R, Linke V, Rakow PEL, Schierholz G, Stuben H (1998) {Is
  there a Landau pole problem in QED?} Phys Rev Lett 80:4119--4122,
  \doi{10.1103/PhysRevLett.80.4119}, \eprint{hep-th/9712244}

\bibitem[{Gell-Mann and Low(1954)}]{Gell-Mann:1954yli}
Gell-Mann M, Low FE (1954) {Quantum electrodynamics at small distances}. Phys
  Rev 95:1300--1312, \doi{10.1103/PhysRev.95.1300}

\bibitem[{Gockeler et~al(1998)Gockeler, Horsley, Linke, Rakow, Schierholz, and
  Stuben}]{Gockeler:1997kt}
Gockeler M, Horsley R, Linke V, Rakow PEL, Schierholz G, Stuben H (1998)
  {Resolution of the Landau pole problem in QED}. Nucl Phys B Proc Suppl
  63:694--696, \doi{10.1016/S0920-5632(97)00875-X}, \eprint{hep-lat/9801004}

\bibitem[{Gies and Jaeckel(2004)}]{Gies:2004hy}
Gies H, Jaeckel J (2004) {Renormalization flow of QED}. Phys Rev Lett
  93:110,405, \doi{10.1103/PhysRevLett.93.110405}, \eprint{hep-ph/0405183}

\bibitem[{Daum et~al(2010)Daum, Harst, and Reuter}]{Daum:2009dn}
Daum JE, Harst U, Reuter M (2010) {Running Gauge Coupling in Asymptotically
  Safe Quantum Gravity}. JHEP 01:084, \doi{10.1007/JHEP01(2010)084},
  \eprint{0910.4938}

\bibitem[{Folkerts et~al(2012)Folkerts, Litim, and Pawlowski}]{Folkerts:2011jz}
Folkerts S, Litim DF, Pawlowski JM (2012) {Asymptotic freedom of Yang-Mills
  theory with gravity}. Phys Lett B 709:234--241,
  \doi{10.1016/j.physletb.2012.02.002}, \eprint{1101.5552}

\bibitem[{Harst and Reuter(2011)}]{Harst:2011zx}
Harst U, Reuter M (2011) {QED coupled to QEG}. JHEP 05:119,
  \doi{10.1007/JHEP05(2011)119}, \eprint{1101.6007}

\bibitem[{Eichhorn and Versteegen(2018)}]{Eichhorn:2017lry}
Eichhorn A, Versteegen F (2018) {Upper bound on the Abelian gauge coupling from
  asymptotic safety}. JHEP 01:030, \doi{10.1007/JHEP01(2018)030},
  \eprint{1709.07252}

\bibitem[{De~Brito et~al(2019)De~Brito, Eichhorn, and
  Pereira}]{DeBrito:2019gdd}
De~Brito GP, Eichhorn A, Pereira AD (2019) {A link that matters: Towards
  phenomenological tests of unimodular asymptotic safety}. JHEP 09:100,
  \doi{10.1007/JHEP09(2019)100}, \eprint{1907.11173}

\bibitem[{de~Brito and Eichhorn(2022)}]{deBrito:2022vbr}
de~Brito GP, Eichhorn A (2022) {Nonvanishing gravitational contribution to
  matter beta functions for vanishing dimensionful regulators}
  \eprint{2201.11402}

\bibitem[{Eichhorn and Held(2018)}]{Eichhorn:2018whv}
Eichhorn A, Held A (2018) {Mass difference for charged quarks from
  asymptotically safe quantum gravity}. Phys Rev Lett 121(15):151,302,
  \doi{10.1103/PhysRevLett.121.151302}, \eprint{1803.04027}

\bibitem[{Eichhorn et~al(2018)Eichhorn, Held, and Wetterich}]{Eichhorn:2017muy}
Eichhorn A, Held A, Wetterich C (2018) {Quantum-gravity predictions for the
  fine-structure constant}. Phys Lett B 782:198--201,
  \doi{10.1016/j.physletb.2018.05.016}, \eprint{1711.02949}

\bibitem[{Robinson and Wilczek(2006)}]{Robinson:2005fj}
Robinson SP, Wilczek F (2006) {Gravitational correction to running of gauge
  couplings}. Phys Rev Lett 96:231,601, \doi{10.1103/PhysRevLett.96.231601},
  \eprint{hep-th/0509050}

\bibitem[{Pietrykowski(2007)}]{Pietrykowski:2006xy}
Pietrykowski AR (2007) {Gauge dependence of gravitational correction to running
  of gauge couplings}. Phys Rev Lett 98:061,801,
  \doi{10.1103/PhysRevLett.98.061801}, \eprint{hep-th/0606208}

\bibitem[{Toms(2007)}]{Toms:2007sk}
Toms DJ (2007) {Quantum gravity and charge renormalization}. Phys Rev D
  76:045,015, \doi{10.1103/PhysRevD.76.045015}, \eprint{0708.2990}

\bibitem[{Ebert et~al(2008)Ebert, Plefka, and Rodigast}]{Ebert:2007gf}
Ebert D, Plefka J, Rodigast A (2008) {Absence of gravitational contributions to
  the running Yang-Mills coupling}. Phys Lett B 660:579--582,
  \doi{10.1016/j.physletb.2008.01.037}, \eprint{0710.1002}

\bibitem[{Tang and Wu(2010)}]{Tang:2008ah}
Tang Y, Wu YL (2010) {Gravitational Contributions to the Running of Gauge
  Couplings}. Commun Theor Phys 54:1040--1044, \doi{10.1088/0253-6102/54/6/15},
  \eprint{0807.0331}

\bibitem[{Toms(2010)}]{Toms:2010vy}
Toms DJ (2010) {Quantum gravitational contributions to quantum
  electrodynamics}. Nature 468:56--59, \doi{10.1038/nature09506},
  \eprint{1010.0793}

\bibitem[{Anber et~al(2011)Anber, Donoghue, and El-Houssieny}]{Anber:2010uj}
Anber MM, Donoghue JF, El-Houssieny M (2011) {Running couplings and operator
  mixing in the gravitational corrections to coupling constants}. Phys Rev D
  83:124,003, \doi{10.1103/PhysRevD.83.124003}, \eprint{1011.3229}

\bibitem[{Baldazzi et~al(2021{\natexlab{a}})Baldazzi, Percacci, and
  Zambelli}]{Baldazzi:2020vxk}
Baldazzi A, Percacci R, Zambelli L (2021{\natexlab{a}}) {Functional
  renormalization and the $\overline{\text{MS}}$ scheme}. Phys Rev D
  103(7):076,012, \doi{10.1103/PhysRevD.103.076012}, \eprint{2009.03255}

\bibitem[{Baldazzi et~al(2021{\natexlab{b}})Baldazzi, Percacci, and
  Zambelli}]{Baldazzi:2021guw}
Baldazzi A, Percacci R, Zambelli L (2021{\natexlab{b}}) {Limit of vanishing
  regulator in the functional renormalization group}. Phys Rev D
  104(7):076,026, \doi{10.1103/PhysRevD.104.076026}, \eprint{2105.05778}

\bibitem[{Alkofer et~al(2020)Alkofer, Eichhorn, Held, Nieto, Percacci, and
  Schr\"ofl}]{Alkofer:2020vtb}
Alkofer R, Eichhorn A, Held A, Nieto CM, Percacci R, Schr\"ofl M (2020) {Quark
  masses and mixings in minimally parameterized UV completions of the Standard
  Model}. Annals Phys 421:168,282, \doi{10.1016/j.aop.2020.168282},
  \eprint{2003.08401}

\bibitem[{Kowalska et~al(2022)Kowalska, Pramanick, and
  Sessolo}]{Kowalska:2022ypk}
Kowalska K, Pramanick S, Sessolo EM (2022) {Naturally small Yukawa couplings
  from trans-Planckian asymptotic safety}. JHEP 08:262,
  \doi{10.1007/JHEP08(2022)262}, \eprint{2204.00866}

\bibitem[{Pendleton and Ross(1981)}]{Pendleton:1980as}
Pendleton B, Ross GG (1981) {Mass and Mixing Angle Predictions from Infrared
  Fixed Points}. Phys Lett B 98:291--294, \doi{10.1016/0370-2693(81)90017-4}

\bibitem[{Shaposhnikov and Wetterich(2010)}]{Shaposhnikov:2009pv}
Shaposhnikov M, Wetterich C (2010) {Asymptotic safety of gravity and the Higgs
  boson mass}. Phys Lett B 683:196--200, \doi{10.1016/j.physletb.2009.12.022},
  \eprint{0912.0208}

\bibitem[{Bezrukov and Shaposhnikov(2015)}]{Bezrukov:2014ina}
Bezrukov F, Shaposhnikov M (2015) {Why should we care about the top quark
  Yukawa coupling?} J Exp Theor Phys 120:335--343,
  \doi{10.1134/S1063776115030152}, \eprint{1411.1923}

\bibitem[{Aad et~al(2012)}]{ATLAS:2012yve}
Aad G, et~al (2012) {Observation of a new particle in the search for the
  Standard Model Higgs boson with the ATLAS detector at the LHC}. Phys Lett B
  716:1--29, \doi{10.1016/j.physletb.2012.08.020}, \eprint{1207.7214}

\bibitem[{Chatrchyan et~al(2012)}]{CMS:2012qbp}
Chatrchyan S, et~al (2012) {Observation of a New Boson at a Mass of 125 GeV
  with the CMS Experiment at the LHC}. Phys Lett B 716:30--61,
  \doi{10.1016/j.physletb.2012.08.021}, \eprint{1207.7235}

\bibitem[{Pawlowski et~al(2019)Pawlowski, Reichert, Wetterich, and
  Yamada}]{Pawlowski:2018ixd}
Pawlowski JM, Reichert M, Wetterich C, Yamada M (2019) {Higgs scalar potential
  in asymptotically safe quantum gravity}. Phys Rev D 99(8):086,010,
  \doi{10.1103/PhysRevD.99.086010}, \eprint{1811.11706}

\bibitem[{Wetterich(2021)}]{Wetterich:2019rsn}
Wetterich C (2021) {Effective scalar potential in asymptotically safe quantum
  gravity}. Universe 7(2):45, \doi{10.3390/universe7020045},
  \eprint{1911.06100}

\bibitem[{Eichhorn et~al(2021)Eichhorn, Pauly, and Ray}]{Eichhorn:2021tsx}
Eichhorn A, Pauly M, Ray S (2021) {Towards a Higgs mass determination in
  asymptotically safe gravity with a dark portal}. JHEP 10:100,
  \doi{10.1007/JHEP10(2021)100}, \eprint{2107.07949}

\bibitem[{Kwapisz(2019)}]{Kwapisz:2019wrl}
Kwapisz JH (2019) {Asymptotic safety, the Higgs boson mass, and beyond the
  standard model physics}. Phys Rev D 100(11):115,001,
  \doi{10.1103/PhysRevD.100.115001}, \eprint{1907.12521}

\bibitem[{Wetterich and Yamada(2017)}]{Wetterich:2016uxm}
Wetterich C, Yamada M (2017) {Gauge hierarchy problem in asymptotically safe
  gravity--the resurgence mechanism}. Phys Lett B 770:268--271,
  \doi{10.1016/j.physletb.2017.04.049}, \eprint{1612.03069}

\bibitem[{Bezrukov and Shaposhnikov(2008)}]{Bezrukov:2007ep}
Bezrukov FL, Shaposhnikov M (2008) {The Standard Model Higgs boson as the
  inflaton}. Phys Lett B 659:703--706, \doi{10.1016/j.physletb.2007.11.072},
  \eprint{0710.3755}

\bibitem[{Eichhorn and Pauly(2021)}]{Eichhorn:2020kca}
Eichhorn A, Pauly M (2021) {Safety in darkness: Higgs portal to simple Yukawa
  systems}. Phys Lett B 819:136,455, \doi{10.1016/j.physletb.2021.136455},
  \eprint{2005.03661}

\bibitem[{Reichert and Smirnov(2020)}]{Reichert:2019car}
Reichert M, Smirnov J (2020) {Dark Matter meets Quantum Gravity}. Phys Rev D
  101(6):063,015, \doi{10.1103/PhysRevD.101.063015}, \eprint{1911.00012}

\bibitem[{Hamada et~al(2020)Hamada, Tsumura, and Yamada}]{Hamada:2020vnf}
Hamada Y, Tsumura K, Yamada M (2020) {Scalegenesis and fermionic dark matters
  in the flatland scenario}. Eur Phys J C 80(5):368,
  \doi{10.1140/epjc/s10052-020-7929-3}, \eprint{2002.03666}

\bibitem[{Grabowski et~al(2019)Grabowski, Kwapisz, and
  Meissner}]{Grabowski:2018fjj}
Grabowski F, Kwapisz JH, Meissner KA (2019) {Asymptotic safety and Conformal
  Standard Model}. Phys Rev D 99(11):115,029, \doi{10.1103/PhysRevD.99.115029},
  \eprint{1810.08461}

\bibitem[{Meissner and Nicolai(2007)}]{Meissner:2006zh}
Meissner KA, Nicolai H (2007) {Conformal Symmetry and the Standard Model}. Phys
  Lett B 648:312--317, \doi{10.1016/j.physletb.2007.03.023},
  \eprint{hep-th/0612165}

\bibitem[{Kowalska and Sessolo(2021)}]{Kowalska:2020zve}
Kowalska K, Sessolo EM (2021) {Minimal models for g-2 and dark matter confront
  asymptotic safety}. Phys Rev D 103(11):115,032,
  \doi{10.1103/PhysRevD.103.115032}, \eprint{2012.15200}

\bibitem[{Boos et~al(2022{\natexlab{a}})Boos, Carone, Donald, and
  Musser}]{Boos:2022jvc}
Boos J, Carone CD, Donald NL, Musser MR (2022{\natexlab{a}}) {Asymptotic safety
  and gauged baryon number}. Phys Rev D 106(3):035,015,
  \doi{10.1103/PhysRevD.106.035015}, \eprint{2206.02686}

\bibitem[{Boos et~al(2022{\natexlab{b}})Boos, Carone, Donald, and
  Musser}]{Boos:2022pyq}
Boos J, Carone CD, Donald NL, Musser MR (2022{\natexlab{b}}) {Asymptotically
  safe dark matter with gauged baryon number} \eprint{2209.14268}

\bibitem[{Peccei and Quinn(1977{\natexlab{a}})}]{Peccei:1977hh}
Peccei RD, Quinn HR (1977{\natexlab{a}}) {CP Conservation in the Presence of
  Instantons}. Phys Rev Lett 38:1440--1443, \doi{10.1103/PhysRevLett.38.1440}

\bibitem[{Peccei and Quinn(1977{\natexlab{b}})}]{Peccei:1977ur}
Peccei RD, Quinn HR (1977{\natexlab{b}}) {Constraints Imposed by CP
  Conservation in the Presence of Instantons}. Phys Rev D 16:1791--1797,
  \doi{10.1103/PhysRevD.16.1791}

\bibitem[{Ferreira(2021)}]{Ferreira:2020fam}
Ferreira EGM (2021) {Ultra-light dark matter}. Astron Astrophys Rev 29(1):7,
  \doi{10.1007/s00159-021-00135-6}, \eprint{2005.03254}

\bibitem[{Ringwald(2014)}]{Ringwald:2012cu}
Ringwald A (2014) {Searching for axions and ALPs from string theory}. J Phys
  Conf Ser 485:012,013, \doi{10.1088/1742-6596/485/1/012013},
  \eprint{1209.2299}

\bibitem[{de~Brito et~al(2022)de~Brito, Eichhorn, and Lino~dos
  Santos}]{deBrito:2021akp}
de~Brito GP, Eichhorn A, Lino~dos Santos RR (2022) {Are there ALPs in the
  asymptotically safe landscape?} JHEP 06:013, \doi{10.1007/JHEP06(2022)013},
  \eprint{2112.08972}

\bibitem[{Eichhorn et~al(2020)Eichhorn, Held, and Wetterich}]{Eichhorn:2019dhg}
Eichhorn A, Held A, Wetterich C (2020) {Predictive power of grand unification
  from quantum gravity}. JHEP 08:111, \doi{10.1007/JHEP08(2020)111},
  \eprint{1909.07318}

\bibitem[{Held et~al(2022)Held, Kwapisz, and Sartore}]{Held:2022hnw}
Held A, Kwapisz J, Sartore L (2022) {Grand unification and the Planck scale: an
  SO(10) example of radiative symmetry breaking}. JHEP 08:122,
  \doi{10.1007/JHEP08(2022)122}, \eprint{2204.03001}

\bibitem[{Held(2019)}]{Held:2019vmi}
Held A (2019) {From particle physics to black holes: The predictive power of
  asymptotic safety.} PhD thesis, U. Heidelberg (main),
  \doi{10.11588/heidok.00027607}

\bibitem[{Eichhorn and Held(2022)}]{Eichhorn:2022vgp}
Eichhorn A, Held A (2022) {Dynamically vanishing Dirac neutrino mass from
  quantum scale symmetry} \eprint{2204.09008}

\bibitem[{De~Brito et~al(2019)De~Brito, Hamada, Pereira, and
  Yamada}]{DeBrito:2019rrh}
De~Brito GP, Hamada Y, Pereira AD, Yamada M (2019) {On the impact of Majorana
  masses in gravity-matter systems}. JHEP 08:142,
  \doi{10.1007/JHEP08(2019)142}, \eprint{1905.11114}

\bibitem[{Dom\`enech et~al(2021)Dom\`enech, Goodsell, and
  Wetterich}]{Domenech:2020yjf}
Dom\`enech G, Goodsell M, Wetterich C (2021) {Neutrino masses, vacuum stability
  and quantum gravity prediction for the mass of the top quark}. JHEP 01:180,
  \doi{10.1007/JHEP01(2021)180}, \eprint{2008.04310}

\bibitem[{Kowalska et~al(2021)Kowalska, Sessolo, and
  Yamamoto}]{Kowalska:2020gie}
Kowalska K, Sessolo EM, Yamamoto Y (2021) {Flavor anomalies from asymptotically
  safe gravity}. Eur Phys J C 81(4):272, \doi{10.1140/epjc/s10052-021-09072-1},
  \eprint{2007.03567}

\bibitem[{Chikkaballi et~al(2022)Chikkaballi, Kotlarski, Kowalska, Rizzo, and
  Sessolo}]{Chikkaballi:2022urc}
Chikkaballi A, Kotlarski W, Kowalska K, Rizzo D, Sessolo EM (2022) {Constraints
  on $Z'$ solutions to the flavor anomalies with trans-Planckian asymptotic
  safety} \eprint{2209.07971}

\bibitem[{Cyrol et~al(2016)Cyrol, Fister, Mitter, Pawlowski, and
  Strodthoff}]{Cyrol:2016tym}
Cyrol AK, Fister L, Mitter M, Pawlowski JM, Strodthoff N (2016) {Landau gauge
  Yang-Mills correlation functions}. Phys Rev D 94(5):054,005,
  \doi{10.1103/PhysRevD.94.054005}, \eprint{1605.01856}

\bibitem[{Sakharov(1967)}]{Sakharov:1967dj}
Sakharov AD (1967) {Violation of CP Invariance, C asymmetry, and baryon
  asymmetry of the universe}. Pisma Zh Eksp Teor Fiz 5:32--35,
  \doi{10.1070/PU1991v034n05ABEH002497}

\end{thebibliography}
